\documentclass[10pt]{IEEEtran}
\usepackage{enumitem}
\usepackage{graphicx,multicol} % for pdf, bitmapped graphics files
\usepackage{epsfig} % for postscript graphics files
\usepackage{amsmath,amsthm} % assumes amsmath package installed
\usepackage{amssymb}  % assumes amsmath package installed
\usepackage{float}
\usepackage{cases,setspace,adjustbox}
\usepackage{cite}
\usepackage{algorithmic}
\usepackage[ruled,vlined]{algorithm2e}
\usepackage{array}
\usepackage[update,prepend]{epstopdf}
\usepackage{eqparbox}
\usepackage{subfig}
\usepackage{fixltx2e}
\usepackage{url}
\usepackage{array}
\usepackage{caption}
\usepackage{xcolor}
\usepackage{mathtools}
\usepackage{bbm}
\usepackage{multirow}
\usepackage{multicol}
\usepackage{dblfloatfix}
\usepackage{footnote}

\makesavenoteenv{tabular}
\makesavenoteenv{table}
\newcommand{\vect}[1]{\boldsymbol{#1}}
\usepackage{booktabs}

\DeclareMathAlphabet\mathbfcal{OMS}{cmsy}{b}{n}
\setlist[itemize]{leftmargin=*}
\usepackage[hang,flushmargin]{footmisc}

\newcommand{\pidle}{p_{\text{I}}}
\newcommand{\psucci}[1]{p_{\text{S}}^{(#1)}}
\newcommand{\pbusy}[1]{p_{\text{B}}^{(#1)}}
\newcommand{\pcol}{p_{\text{C}}}
\newcommand{\psucc}{p_{\text{S}}}
\newcommand{\pidleM}[1]{p_{\text{I},{#1}}}
\newcommand{\psucciM}[2]{p_{\text{S},{#1}}^{(#2)}}
\newcommand{\pbusyM}[2]{p_{\text{B},{#1}}^{(#2)}}
\newcommand{\pcolM}[1]{p_{\text{C},{#1}}}
\newcommand{\psuccM}[1]{p_{\text{S},{#1}}}
\newcommand{\lidle}{\sigma_\text{I}}
\newcommand{\lsucc}{\sigma_\text{S}}
\newcommand{\lcol}{\sigma_\text{C}}

\newcommand{\AoI}[1]{\Delta_{#1}}		% age delta_{i}
\newcommand{\initAoIT}[2]{\delta^{-}_{#1{#2}}} 
\newcommand{\AoIT}[2]{{\Delta}_{#1{#2}}}

\newcommand{\AvgAoIT}[2]{\widetilde{\Delta}_{#1{#2}}}
\newcommand{\AvginitAoIT}[2]{\widetilde{\Delta}^{-}{(#1){#2}}}
\newcommand{\AvginitAoI}{\widetilde{\Delta}^{-}}
\newcommand{\AvginitAoITh}[2]{\Theta_{#1{#2}}}

\newcommand{\Thr}[1]{\Gamma_{#1}}

\newcommand{\AvgThr}[1]{\widetilde{\Gamma}_{#1}}

\newcommand{\tauD}{\tau_{\mathrm{A}}}
\newcommand{\tauW}{\tau_{\mathrm{T}}}
\newcommand{\tauDC}{\widehat{\tau}_{A}}
\newcommand{\tauWC}{\widehat{\tau}_{T}}
\newcommand{\AO}{\textrm{AON}}
\newcommand{\TO}{\textrm{TON}}
\newcommand{\uD}[1]{u^{\mathrm{A}}_{#1}}
\newcommand{\uW}[1]{u^{\mathrm{T}}_{#1}}
\newcommand{\nD}{\mathcal{N}_{\mathrm{A}}}
\newcommand{\nW}{\mathcal{N}_{\mathrm{T}}}
\newcommand{\ND}{\mathrm{N}_{\mathrm{A}}}
\newcommand{\NW}{\mathrm{N}_{\mathrm{T}}}
\newcommand{\UD}[1]{\mathrm{U}^{\mathrm{A}}_{#1}}
\newcommand{\UW}[1]{\mathrm{U}^{\mathrm{T}}_{#1}}
\newcommand{\UDC}[1]{\mathrm{U}^{\mathrm{A}}_{#1}}
\newcommand{\UWC}[1]{\mathrm{U}^{\mathrm{T}}_{#1}}
\newcommand{\T}{\mathcal{T}}
\newcommand{\I}{\mathcal{I}}
\newcommand{\PR}{\text{P}_{\text{R}}}

\newtheorem{proposition}{Proposition}
\newtheorem{observation}{Observation}
\newtheorem{statement}{Statement}
\newtheorem{corollary}{Corollary}
\newcommand{\SG}[1]{{\color{black} {#1}}}
\newcommand{\sg}[1]{{\color{black} {#1}}}

\begin{document}
\renewcommand{\arraystretch}{1.15}
\renewcommand{\vec}[1]{\mathbf{#1}}
\title{Coexistence of Age and Throughput Optimizing Networks: A Spectrum Sharing Game}
\author{Sneihil Gopal$^{*}$, Sanjit K. Kaul$^{*}$, Rakesh Chaturvedi$^{\dagger}$ and Sumit Roy$^{\ddagger}$\\
$^{*}$Wireless Systems Lab, IIIT-Delhi, India,
$^{\dagger}$Department of Social Sciences \& Humanities, IIIT-Delhi, India\\
$^{\ddagger}$University of Washington, Seattle, WA\\
\{sneihilg, skkaul, rakesh\}@iiitd.ac.in, sroy@uw.edu}
\maketitle
\begin{abstract}
We investigate the coexistence of an age optimizing network ($\AO$) and a throughput optimizing network ($\TO$) that share a common spectrum band. We consider two modes of long run coexistence: (a) networks {\em compete} with each other for spectrum access, causing them to interfere and (b) networks {\em cooperate} to achieve non-interfering access.

To model competition, we define a non-cooperative stage game parameterized by the average age of the $\AO$ at the beginning of the stage, derive its mixed strategy Nash equilibrium (MSNE), and analyze the evolution of age and throughput over an infinitely repeated game in which each network plays the MSNE at every stage. Cooperation uses a coordination device that performs a coin toss during each stage to select the network that must access the medium. Networks use the grim trigger punishment strategy, reverting to playing the MSNE every stage forever if the other disobeys the device. We determine if there exists a subgame perfect equilibrium, i.e., the networks obey the device forever as they find cooperation beneficial. We show that networks choose to cooperate only when they consist of a sufficiently small number of nodes, otherwise they prefer to disobey the device and compete.
\IEEEpeerreviewmaketitle

\end{abstract}
\section{Introduction}
\label{sec:intro}
The emerging Internet-of-Things (IoT) will require large number of (non-traditional) devices to sense and communicate information (either their own status or that of their proximate environment) to a network coordinator/aggregator or other devices. Applications include real-time monitoring for disaster management, environmental monitoring, industrial control and surveillance~\cite[references therein]{fanet}, which require timely delivery of updates to a central station. Another set of popular applications include vehicular networking for future autonomous operations where each vehicular node broadcasts a vector (e.g. position, velocity and other status information) to enable applications like collision avoidance and vehicle coordination like platooning~\cite{vanet}. 

\begin{figure}[t]
\begin{center}
\includegraphics[width=0.90\columnwidth]{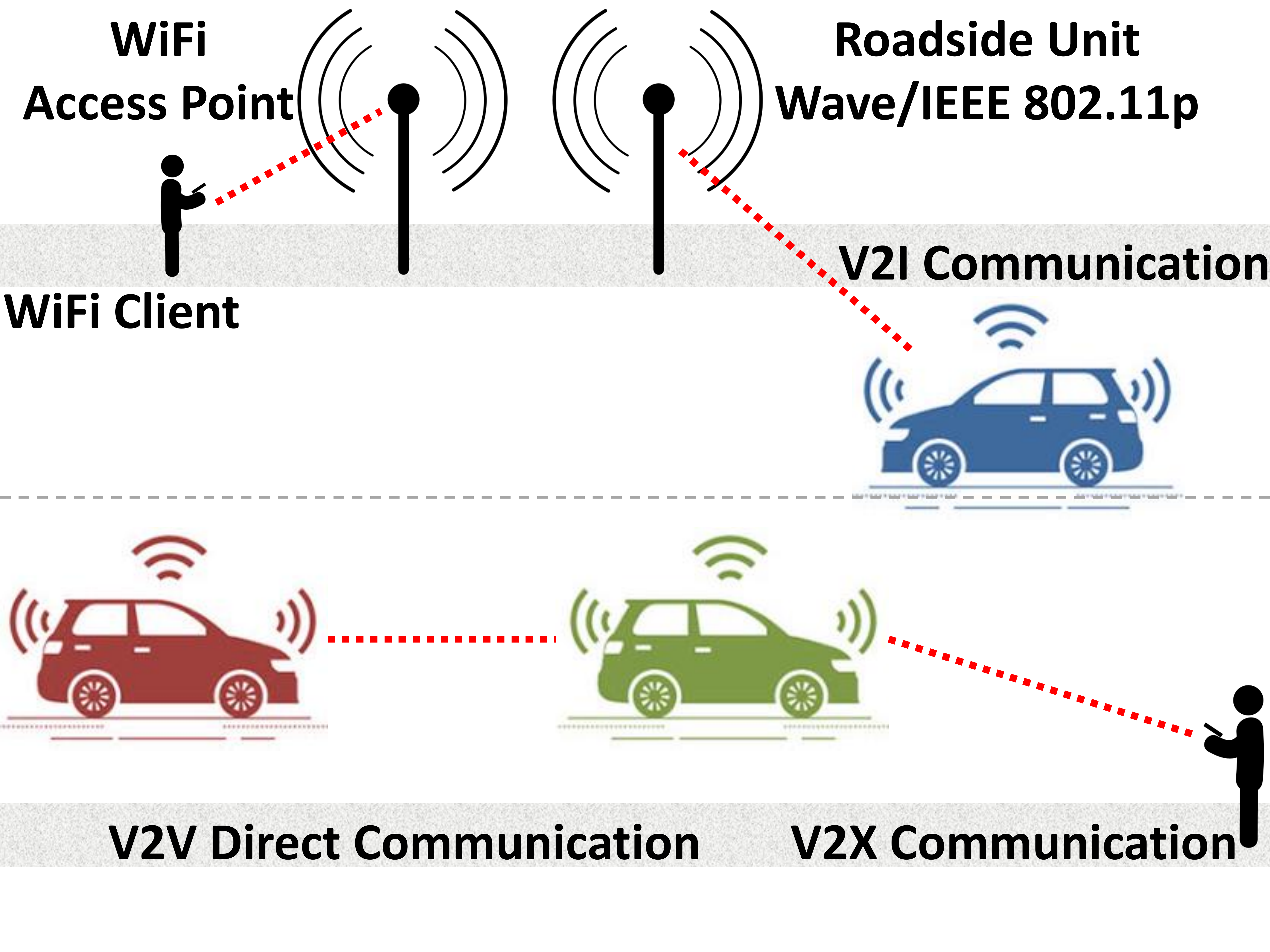}
\caption{\small Example spectrum sharing scenario where a WiFi AP-client link shares the $5.85-5.925$ GHz band with a DSRC-based vehicular network. The band previously reserved for vehicular communication was recently opened by the FCC in the US for use by WiFi ($802.11$ ac) devices~\cite{liu2017}.}
\label{fig:example}%
\end{center}
\end{figure}

In many scenarios, such new IoT networks will use existing (and potentially newly allocated) unlicensed bands, and hence be required to share the spectrum with incumbent networks. For instance, the U.S. Federal Communications Commission (FCC) recently opened up the $5.85-5.925$ GHz band, previously reserved for vehicular dedicated short range communication (DSRC) for use by high throughput WiFi, leading to the need for spectrum sharing between WiFi and vehicular networks~\cite{liu2017}. Figure~\ref{fig:example} provides an illustration in which a WiFi access point communicates with its client in the vicinity of a DSRC-based vehicular network. Similarly, Unmanned Aerial Vehicles (UAVs)~\cite{fanet} equipped with WiFi technology used for (wide-area) environmental monitoring will need to share spectrum with regular terrestrial WiFi networks. 

\begin{figure*}[t]
\centering
\subfloat[Competitive Mode of Coexistence]{\includegraphics[scale=0.24]{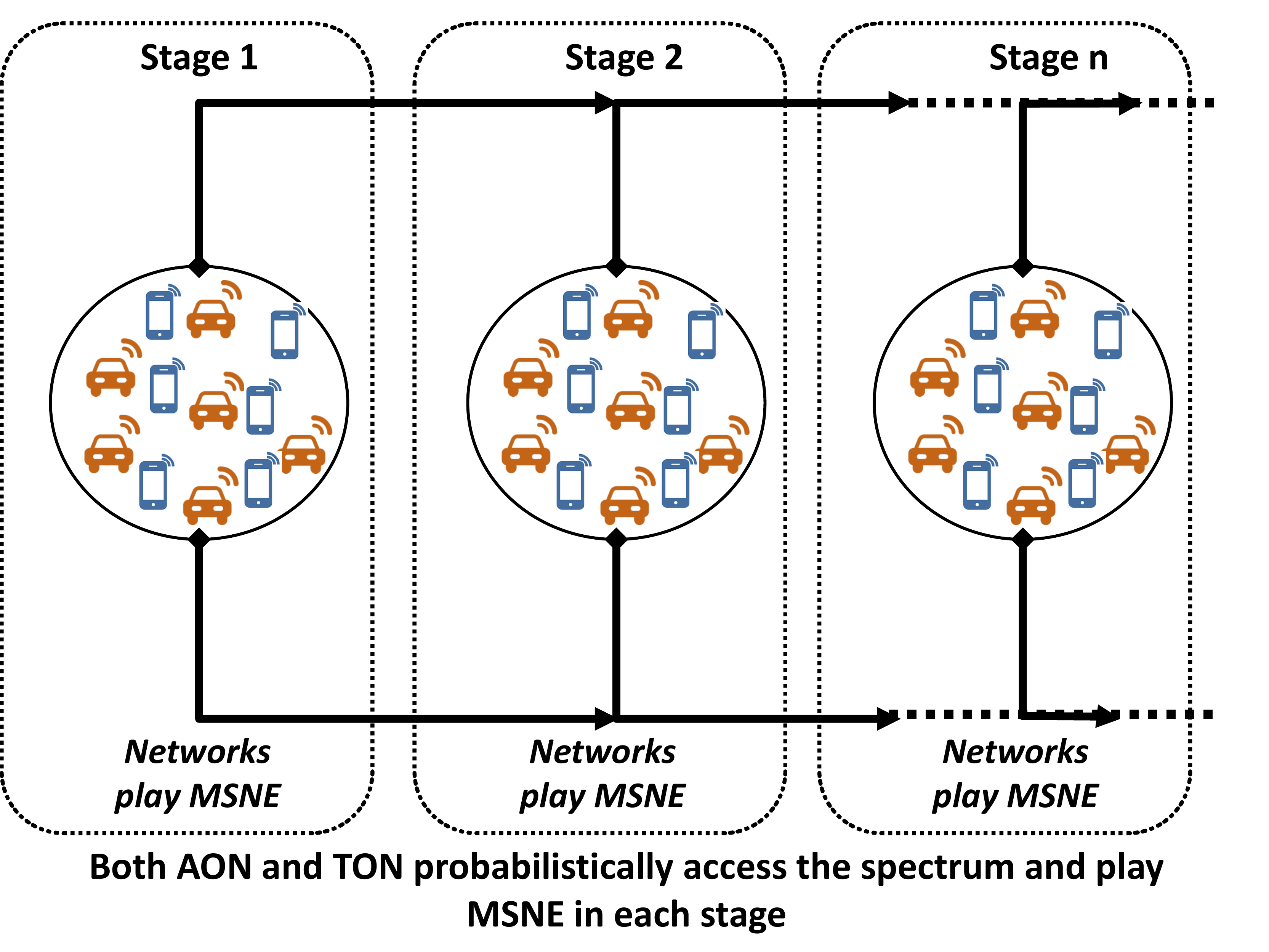}\label{fig:comp_and_coop_a}}
\qquad
\subfloat[Cooperative Mode of Coexistence]{\includegraphics[scale=0.24]{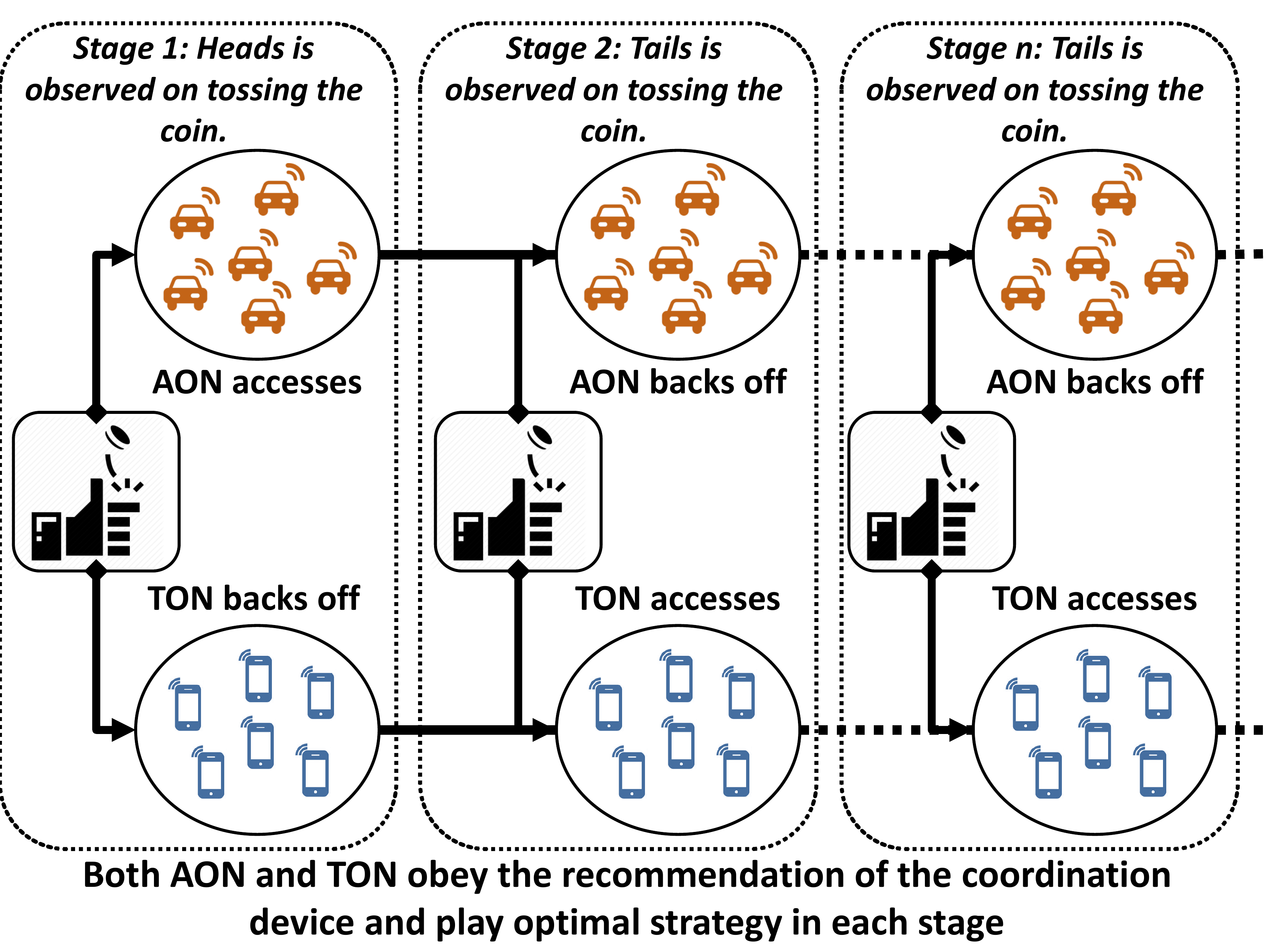}\label{fig:comp_and_coop_b}}
\caption{\small Illustration of different modes of coexistence. (a) Networks compete and probabilistically access the shared spectrum in every stage of the repeated game. (b) Networks cooperate and cooperation is enabled using a coordination device which tosses a coin in every stage of the repeated game and recommends the $\AO$ (resp. the $\TO$) to access the shared spectrum when heads (resp. tails) is observed on tossing the coin and the $\TO$ (resp. the $\AO$) to backoff.}
\label{fig:comp_and_coop}
\end{figure*}

\begin{figure}[t]
\begin{center}
\includegraphics[scale = 0.24]{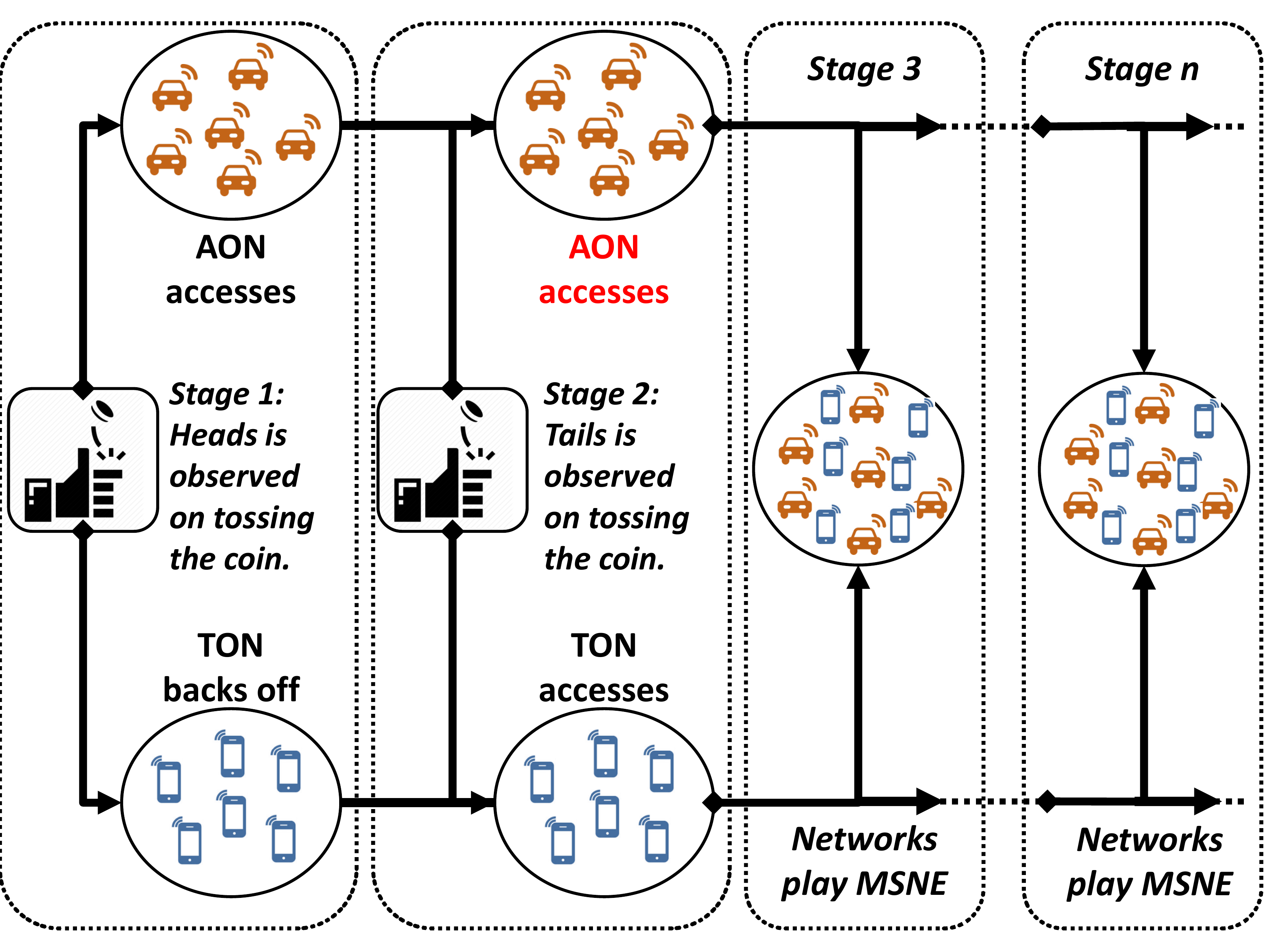}
\caption{\small Illustration of the proposed coexistence etiquette, where, the $\AO$ disobeys the recommendation of the device in stage $2$ such that the grim trigger comes into play, and the networks revert to using the MSNE from stage $3$ onward.}
\label{fig:coexistence_mech}
\vspace{-1em}
\end{center}
\end{figure}

Networks of such IoT devices would like to optimize {\em freshness of status}. In our work, we measure freshness using the age of information~\cite{kaul2012real} metric. \SG{Age of information (AoI) is a newly introduced metric that measures the time elapsed since the last update received at the destination was generated at the source~\cite{kaul2012real}. It is, therefore, a destination-centric metric, and is suitable for networks that care about timely delivery of updates. A typical example is the DSRC-based vehicular network shown in Figure~\ref{fig:example}, where, each vehicle desires fresh status updates (position, velocity etc.) from other vehicles, to enable applications such as collision avoidance, platooning, etc.} Such networks, hereafter referred to as age optimizing networks ($\AO$) will need to co-exist with traditional data networks such as WiFi designed to provide high throughput for its users, hereafter, throughput optimizing networks ($\TO$). This work explores strategies for their coexistence using a repeated game theoretic approach. For symmetry, we assume that both networks use a WiFi-like CSMA/CA (Carrier Sense Multiple Access with Collision Avoidance) based medium access protocol. Each CSMA/CA slot represents a stage game whereby all networks are assumed to be selfish players that optimize their own long-run utility. While an $\AO$ wants to minimize the discounted sum average age of updates of its nodes (at a monitor), a $\TO$ wants to maximize the discounted sum average throughput. 

We consider two modes of coexistence namely \emph{competition} and \emph{cooperation}. When \emph{competing}, as shown in Figure~\ref{fig:comp_and_coop_a}, nodes in the networks probabilistically interfere with those of the other as they access the shared medium. We model the interaction between an $\AO$ and a $\TO$ in each CSMA/CA slot as a non-cooperative stage game and derive its mixed strategy Nash equilibrium (MSNE). We study the evolution of the equilibrium strategy over time, when players play the MSNE in each stage of the repeated game, and the resulting utilities of the networks. 

When cooperating, as shown in Figure~\ref{fig:comp_and_coop_b}, a coordination device schedules the networks to access the medium such that nodes belonging to different networks don't interfere with each other. The coordination device uses a coin toss in every stage to recommend who between the $\AO$ and $\TO$ must access the medium during the slot. Similar to the competitive mode, we define the stage game and derive the optimal strategy that networks would play in a stage, if chosen by the device to access the medium. 

Next, we check whether networks prefer cooperation to competition over the long run. To do so, we propose a coexistence etiquette, where, if in any stage a network doesn't follow the device's recommendation, networks revert to using the MSNE forever. In other words, if a network doesn't cooperate in any stage, networks stop cooperating and start competing in the stages thereafter. Such a strategy is commonly referred to as \emph{grim trigger}~\cite{bookrepeated} because it includes a trigger: once a network deviates from the device's recommendation, this is the trigger that causes the networks to revert their behavior to playing the MSNE forever. Figure~\ref{fig:coexistence_mech} illustrates an example scenario where the coexistence etiquette is employed. The $\AO$ disobeys the recommendation of the device in stage $2$ such that the grim trigger comes into play, and networks revert to using the MSNE forever from stage $3$ onward. One would expect that grim trigger will have networks always obey the device if in fact they preferred cooperation to competition in the long run. We identify when networks prefer cooperation by checking if the strategy profile that results by obeying the device forms a subgame-perfect equilibrium (SPE)~\cite{bookrepeated}. 

Further, we employ the proposed coexistence etiquette to two cases of practical interest (a) when collision slots (more than one node accesses the channel leading to all transmissions received in error) are at least as large as slots that see a successful (interference free) data transmission by exactly one node, and (b) collision slots are smaller than a successful data transmission slot. To exemplify, while the former holds when networks use the basic access mechanism defined for the $802.11$ MAC~\cite{bianchi}, the latter is true for networks employing the RTS/CTS\footnote{In RTS/CTS based access mechanism, under the assumption of perfect channel sensing, collisions occur only when RTS frames are transmitted, which are much smaller than data payload frames, and hence a collision slot is smaller than a successful transmission slot.} based access mechanism~\cite{bianchi}. 

We show that in both cases networks prefer cooperation when they have a small number of nodes. However, for large numbers of nodes, networks end up competing, as disobeying the coordination device benefits one of them. Specifically, when collision slots are at least as large as successful transmission slots, the $\TO$ finds competition more favorable, i.e., sees higher throughput, and the $\AO$ finds cooperation more beneficial, i.e., sees smaller age, whereas, when collision slots are smaller than successful transmission slot, the $\TO$ prefers cooperation and the $\AO$ competition. Our analysis shows that in the former, occasionally the $\AO$ refrains from transmitting during a slot. If competing, such slots allow the $\TO$ interference free access to the medium. If cooperating, such slots are not available to the $\TO$. Thus, competing improves $\TO$'s payoff. In contrast, in the latter, the $\AO$ sees benefit in accessing the medium aggressively. Competition improves the $\AO$ payoff.

Next, in Section~\ref{sec:related}, we give an overview of related works. In Section~\ref{sec:model} we describe the network model. This is followed by Section~\ref{sec:game} in which we discuss the formulation of the non-cooperative stage game, derive the mixed strategy Nash Equilibrium (MSNE) and analyze the repeated game with competition. In Section~\ref{sec:coop} we discuss the stage game with cooperation, derive the optimal strategies that networks would play and analyze the repeated game. We describe the proposed coexistence etiquette in detail in Section~\ref{sec:coex_etiquette}. Computational analysis is carried out in Section \ref{sec:results} where we describe the evaluation setup and also state our main results. We conclude in Section~\ref{sec:conclusion}.
\section{Related Work}
\label{sec:related}
Recent works such as~\cite{marco2017,liu2017,naik2017,khan2017} studied the coexistence of DSRC based vehicular networks and WiFi. In these works authors provided an in-depth study of the inherent differences between the two technologies, the coexistence challenges and proposed solutions to improve coexistence. However, the aforementioned works looked at the coexistence of DSRC and WiFi as the coexistence of two CSMA/CA based networks, with different MAC parameters, where the packets of the DSRC network took precedence over that of the WiFi network. Also, in~\cite{marco2017,liu2017,naik2017,khan2017} authors proposed tweaking the MAC parameters of the WiFi network in order to protect the DSRC network. In contrast to~\cite{marco2017,liu2017,naik2017,khan2017}, we look at the coexistence problem as that of coexistence of networks which have equal access rights to the spectrum, use similar access mechanisms but have different objectives. While the WiFi network ($\TO$) aims to maximize throughput and the DSRC network ($\AO$) desires to minimize age.

In~\cite{mario2005,ma2006,inaltekin2008,chen2010} authors employed game theory to study the behavior of nodes in wireless networks. In~\cite{mario2005} authors studied the selfish behavior of nodes in CSMA/CA networks and proposed a distributed protocol to guide multiple selfish nodes to operate at a Pareto-optimal Nash equilibrium. In~\cite{ma2006} authors studied user behavior under a generalized slotted-Aloha protocol, identified throughput bounds for a system of cooperative users and explored the trade-off between user throughput and short-term fairness. In~\cite{inaltekin2008} authors analyzed Nash equilibria in multiple access with selfish nodes and in~\cite{chen2010} authors developed a game-theoretic model called random access game for contention control and proposed a novel medium access method derived from CSMA/CA that could stabilize the network around a steady state that achieves optimal throughput. 

While throughput as a payoff function has been extensively studied from the game theoretic point of view (see ~\cite{mario2005,ma2006,inaltekin2008,chen2010}), age as a payoff function has not garnered much attention yet. In~\cite{kaul2011minimizing}, the authors investigated minimizing the age of status updates sent by vehicles over a CSMA network. The concept was further investigated in the context of wireless networks in~\cite{sun2017update,yates2017status,yates2015lazy,kadota2018}. \SG{In~\cite{impact2017,YinAoI2018,garnaev2019,gac2018,nguyen2018,SGAoI2020Non,GameofAges2020,zheng2019,SGAoI2018,SGAoI2019} authors studied games with age as the payoff function. In~\cite{impact2017, YinAoI2018,gac2018,garnaev2019}, authors studied an adversarial setting where one player aims to maintain the freshness of information updates while the other player aims to prevent this. In~\cite{nguyen2018}, authors formulated a two-player game to model the interaction between transmitter-receiver pairs over an interference channel in a time-critical system and derived Nash and Stackelberg strategies. In~\cite{SGAoI2020Non} and \cite{GameofAges2020}, authors studied the coexistence of nodes that value timeliness of their information at others and provided insights into how competing nodes would coexist. In~\cite{zheng2019}, authors proposed a Stackelberg game between an access point and its helpers for a wireless powered network with an AoI-based utility for the former and a profit-based utility for the latter. 

In~\cite{hao2019economics,zhang2019price,wang2019dynamic}, authors considered the economic issues related to age in content centric networks. In~\cite{hao2019economics}, authors studied the economic issues related to managing age of selfish content platforms and modeled their interactions as a non-cooperative game under various information scenarios. In~\cite{zhang2019price}, authors studied the pricing mechanism design for fresh data and proposed a time-dependent and a quantity-based pricing scheme. %Authors showed that on an average optimal quantity-based pricing is more profitable and incurs less social cost than optimal time-dependent pricing. 
In~\cite{wang2019dynamic}, authors studied dynamic pricing that minimized the discounted age and payment over time for the content provider.}

In earlier work~\cite{SGAoI2018}, we proposed a game theoretic approach to study the coexistence of DSRC and WiFi, where the DSRC network desires to minimize the time-average age of information and the WiFi network aims to maximize the average throughput. We studied the one-shot game and evaluated the Nash and Stackelberg equilibrium strategies. However, the model in~\cite{SGAoI2018} did not capture well the interaction of networks, evolution of their respective strategies and payoffs over time, which the repeated game model allowed us to capture in~\cite{SGAoI2019}. In~\cite{SGAoI2019}, via the repeated game model we were able to shed better light on the $\AO$-$\TO$ interaction and how their different utilities distinguish their coexistence from the coexistence of utility maximizing CSMA/CA based networks. In this work, we extend the work in~\cite{SGAoI2019} and explore the possibility of cooperation between an $\AO$ and a $\TO$.

Works such as~\cite{repeatedTse2007,repeatedWu2009,repeatedSingh2015} employed repeated games in the context of coexistence. Since repeated games might foster cooperation, authors in~\cite{repeatedTse2007} studied a punishment-based repeated game to model cooperation between multiple networks in an unlicensed band and illustrated that under certain conditions %selfish behavior incur negligible losses and 
whether the systems cooperate or not does not have much influence on the performance. Similar to~\cite{repeatedTse2007}, authors in~\cite{repeatedWu2009} studied a punishment-based repeated game to incorporate cooperation, however, they also proposed mechanisms to ensure user honesty. Contrary to the above works, where coexisting networks have similar objectives and the equilibrium strategies are static in each stage, networks in our work have different objectives and the equilibrium strategy of the $\AO$, as we show later, is dynamic and evolves over stages.
\section{Network Model}
\label{sec:model}
Let $\nD = \{1,2,\dots,\ND\}$ and $\nW = \{1,2,\dots,\NW\}$ denote the set of nodes in the $\AO$ and the $\TO$, respectively, that contend for access to the shared wireless medium. Both $\AO$ and $\TO$ nodes use a CSMA/CA based access mechanism. For the purposes of this section, \textit{network} represents a group of nodes that contend for the medium without reference to whether the nodes belong to the $\AO$ or the $\TO$. %We define events and probabilities that will appear in the context of both competition and cooperation. 
\SG{Contention for the shared wireless medium results in interference between nodes which may cause transmitted packets to be decoded in error. The impact of interference is often captured either by employing the SINR model~\cite{yang2013capacity} or by using a collision channel model~\cite{mario2005,chen2010,bianchi}. In this work, we employ a collision channel model. Specifically, we assume that all nodes can sense each other's packet transmissions and model the CSMA/CA mechanism as a slotted access mechanism. A slot which has no node transmit in it is an idle slot. In case exactly one node transmits a packet in a slot, the transmission is always successfully decoded. If more than one node transmits, none of the transmissions in the slot are successfully decoded and we say that a collision slot occurred.}

\SG{We assume a generate-at-will model~\cite{kadota2018GenerateAtWill,bacinoglu2015GenerateAtWill}, wherein a node is able to generate a fresh update at will. The consequence of this assumption is that a node that transmits a packet always sends a freshly generated update (age $0$ at the beginning of the transmission) in it.} We give the definitions of the parameters used in this paper in Table~\ref{Table:glossary}. Let $\pidle$ be the probability of an idle slot, which is a slot in which no node transmits. Let $\psucci{i}$ be the probability of a successful transmission by node $i$ in a slot and let $\psucc$ be the probability of a successful transmission in a slot. We say that node $i$ sees a busy slot if in the slot node $i$ doesn't transmit and exactly one other node transmits. Let $\pbusy{i}$ be the probability that a busy slot is seen by node $i$. Let $\pcol$ be the probability that a collision occurs in a slot.

Let $\sigma_I, \sigma_S$ and $\sigma_C$ denote the lengths of an idle, successful, and collision slot, respectively. Next, we define the throughput of a $\TO$ node and the age of an $\AO$ node, respectively, in terms of the above probabilities and slot lengths. We will detail the calculation of these probabilities for the competitive and the cooperative mode in Section~\ref{sec:game} and Section~\ref{sec:coop}, respectively. 

\begin{table}[t]
\footnotesize
\caption{Glossary of Terms}
\begin{tabular}{|p{2cm}|p{6cm}|}
\hline
Parameter & Definition\\
\hline\hline
\vspace{.01em}
$\mathcal{N}$ & Set of players. $\mathcal{N} = \{\mathrm{A},\mathrm{T}\}$, $\mathrm{A}$ denotes $\AO$ and $\mathrm{T}$ denotes $\TO$.\\\hline\vspace{.01em}
$\mathcal{S}_k$ & Set of pure strategies of player $k\in\mathcal{N}$. $\mathbb{S}_{k} = \{\T,\I\}$, $\T$ denotes transmit and $\I$ denotes idle.\\\hline
$\ND$, $\NW$ & Number of nodes in the $\AO$ and the $\TO$.\\\hline
$\nD, \nW$ & Set of nodes in the $\AO$ and the $\TO$. $\nD = \{1,2,\dots,\ND\}$ and $\nW = \{1,2,\dots,\NW\}$.\\\hline 
$\lidle, \lsucc, \lcol$ & Length of an idle slot, successful transmission slot and collision slot.\\\hline\vspace{.01em}
$\pidleM{\mathbbm{NC}}, \pidleM{\mathbbm{C}}$ & Probability of an idle slot in competitive or non-cooperative ($\mathbbm{NC}$) mode and cooperative ($\mathbbm{C}$) mode of coexistence.\\\hline\vspace{.01em}
$\psucciM{\mathbbm{NC}}{i},\psucciM{\mathbbm{C}}{i}$ & Probability of a successful transmission by node $i$ in a slot in competitive or non-cooperative ($\mathbbm{NC}$) mode and cooperative ($\mathbbm{C}$) mode of coexistence.\\\hline\vspace{.01em}
$\psuccM{\mathbbm{NC}},\psuccM{\mathbbm{C}}$ & Probability of a successful transmission in a slot in competitive or non-cooperative ($\mathbbm{NC}$) mode and cooperative ($\mathbbm{C}$) mode of coexistence.\\\hline\vspace{.01em}
$\pbusyM{\mathbbm{NC}}{i},\pbusyM{\mathbbm{C}}{i}$ & Probability that a busy slot is seen by node $i$ in competitive or non-cooperative ($\mathbbm{NC}$) mode and cooperative ($\mathbbm{C}$) mode of coexistence.\\\hline\vspace{.01em}
$\pcolM{\mathbbm{NC}},\pcolM{\mathbbm{C}}$ & Probability that a collision occurs in a slot in competitive or non-cooperative ($\mathbbm{NC}$) mode and cooperative ($\mathbbm{C}$) mode of coexistence.\\\hline
$\AvgThr{}$ & Average throughput of the $\TO$ in a slot.\\\hline\vspace{.01em}
$\AvginitAoI, \AvgAoIT{}{}$ & Respectively, age of status updates, averaged over nodes in a $\AO$, at a slot beginning and the network age at the end of a stage.\\\hline\vspace{.01em}
$\tauD^{*}, \tauW^{*}$ & Access probability of an $\AO$ node and a $\TO$ node in the competitive mode.\\\hline\vspace{.01em}
$\tauDC, \tauWC$ & Access probability of an $\AO$ node and a $\TO$ node in the cooperative mode.\\\hline
$\phi^{*}(\tauD^{*},\tauW^{*})$ & Mixed strategy Nash Equilibrium.\\\hline
$\alpha$ & Discount factor, $\alpha \in (0,1)$.\\\hline\vspace{.01em}
$\PR$ & Probability of obtaining heads ($\mathbbm{H}$) on coin toss by the coordination device.\\\hline\vspace{.01em}
$u^{k}_{\mathbbm{NC}},u^{k}_{\mathbbm{C}}$ & Stage game payoff for player $k\in\mathcal{N}$ for the competitive or non-cooperative ($\mathbbm{NC}$) mode and cooperative ($\mathbbm{C}$) mode of coexistence.\\\hline\vspace{.01em}
$U^{k}_{\mathbbm{NC}}, U^{k}_{\mathbbm{C}}$ & Average discounted payoff for player $k\in\mathcal{N}$ for the competitive or non-cooperative ($\mathbbm{NC}$) mode and cooperative ($\mathbbm{C}$) mode of coexistence.\\\hline
\end{tabular}
\label{Table:glossary}
\normalsize
\end{table} 

\subsection{Throughput of a $\TO$ node over a slot}
\label{sec:thr}
Let the rate of transmission be fixed to $r$ bits/sec in any slot. Define the throughput $\Thr{i}$ of any $\TO$ node $i\in \nW$, in a slot as the number of bits transmitted successfully in the slot. This is a random variable with probability mass function (PMF)
\begin{align}
P[\Thr{i} = \gamma] =
    \begin{cases}
      \psucci{i} &\gamma = \lsucc r, \\
      1-\psucci{i} &\gamma = 0,\\
      0 & \text{otherwise.}
    \end{cases}
\label{Eq:ThrPMF}
\end{align}
Thus the \SG{throughput} $\AvgThr{i}$ of node $i$ is
\begin{align}
\AvgThr{i} = \psucci{i}\lsucc r.
\label{Eq:Thr}
\end{align}
The \SG{network throughput} of the $\TO$ in a slot is
\begin{align}
\AvgThr{} = \frac{1}{\NW}\sum\limits_{i=1}^{\NW}  \AvgThr{i}.
\label{Eq:NetThr}
\end{align}
We assume that the throughput in a slot is independent of that in the previous slots\footnote{Our assumption is based on the analysis in~\cite{bianchi}, where the author assumes that at each transmission attempt, regardless of the number of retransmissions suffered, the probability of a collision seen by a packet being transmitted is constant and independent.}. 
\subsection{Age of an $\AO$ node over a slot}
\label{sec:age}
\begin{figure}[t]
\begin{center}
\includegraphics[width=0.95\columnwidth]{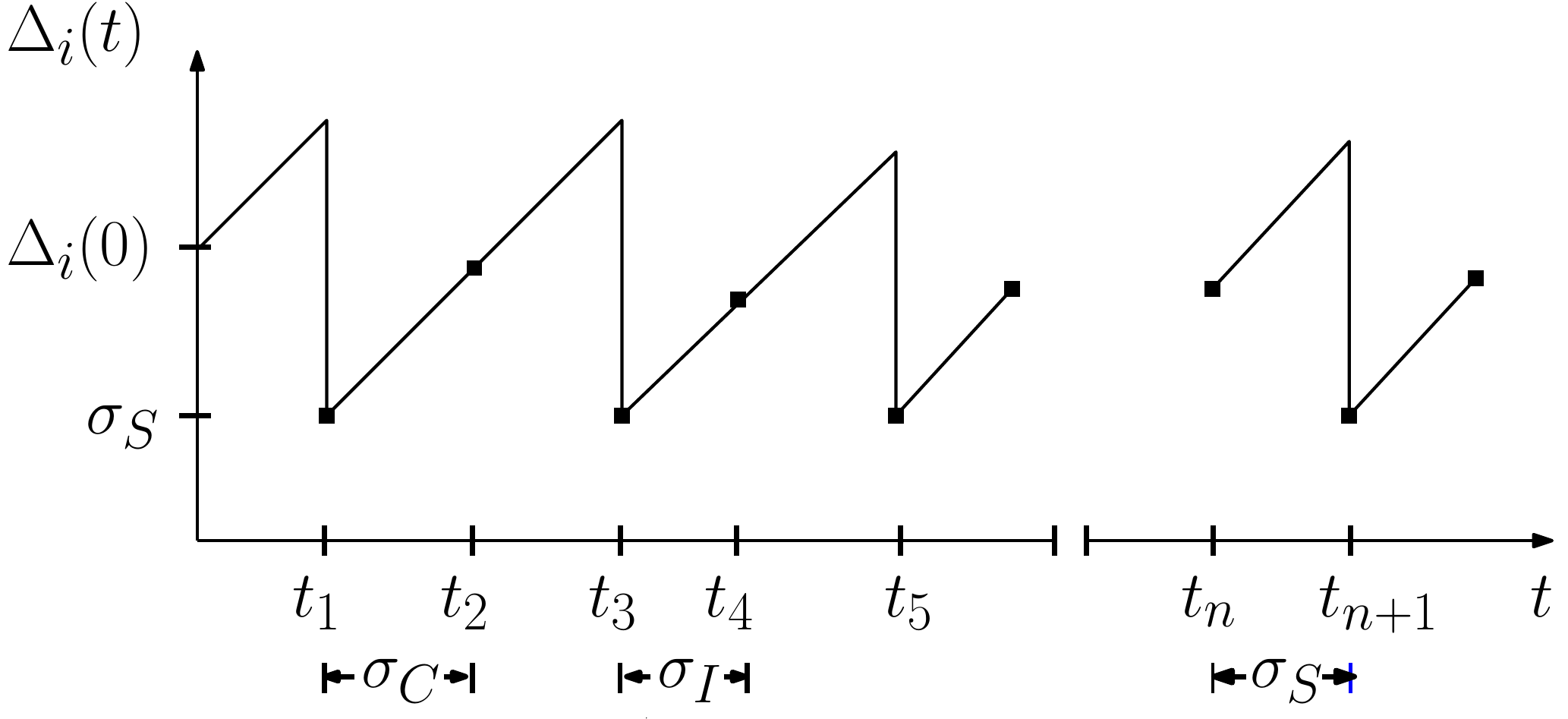}
\end{center}
\caption{\small \SG{Sample path of age $\AoI{i}(t)$ of $\AO$ node $i$'s update at other $\AO$ nodes. $\AoI{i}(0)$ is the initial age. A successful transmission by node $i$ resets its age to $\lsucc$. Otherwise its age increases either by $\lsucc$, $\lcol$ or $\lidle$ depending on whether the slot is a busy slot, a collision slot or an idle slot.  The time instants $t_{n}$, where, $n \in \{1,2,\dots\}$, show the slot boundaries. In the figure, a collision slot starts at $t_1$, an idle slot at $t_3$, and a slot in which the node $i$ transmits successfully starts at $t_n$. Note that while the age $\AoI{i}(t)$ evolves in continuous time, stage payoffs~(\ref{Eq:thrpayoff}),~(\ref{Eq:agepayoff}),~(\ref{Eq:thrpayoff_coop}),~(\ref{Eq:agepayoff_coop}) are calculated only at slot (stage) boundaries.}}
\label{fig:instaAoI}%
\vspace{-0.2in}
\end{figure}
Let $u_{i}(t)$ be the timestamp of the most recent status update of any $\AO$ node $i \in \nD$, at other nodes in the AON at time $t$. The status update age of node $i$ at $\AO$ node $j \in \nD\setminus i$ at time $t$ is the stochastic process $\Delta_{i}(t) = t - u_{i}(t)$. \SG{Given the generate-at-will model, node $i$'s age at any other node $j$ either resets to $\lsucc$ if a successful transmission occurs or increases by $\lidle$, $\lcol$ or $\lsucc$ at all other nodes in the $\AO$, respectively, when an idle slot, collision slot or a busy slot occurs.} \SG{Figure~\ref{fig:instaAoI} shows an example sample path of the age $\AoI{i}(t)$.} In what follows we will drop the explicit mention of time $t$ and let $\AoIT{i}{}$ be the age of node $i$'s update at the end and $\Delta^{-}_{i}$ be the age at the beginning of a given slot.

The age $\AoIT{i}{}$ at the end of a slot is thus a random variable with PMF conditioned on age at the beginning of a slot, given by
\begin{align}
P[\AoIT{i}{} = \delta_{i}|\Delta^{-}_{i} = \initAoIT{i}{}] = 
	\begin{cases}
	\pidle & \delta_{i} = \initAoIT{i}{}+\lidle ,\\
	\pcol & \delta_{i} = \initAoIT{i}{}+\lcol,\\
	\pbusy{i} & \delta_{i} = \initAoIT{i}{}+\lsucc,\\
	\psucci{i} & \delta_{i} = \lsucc,\\
	0 & \text{otherwise.}
	\end{cases}
	\label{Eq:AoIPMF}
\end{align}
Using~(\ref{Eq:AoIPMF}), we define the conditional expected age \SG{of $\AO$ node $i$ as}
\SG{\begin{align}
\AvgAoIT{i}{} &\overset{\Delta}{=} E[\AoIT{i}{}|\Delta^{-}_{i} = \initAoIT{i}{}].\nonumber\\
&=(1-\psucci{i})\initAoIT{i}{}+(\pidle\lidle+\psucc\lsucc+\pcol\lcol).
\label{Eq:AoI}
\end{align}}
The \SG{network age of %status updates of nodes in an 
$\AO$ at the end of the slot}, is
\begin{align}
\AvgAoIT{}{}& = \frac{1}{\ND}\sum\limits_{i=1}^{\ND}\AvgAoIT{i}{}.
\label{Eq:NetAoI}
\end{align}
\section{Competition between an $\AO$ and a $\TO$}
\label{sec:game}
We define a repeated game to model the competition between an $\AO$ and a $\TO$. In every CSMA/CA slot, networks must contend for access with the goal of maximizing their expected payoff over an infinite horizon (a countably infinite number of slots). We capture the interaction in a slot as a non-cooperative stage game $G_{\mathbbm{NC}}$, where $\mathbbm{NC}$ stands for non-cooperation or competition and derive it's mixed strategy Nash equilibrium (MSNE). The interaction over the infinite horizon is modeled as the stage game $G_{\mathbbm{NC}}$ played repeatedly in every slot and is denoted by $G^{\infty}_{\mathbbm{NC}}$. Next, we discuss the games $G_{\mathbbm{NC}}$ and $G^{\infty}_{\mathbbm{NC}}$ in detail.
\subsection{Stage game}
\label{sec:one-shot}
We define a parameterized strategic one-shot game~\cite{zuhan} $G_{\mathbbm{NC}} = (\mathcal{N},(\mathcal{S}_k)_{k\in\mathcal{N}},(u_k)_{k\in\mathcal{N}},\AvginitAoI)$, where $\mathcal{N}$ is the set of players, $\mathcal{S}_k$ is the set of pure strategies of player $k$, $u_k$ is the payoff of player $k$ and $\AvginitAoI$ is the additional parameter input to the game $G_{\mathbbm{NC}}$ \SG{given by $\AvginitAoI = (1/\ND)\sum\limits_{i=1}^{\ND}\Delta^{-}_{i}$}. 
\begin{itemize}
\item \textbf{Players:} The $\AO$ and the $\TO$ are the players. We denote the former by $\mathrm{A}$ and the latter by $\mathrm{T}$. We have $\mathcal{N} = \{\mathrm{A},\mathrm{T}\}$.

\item \textbf{Strategy:} \SG{Let $\T$ denote transmit and $\I$ denote idle. For an $\AO$ comprising of $\ND$ nodes, the set of pure strategies is $\mathcal{S}_{\mathrm{A}} \triangleq \mathbb{S}_{1}\times \mathbb{S}_{2} \times \dots \times \mathbb{S}_{\ND}$, where $\mathbb{S}_{i} = \{\T,\I\}$, $\forall i$, is the set from which an action must be assigned to node $i$ in the $\AO$. That is a pure strategy requires the $\AO$ to select for each node in the set $\nD$ either transmit or idle. Similarly, for a $\TO$ comprising of $\NW$ nodes, the set of pure strategies is $\mathcal{S}_{\mathrm{T}} \triangleq \mathbb{S}_{1}\times \mathbb{S}_{2} \times \dots \times \mathbb{S}_{\NW}$.} 

We allow networks to play mixed strategies. For the strategic game $G_{\mathbbm{NC}}$ define $\boldsymbol{\Phi}_{k}$ as the set of probability distributions over the set of strategies $\mathcal{S}_{k}$ of player $k\in \mathcal{N}$. A mixed strategy for player $k$ is an element $\phi_{k} \in \boldsymbol{\Phi}_{k}$, where $\phi_{k}$ is a probability distribution over $\mathcal{S}_{k}$. For example, for an $\AO$ with $\ND = 2$, the set of pure strategies is $\mathcal{S}_{\mathrm{A}} = \mathbb{S}_{1} \times \mathbb{S}_{2} = \{(\T,\T),(\T,\I),(\I,\T),(\I,\I)\}$ and the probability distribution over $\mathcal{S}_{\mathrm{A}}$ is $\phi_{\mathrm{A}}$, such that $\phi_{\mathrm{A}}(s_{\mathrm{A}})\geq 0$ for all $s_\mathrm{A} \in \mathcal{S}_{\mathrm{A}}$ and $\sum_{s_\mathrm{A} \in \mathcal{S}_{\mathrm{A}}} \phi_{\mathrm{A}}(s_{\mathrm{A}}) = 1$.  
\vspace{0.5em}

\SG{Note that the size of the set of pure strategies increases exponentially in the number of nodes in the networks. In general, a probability mass function (PMF) would assign probabilities to each pure strategy in the set. That is the number of probabilities that a PMF must capture increases exponentially in the number of nodes in the network.} 

\SG{Given this seemingly intractable space of PMF(s), in this work, we restrict ourselves to the space of PMF(s) such that the mixed strategies of the $\AO$ are a function of $\tauD$ and that of the $\TO$ are a function of $\tauW$, where $\tauD$ and $\tauW$, are the probabilities with which nodes in an $\AO$ and a $\TO$, respectively, attempt transmission in a slot\footnote{\SG{This forces all nodes in a given network to have the same probability of access. We believe that this is not too restrictive, given that nodes in a network have no intrinsic reason (they all can sense each other's transmissions and those of nodes in the other network, and contribute equally to the network payoff) to experience a different access to the shared spectrum.}}.} As a result, the probability distribution for an $\AO$ with $\ND = 2$, parameterized by $\tauD$, is $\phi_{\mathrm{A}}(\tauD) = \{\tauD^2, \tauD(1-\tauD), (1-\tauD)\tauD, (1-\tauD)^{2}\}$. Similarly, for a $\TO$ with $\NW = 2$, the probability distribution parameterized by $\tauW$, is $\phi_{\mathrm{T}}(\tauW) = \{\tauW^2, \tauW(1-\tauW), (1-\tauW)\tauW, (1-\tauW)^{2}\}$.

\label{sec:MSNE}
\begin{figure}[t]
  \centering
  \subfloat[]{\includegraphics[width=0.48\columnwidth]{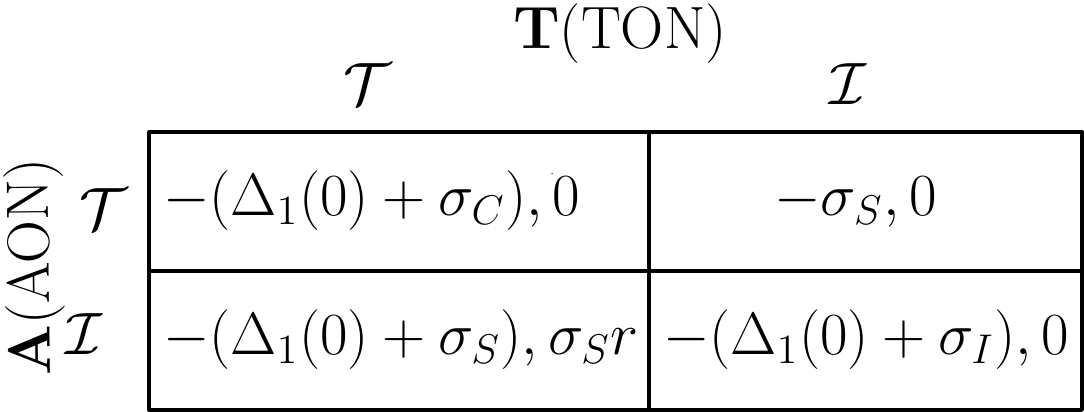}\label{fig:payoff_1a}}
  \quad
  \subfloat[]{\includegraphics[width=0.48\columnwidth]{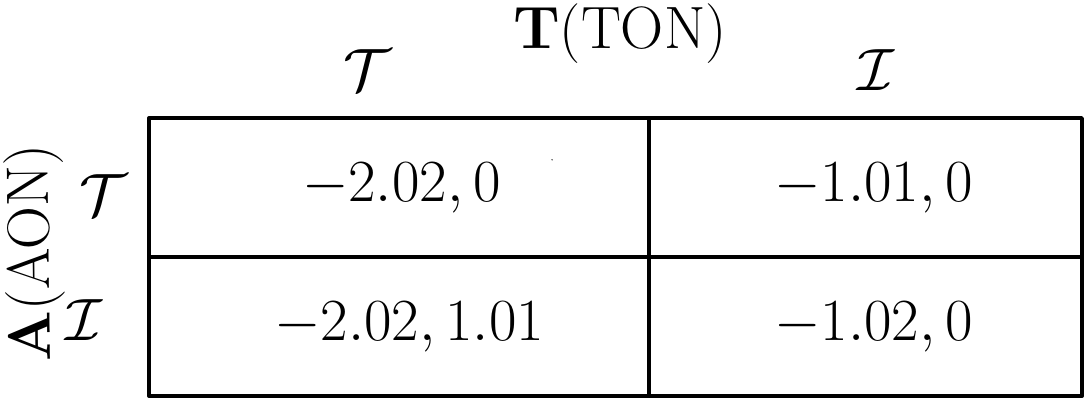}\label{fig:payoff_1b}}
\caption{\small Payoff matrix for the game $G_{\mathbbm{NC}}$ when the $\AO$ and the $\TO$ have one node each. We use negative payoffs for player $\mathrm{A}$ ($\AO$), since it desires to minimize age. (a) Shows the payoff matrix with slot lengths and AoI value at the end of the stage 1. (b) Shows the payoff matrix obtained by substituting $\lsucc = \lcol = 1+\beta$, $\lidle = \beta$\protect\footnotemark, $\AoI{1}(0) = 1+\beta$ and $\beta = 0.01$. ($\mathcal{T},\mathcal{T}$), ($\mathcal{T},\mathcal{I}$) and ($\mathcal{I},\mathcal{T}$) are the pure strategy Nash equilibria.}
\label{fig:payoff_mat}
\end{figure}

\footnotetext{We set the values of $\lidle$, $\lsucc$ and $\lcol$ based on the analysis of CSMA slotted Aloha in~\cite{gallager}, where the authors assume that idle slots have a duration $\beta$ and all data packets have unit length. Nodes in CSMA are allowed to transmit only after detecting an idle slot, i.e., each successful transmission slot and collision slot is followed by an idle slot. Hence, $\lsucc = \lcol = (1+\beta)$.}

\item \textbf{Payoffs:} We have $\NW$ throughput optimizing nodes that attempt transmission with probability $\tauW$ and $\ND$ age optimizing nodes that attempt transmission with probability $\tauD$. As defined in Section~\ref{sec:model}, for the non-cooperative game $G_{\mathbbm{NC}}$, let $\pidleM{\mathbbm{NC}}$ be the probability of an idle slot, $\psuccM{\mathbbm{NC}}$ be the probability of a successful transmission in a slot, $\psucciM{\mathbbm{NC}}{i}$ be the probability of a successful transmission by node $i$, $\pbusyM{\mathbbm{NC}}{i}$ be the probability of a busy slot seen by node $i$ and $\pcolM{\mathbbm{NC}}$ be the probability of collision. We have
\begin{subequations}
\begin{align}
\pidleM{\mathbbm{NC}} &= (1-\tauD)^{\ND}(1-\tauW)^{\NW},
\label{Eq:probidle}\\
\psuccM{\mathbbm{NC}}& = \ND\tauD(1-\tauD)^{(\ND-1)}(1-\tauW)^{\NW}\nonumber\\
& \quad + \NW\tauW(1-\tauW)^{(\NW-1)}(1-\tauD)^{\ND},\label{Eq:probsucc}\\
\psucciM{\mathbbm{NC}}{i}& = 
\begin{cases}
\begin{aligned}
\tauD(1-\tauD)^{(\ND-1)}(1-\tauW)^{\NW},\enspace\forall i\in \nD,\\
\end{aligned}\\
\begin{aligned}
\tauW(1-\tauW)^{(\NW-1)}(1-\tauD)^{\ND},\enspace\forall i\in \nW,\\
\end{aligned}
\end{cases}\label{Eq:probsucci}\\
\pbusyM{\mathbbm{NC}}{i}&= 
\begin{cases}
\begin{aligned}
&(\ND-1)\tauD(1-\tauD)^{(\ND-1)}(1-\tauW)^{\NW}\\
&+\NW\tauW(1-\tauW)^{(\NW-1)}(1-\tauD)^{\ND},
\enspace\forall i\in \nD,\\
\end{aligned}\\
\begin{aligned}
&(\NW-1)\tauW(1-\tauW)^{(\NW-1)}(1-\tauD)^{\ND}\\
&+\ND\tauD(1-\tauD)^{(\ND-1)}(1-\tauW)^{\NW},
\enspace\forall i\in \nW,
\end{aligned}
\end{cases}\label{Eq:probbusy}\\
\pcolM{\mathbbm{NC}}& = 1-\psuccM{\mathbbm{NC}}-\pidleM{\mathbbm{NC}}.
\label{Eq:probcol}
\end{align}
\end{subequations}
\sg{Note that the probabilities~(\ref{Eq:probidle})-(\ref{Eq:probcol}) are independent of the specific node $i$ being considered. This is expected given the mixed strategies we are considering.} The probabilities can be substituted in~(\ref{Eq:ThrPMF})-(\ref{Eq:Thr}) and~(\ref{Eq:AoIPMF})-(\ref{Eq:AoI}), respectively, to calculate the \SG{network throughput}~(\ref{Eq:NetThr}) and age~(\ref{Eq:NetAoI}). 
We use these to obtain the stage payoffs $\uW{\mathbbm{NC}}$ and $\uD{\mathbbm{NC}}$ of the $\TO$ and the $\AO$. They are
\begin{align}
\uW{\mathbbm{NC}}(\tauD,\tauW) &= \AvgThr{}{}(\tauD,\tauW)\label{Eq:thrpayoff},\\
\uD{\mathbbm{NC}}(\tauD,\tauW) &= -\AvgAoIT{}{}(\tauD,\tauW)\label{Eq:agepayoff}.
\end{align}
The networks would like to maximize their payoffs.
\end{itemize}

\subsection{Mixed Strategy Nash Equilibrium}
Figure~\ref{fig:payoff_mat} shows the payoff matrix when each network consists of a single node. As stated in~\cite{nash}, every finite non-cooperative game has a mixed strategy Nash equilibrium (MSNE). 
For the game $G_{\mathbbm{NC}}$ defined in Section~\ref{sec:one-shot}, a mixed-strategy profile $\phi^{*}(\tauD^{*},\tauW^{*}) = (\phi_{\mathrm{A}}^{*}(\tauD^{*}),\phi_{\mathrm{T}}^{*}(\tauW^{*}))$ is a Nash equilibrium~\cite{nash}, if $\phi_{A}^{*}(\tauD^{*})$ and $\phi_{T}^{*}(\tauW^{*})$ are the best responses of player $\mathrm{A}$ and player $\mathrm{T}$, to their respective opponents' mixed strategy. We have
\begin{align*}
\uW{\mathbbm{NC}}(\phi_{A}^{*},\phi_{T}^{*}) \geq \uW{\mathbbm{NC}}(\phi_{A}^{*},\phi_{T}),\quad \forall \phi_{T} \in \boldsymbol{\Phi}_T,\\
\uD{\mathbbm{NC}}(\phi_{A}^{*},\phi_{T}^{*}) \geq \uD{\mathbbm{NC}}(\phi_{A},\phi_{T}^{*}),\quad \forall \phi_{A} \in \boldsymbol{\Phi}_A,
\end{align*}
where, $\phi^{*}(\tauD^{*},\tauW^{*}) \in \boldsymbol{\Phi}$ and $\boldsymbol{\Phi} = \boldsymbol{\Phi}_T \times \boldsymbol{\Phi}_A$ is the profile of mixed strategy. Recall that the probability distributions $\phi_{\mathrm{A}}(\tauD)$ and $\phi_{\mathrm{T}}(\tauW)$ are parameterized by $\tauD$ and $\tauW$, respectively. Proposition~\ref{prop:access} gives the mixed strategy Nash equilibrium.
\begin{proposition}
The mixed strategy Nash equilibrium for the game $G_{\mathbbm{NC}}$ is given by the probabilities $\tauD^{*}$ and $\tauW^{*}$, where 
\begin{subequations}
\scriptsize{
\begin{align}
\tauD^{*}& = 
    \begin{cases}
     \hspace{-0.5em}
      \begin{aligned}
 	  \frac{(1-\tauW^{*})(\AvginitAoI-\ND(\lsucc-\lidle))+\ND\NW\tauW^{*}(\lsucc-\lcol)}{\left(\splitfrac{(1-\tauW^{*})\ND(\AvginitAoI+(\lidle-\lcol)-\ND(\lsucc-\lcol))}{+\ND\NW\tauW^{*}(\lsucc-\lcol)}\right)}
 	  \end{aligned}
	  \vspace{1em}
      &\hspace{-0.85em}
      \begin{aligned}
      &\AvginitAoI> \AvginitAoITh{}{\text{th}},
      \end{aligned}\\
      1&\hspace{-8.5em}
      \begin{aligned}
      \AvginitAoI\leq \AvginitAoITh{}{\text{th}}\text{ \& }\AvginitAoITh{}{\text{th}} = \AvginitAoITh{}{\text{th},1},
      \end{aligned}\\
      0&\hspace{-8.5em}
      \begin{aligned}
      \AvginitAoI\leq \AvginitAoITh{}{\text{th}}\text{ \& }\AvginitAoITh{}{\text{th}} = \AvginitAoITh{}{\text{th},0}.
      \end{aligned}\\
    \end{cases}\label{Eq:Age_MSNE_RTS_CTS}\\
\tauW^{*} &= \frac{1}{\NW\label{Eq:Thr_MSNE_RTS_CTS}
}.
\end{align}}
\normalsize
\label{Eq:MSNE}
\end{subequations}
where, $\AvginitAoITh{}{\text{th}} = \max\{\AvginitAoITh{}{\text{th},0},\AvginitAoITh{}{\text{th},1}\}$, $\AvginitAoITh{}{\text{th},0} = \ND(\lsucc - \lidle) - \frac{\ND\NW\tauW^{*}(\lsucc-\lcol)}{(1-\tauW^{*})}$ and $\AvginitAoITh{}{\text{th},1} = \ND(\lsucc - \lcol)$.\\
\textbf{Proof:} The proof is given in Appendix~\ref{sec:appendix_1}. 
\label{prop:access}
\end{proposition}

Note in~(\ref{Eq:Age_MSNE_RTS_CTS}) and~(\ref{Eq:Thr_MSNE_RTS_CTS}) that $\tauD^{*}$ is a function of \SG{network age} $\AvginitAoI$ observed at the beginning of the slot and the number of nodes in both the networks, whereas, $\tauW^{*}$ is only a function of number of nodes in the $\TO$. The threshold value $\AvginitAoITh{}{\text{th}}$ can either take a value equal to $\AvginitAoITh{}{\text{th},0}$ or $\AvginitAoITh{}{\text{th},1}$. For instance, when $\ND = 1$, $\NW = 1$, and $\lsucc > \lcol$, the threshold value $\AvginitAoITh{}{\text{th}}$ is equal to $\AvginitAoITh{}{\text{th},1} = (\lsucc-\lcol)$ resulting in $\tauD^{*} = 1$. In contrast, when $\lsucc < \lcol$ for $\ND = 1$, $\NW = 1$ the threshold value $\AvginitAoITh{}{\text{th}}$ is equal to $\AvginitAoITh{}{\text{th},0} = \infty$, and since $\AvginitAoI\leq \infty$, $\tauD^{*}$ in this case is $0$. Note that while the parameter $\tauW^{*}$ corresponding to the $\TO$ is equal to $1$, for all selections of $\lcol$, the $\AO$ chooses $\tauD^{*} = 1$ when $\lsucc>\lcol$, and $\tauD^{*} = 0$ when $\lsucc<\lcol$. This is because when $\lsucc<\lcol$ the increase in age due to a successful transmission by the $\TO$, which has $\tauW^{*} = 1$, is less than that due to a collision that would have happened if the $\AO$ chose $\tauD=1$. We discuss this in detail in Section~\ref{sec:repeated_game}. 

\begin{figure}[t]
  \centering
  \subfloat[$\TO$ access probability $\tauW^{*}$]{\includegraphics[width=0.49\columnwidth]{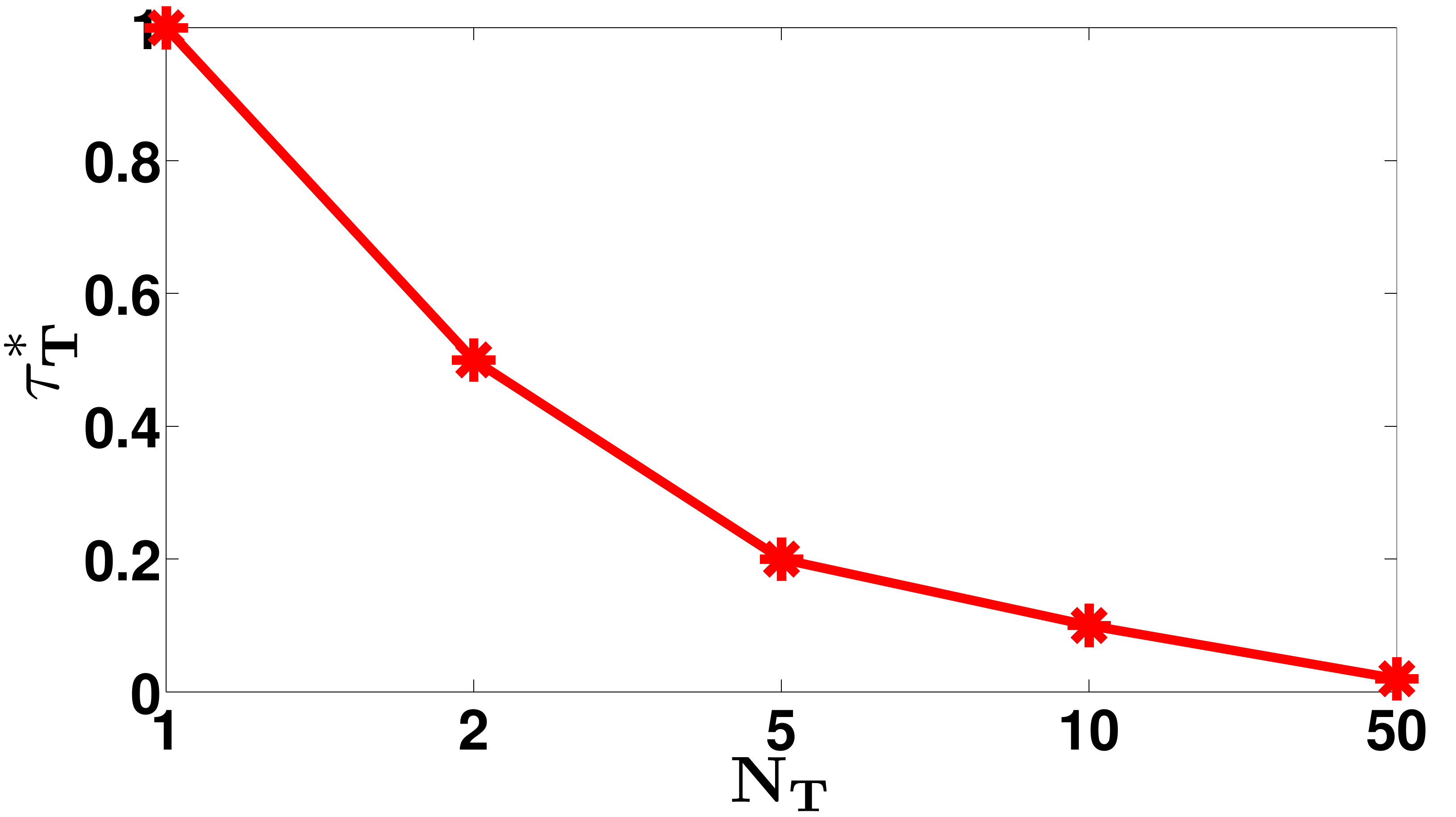}\label{fig:tw_msne}}
  \subfloat[$\AO$ access probability $\tauD^{*}$]{\includegraphics[width=0.49\columnwidth]{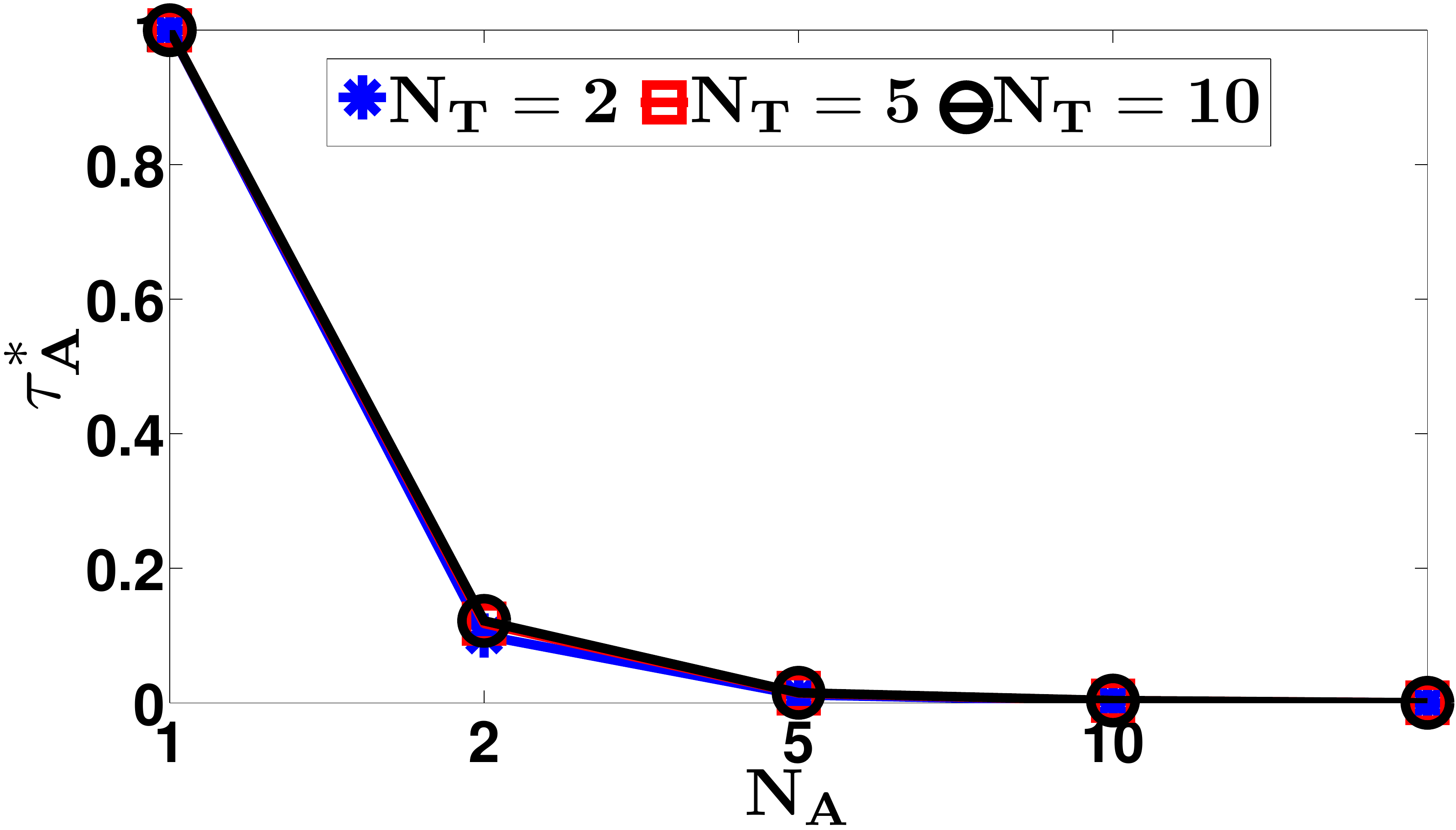}\label{fig:ta_msne}}\\
  \subfloat[$\TO$ stage payoff]{\includegraphics[width=0.49\columnwidth]{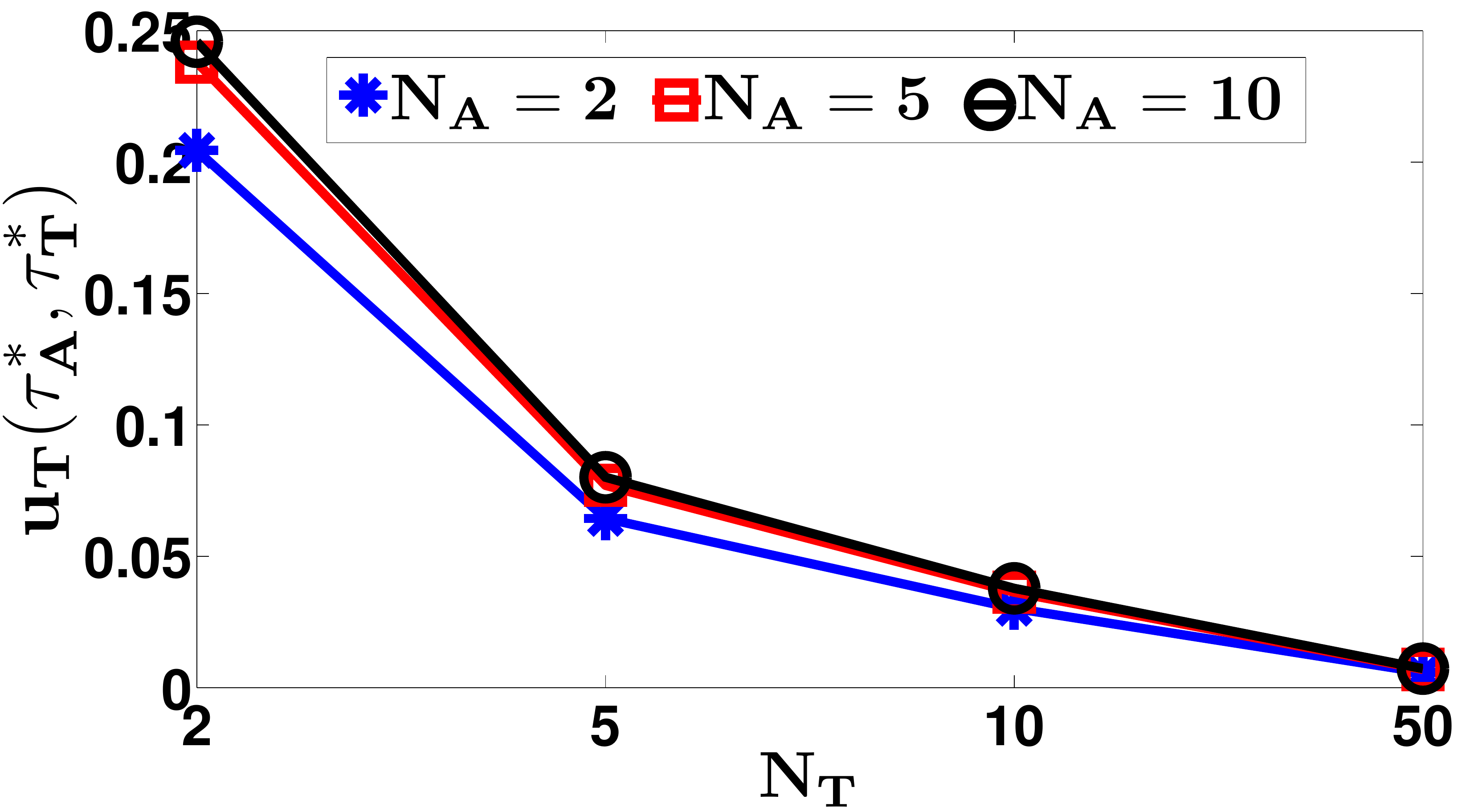}\label{fig:uw_msne}}
  \subfloat[$\AO$ stage payoff]{\includegraphics[width=0.49\columnwidth]{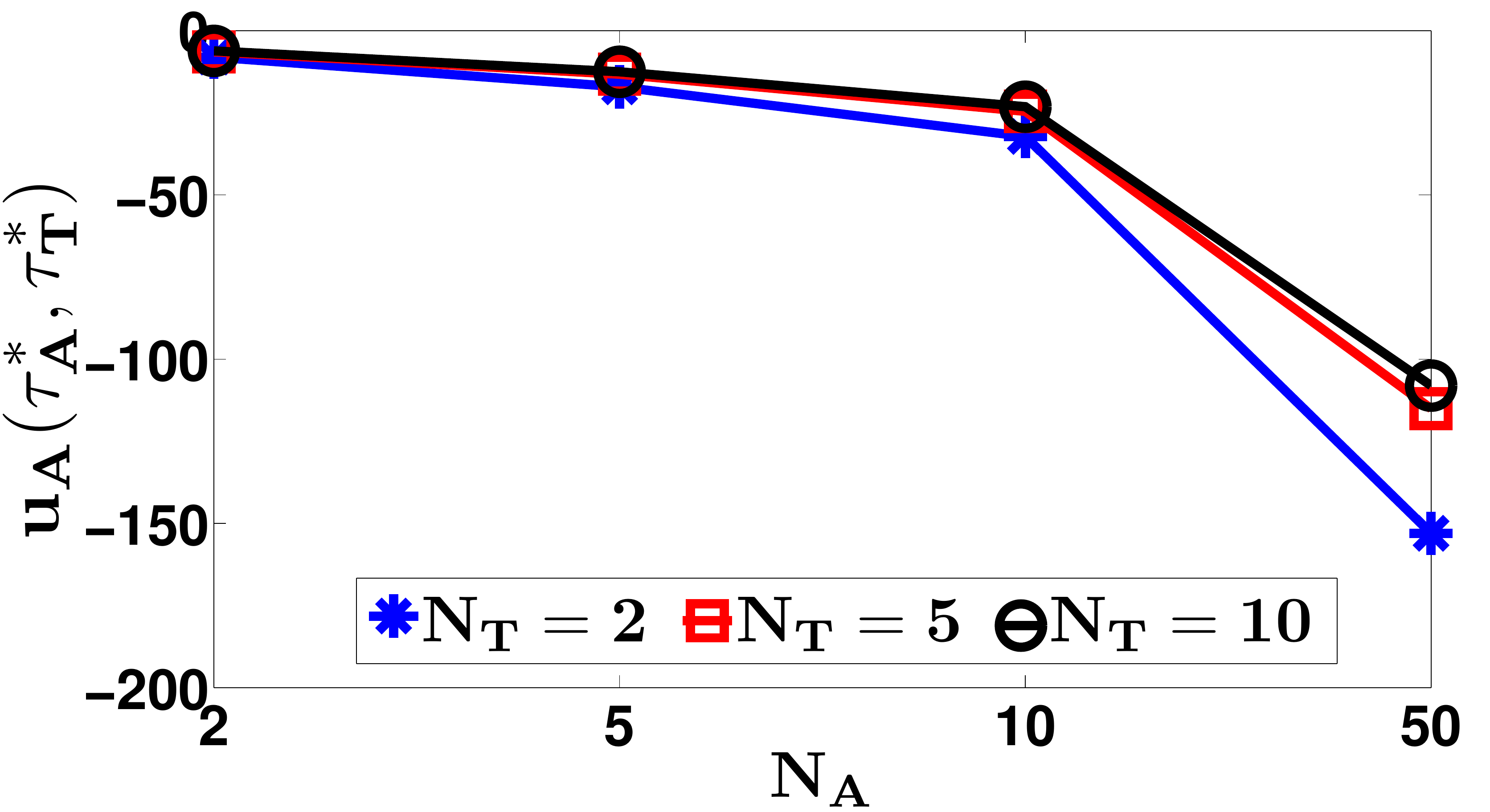}\label{fig:ua_msne}}
\caption{\small Access probabilities and stage payoff of the $\TO$ and the $\AO$ for different selections of $\NW$ and $\ND$ when networks choose to play the MSNE. The stage payoff corresponds to $\AvginitAoI = \AvginitAoITh{}{\text{th},0} + \lsucc$, $\lsucc = 1+\beta$, $\lcol = 2(1+\beta)$, $\lidle = \beta$ and $\beta = 0.01$.}
\label{fig:PayoffvsNodes}
\end{figure}

A distinct feature of the stage game is the effect of \textit{self-contention} and \textit{competition} on the network utilities\footnote{We had earlier observed self-contention and competition in~\cite{SGAoI2018} where we considered an alternate one-shot game and in~\cite{SGAoI2019} where we studied a repeated game with competing networks.}. We define self-contention as the impact of nodes within one's own network and competition as the impact of nodes in the other network, respectively, on the network utilities. Figure~\ref{fig:PayoffvsNodes} shows the affect of self-contention and competition on the access probabilities and stage payoffs. We choose $\AvginitAoI= \AvginitAoITh{}{\text{th},0} + \lsucc$ as it gives $\tauD^{*}\in (0,1)$ (see~(\ref{Eq:Age_MSNE_RTS_CTS})). As shown in Figure~\ref{fig:tw_msne} and Figure~\ref{fig:uw_msne}, while the access probability $\tauW^{*}$ for the $\TO$ is independent of the number of nodes in the $\AO$, the payoff of the $\TO$ increases as the number of nodes in the $\AO$ increase. Intuitively, since increase in the number of $\AO$ nodes results in increase in competition, the payoff of the $\TO$ should decrease. However, the payoff of the $\TO$ increases. For example, for $\NW = 2$, as shown in Figure~\ref{fig:uw_msne}, the payoff of the $\TO$ increases from $0.2044$ to $ 0.2451$ as $\ND$ increases from $2$ to $10$. This increase is due to increase in self-contention within the $\AO$ which forces it to be conservative. Specifically, as shown in Figure~\ref{fig:ta_msne}, the access probability $\tauD^{*}$ decreases with increase in the number of nodes in the $\AO$. For $\NW = 2$, $\tauD^{*}$ decreases from $1$ to $0.0001$ as $\ND$ increases from $1$ to $50$. This benefits the $\TO$. Similarly as shown in Figure~\ref{fig:ua_msne}, as the number of $\TO$ nodes increases the payoff of the $\AO$ improves, since the access probability of the $\TO$ decreases (see Figure~\ref{fig:tw_msne}). 

For the game $G_{\mathbbm{NC}}$, when $\lsucc = \lcol$, the access probabilities $\tauD^{*}$ and $\tauW^{*}$ are shown in Corollary~\ref{cor:BA}. They are independent of the number of nodes in the other network and their access probability.  %As a result, the equilibrium strategy of each network is also its dominant strategy for the case when $\lsucc = \lcol$. 
\begin{corollary}
The mixed strategy Nash equilibrium for the game $G_{\mathbbm{NC}}$ when $\lsucc = \lcol$ is obtained using~(\ref{Eq:Age_MSNE_RTS_CTS}) and is given by
\begin{subequations}
\begin{align}
\tauD^{*} & = 
    \begin{cases}
      \frac{\ND(\lidle-\lsucc)+\AvginitAoI}{\ND(\lidle-\lcol+\AvginitAoI)}&\AvginitAoI > \ND(\lsucc-\lidle), \\
      0 &\text{ otherwise }.
    \end{cases}\label{Eq:Age_MSNE_BA}\\
\tauW^{*} &= \frac{1}{\NW\label{Eq:Thr_MSNE_BA}
}.
\end{align}
\label{Eq:MSNE_BA}
\end{subequations}
This equilibrium strategy of each network is also its dominant strategy.
\label{cor:BA}
\end{corollary}
%\SG{We provide further insights into the differences in the behavior of the $\TO$ and the $\AO$ as captured by Proposition~\ref{prop:access} and corollary~\ref{cor:BA} in Appendix~\ref{sec:appendix_1_discussion}.}
\subsection{Discussion on Mixed Strategy Nash Equilibrium (MSNE)}
\label{sec:appendix_1_discussion}
\SG{The $\TO$ is indifferent to the presence of the $\AO$. This can be explained via the stage payoff of the $\TO$ given in~(\ref{Eq:stage_payoff_ton}). Clearly, the $\tauW$ that optimizes the stage payoff is independent of $\ND$ and $\tauD$. 

One may intuitively explain the indifference of the $\TO$ to the presence of the $\AO$ in the following manner. Recall that the $\TO$ has a node see a throughput greater than zero only when the node transmits successfully. Else, it sees a throughput of $0$. We argue that there is no reason for the $\TO$ to choose an access probability, in the presence of the $\AO$, that is larger than what it would choose in the absence of the $\AO$. This is because a larger probability of access will simply increase the self-contention amongst the nodes in the $\TO$ resulting in a larger fraction of collision slots and a smaller throughput. In case, in the presence of the $\AO$, the $\TO$ chooses a smaller probability of access than it would choose in the absence of the $\AO$, its nodes will have fewer successful transmissions and will see more idle slots and slots with successful transmissions by nodes in the $\AO$. In summary, choosing neither a larger nor a smaller probability of access than it would choose in the absence of the $\AO$ increases the throughput of the $\TO$.

Now consider the $\AO$. It sees an increase in age in an idle slot, in a slot that sees a successful transmission by the $\TO$ and a collision slot. A reduction occurs only if a node in the $\AO$ transmits successfully. The equilibrium access probability of the $\AO$ is impacted by the relative lengths of the collision and successful transmission slots. When collision slots are shorter than successful transmission slots, the $\AO$ picks larger probabilities of access and in fact may have its nodes transmit with probability $1$ (see~(\ref{Eq:KKT_OPT1_temp})). On the other hand, when the successful transmission slots are smaller than collision slots, the $\AO$ picks relatively smaller probabilities of access and in fact may have its nodes access with probability $0$~(\ref{Eq:KKT_OPT1_temp}). 

The above choices by the $\AO$ capture the fact that when competing for spectrum, the $\AO$ adapts to the $\TO$ by pushing for either relatively more collision slots or more slots in which a node in the $\TO$ transmits successfully. For when the length of a collision slot is equal to that of a successful transmission slot, the $\AO$ is indifferent to any change in the balance between collision slots and slots in which a node in the $\TO$ transmits successfully that occurs due to the competing $\TO$. The access probability of the $\AO$ becomes independent of $\NW$ and $\tauW$.}

\begin{figure*}[t]
\subfloat[]{\includegraphics[width = 0.32\textwidth]{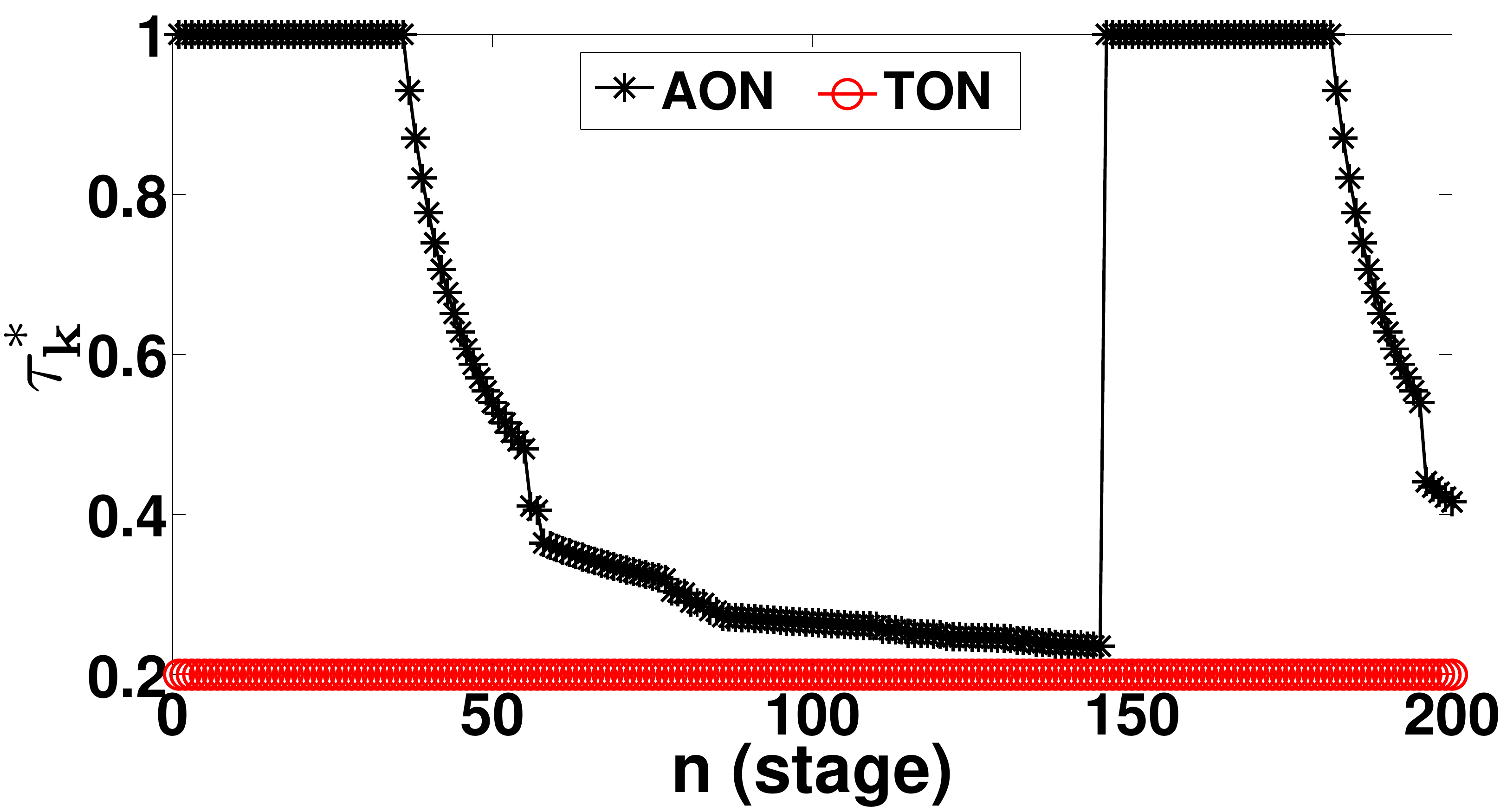}\label{fig:Ex_NE_10lc_eq_ls}}
\subfloat[]{\includegraphics[width = 0.32\textwidth]{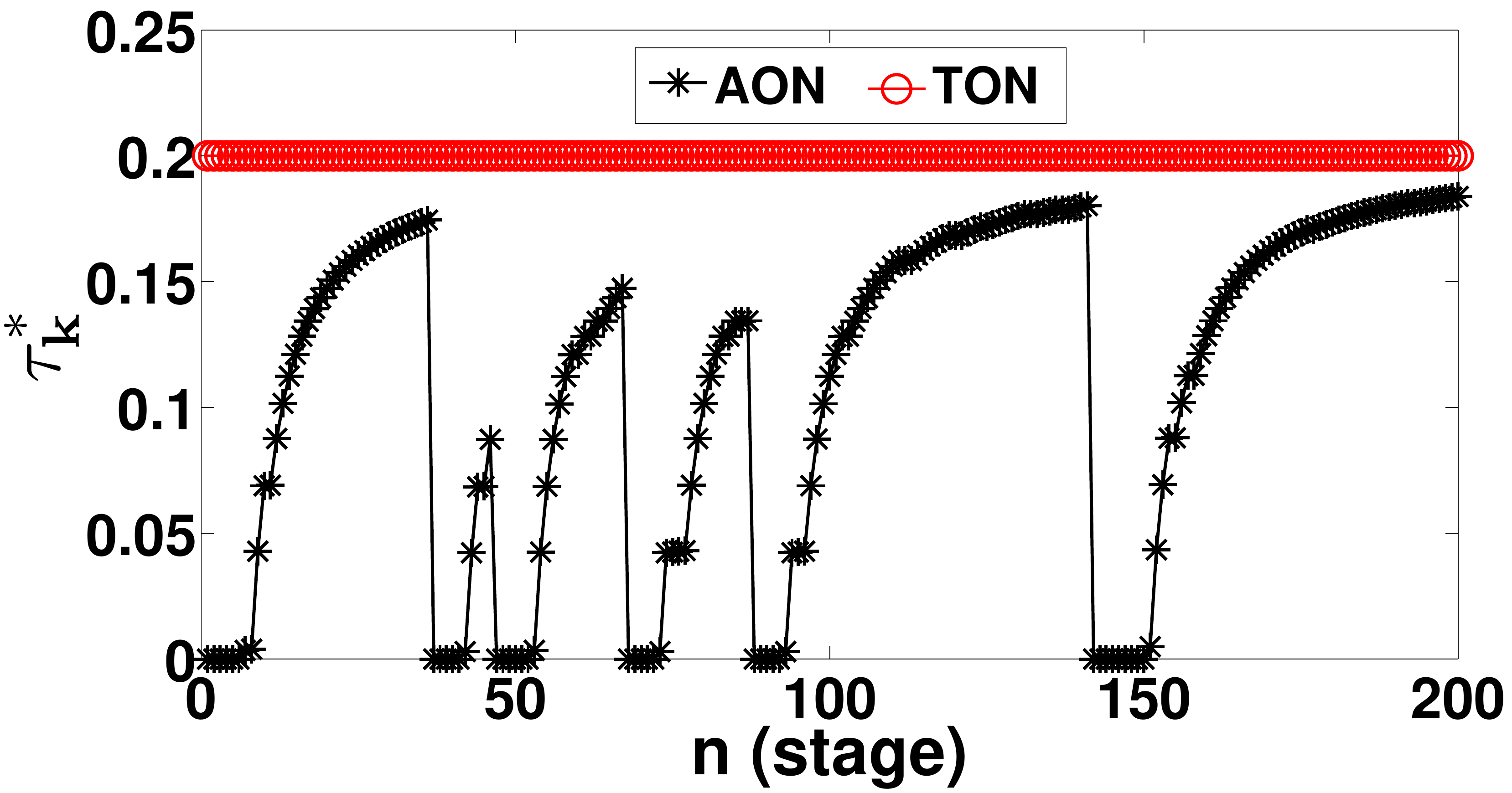}\label{fig:Ex_NE_lc_eq_ls}}
\subfloat[]{\includegraphics[width = 0.32\textwidth]{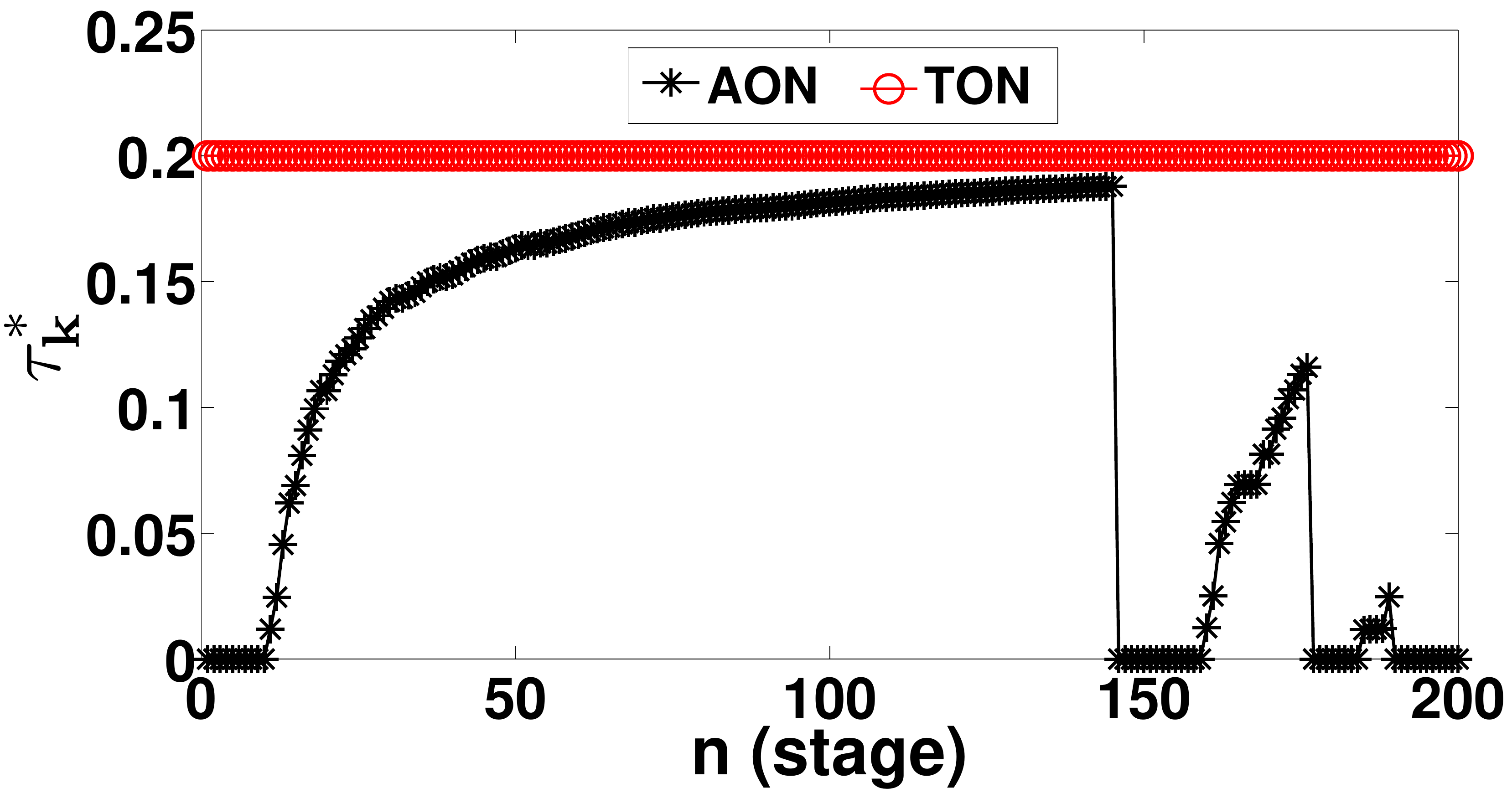}\label{fig:Ex_NE_lc_eq_2ls}}
\caption{Illustration of per stage access probability $\tau_k^{*}$ ($k\in\{\mathrm{A},\mathrm{T}\}$) of the $\AO$ and the $\TO$ when (a) $\lcol = 0.1\lsucc$, (b) $\lcol = \lsucc$, and (c) $\lcol = 2\lsucc$. The results correspond to $\ND = 5, \NW = 5, \lsucc = 1+\beta, \lidle = \beta$ and $\beta = 0.01$.}
\label{fig:ExRun}
\end{figure*}

\subsection{Repeated game}
\label{sec:repeated_game}
We consider an infinitely repeated game, defined as $G^{\infty}_{\mathbbm{NC}}$, in which the one-shot game $G_{\mathbbm{NC}}$, where, players play the MSNE~(\ref{Eq:MSNE}), is played in every stage (slot) $n\in\{1,2,\dots\}$. We consider perfect monitoring~\cite{zuhan}, i.e., at the end of each stage, all players observe the action profile chosen by every other player.\footnote{We leave the study of more realistic assumptions of imperfect and private monitoring to the future.} 

\SG{The essential components of a repeated game include the state variable, the constituent stage game, and the state transition function. For our repeated game $G^{\infty}_{\mathbbm{NC}}$, the state at the beginning of stage $n$ consists of the ages $\Delta^{-}_{i}(n)$, at the beginning of the stage, for all nodes $i$ in the $\AO$. The constituent stage game is the parameterized game $G_{\mathbbm{NC}}$, defined in Section~\ref{sec:one-shot}, where in the parameter at the beginning of stage $n$ is $\AvginitAoI(n)$, which, given the definition in Section~\ref{sec:one-shot}, is $\AvginitAoI(n) = (1/\ND)\sum\limits_{i=1}^{\ND}\Delta^{-}_{i}(n)$. The ages $\AoIT{i}{}(n)$ at the end of a stage (which is also the beginning of the next stage), given $\Delta^{-}_{i}(n)$, for all nodes $i$ in the $\AO$, are governed by the conditional PMF given in Equation~(\ref{Eq:AoIPMF}), with the probabilities of idle, successful transmission, busy, and collision in the equation, appropriately substituted by those corresponding to the stage game and given by~(\ref{Eq:probidle})-(\ref{Eq:probcol}).}
%\SG{In addition to the action profiles, players in the beginning of any stage $n$ also observe $\AvginitAoIT{n}{}$, which is the \SG{network age} of the $\AO$ at the end of stage $n-1$. This similar to the one-shot game $G_{\mathbbm{NC}}$, is the additional parameter observed in the beginning of each stage of the infinitely repeated game $G^{\infty}_{\mathbbm{NC}}$, which along with the action profile impacts the payoff, i.e., age in each stage, since the $\AO$ equilibrium strategy as shown in~(\ref{Eq:Age_MSNE_RTS_CTS}) in any stage is a function of $\AvginitAoIT{n}{}$.}
%, the utilities of the networks are also intertwined.}

Player $k$'s average discounted payoff for the game $G^{\infty}_{\mathbbm{NC}}$, where $k \in \mathcal{N}$ is
\SG{\begin{align}
\mathrm{U}^{k}_{\mathbbm{NC}}= E_{\phi}\left\{(1-\alpha)\sum\limits_{n=1}^{\infty}\alpha^{n-1}u^{k,n}_{\mathbbm{NC}}(\phi)\right\},
\label{Eq:AvgDisPayoff}
\end{align}}
where, the expectation is taken with respect to the strategy profile $\phi$, \SG{$u^{k,n}_{\mathbbm{NC}}(\phi)$} is player $k$'s payoff in stage $n$ and $0<\alpha<1$ is the discount factor. A discount factor $\alpha$ closer to $1$ means that the player values not only the stage payoff but also the impact of its action on payoffs in the future, i.e., the player is far-sighted, whereas $\alpha$ closer to $0$ means that the player is myopic and values more the payoffs in the short-term. By substituting~(\ref{Eq:thrpayoff}) and~(\ref{Eq:agepayoff}) in~(\ref{Eq:AvgDisPayoff}), we can obtain the average discounted payoffs $\UW{\mathbbm{NC}}$ and $\UD{\mathbbm{NC}}$, of the $\TO$ and the $\AO$, respectively.

Figure~\ref{fig:ExRun} shows the access probabilities of the $\TO$ and the $\AO$ for the repeated game $G^{\infty}_{\mathbbm{NC}}$ when (a) $\lsucc > \lcol$ (see Figure~\ref{fig:Ex_NE_10lc_eq_ls}), and (b) $\lsucc \leq \lcol$ (see Figure~\ref{fig:Ex_NE_lc_eq_ls} for $\lsucc = \lcol$ and Figure~\ref{fig:Ex_NE_lc_eq_2ls}
for $\lsucc<\lcol$). We set $\ND = \NW = 5$, $\lsucc = 1+\beta$, $\lcol = 0.1\lsucc$, $\lidle = \beta$ and $\beta = 0.01$. As a result, the threshold values in~(\ref{Eq:Age_MSNE_RTS_CTS}), i.e., $\AvginitAoITh{}{\text{th},0}$ and $\AvginitAoITh{}{\text{th},1}$, are $-0.6812$ and $4.5450$, respectively, resulting in $\AvginitAoITh{}{\text{th}} = \max\{\AvginitAoITh{}{\text{th},0},\AvginitAoITh{}{\text{th},1}\} = 4.5450$. Since $\AvginitAoITh{}{\text{th},0}$ and $\AvginitAoITh{}{\text{th},1}$ are independent of $\AvginitAoI$ (see Proposition~\ref{prop:access}), the resulting $\AvginitAoITh{}{\text{th}}$ is constant across all stages of the repeated game $G^{\infty}_{\mathbbm{NC}}$. As a result, as shown in Figure~\ref{fig:Ex_NE_10lc_eq_ls}, $\tauD^{*} = 1$ for $n \in \{1,\dots,36\}$ since $\AvginitAoIT{n}{}< 4.5450$. However, for $n =37$, $\tauD^{*} =  0.9295$ as $\AvginitAoIT{37}{} =  4.6460$ exceeds the threshold value. Similarly, the threshold value in~(\ref{Eq:Age_MSNE_BA}), when $\lsucc = \lcol$, is $\ND(\lsucc - \lidle) = 5$. As a result, as shown in Figure~\ref{fig:Ex_NE_lc_eq_ls}, nodes in the $\AO$ access the medium with $\tauD^* \in (0,1)$ in any stage $n$ only if the \SG{network age} in the $(n-1)^{th}$ stage exceeds the threshold value, i.e., $\AvginitAoIT{n}{}> 5$, otherwise $\tauD^{*} = 0$.
 
Note that when $\lsucc \leq \lcol$, nodes in the $\AO$ as shown in Figure~\ref{fig:Ex_NE_lc_eq_ls} and Figure~\ref{fig:Ex_NE_lc_eq_2ls}, occasionally refrain from transmission, i.e., choose $\tauD^{*} = 0$ during a stage. In contrast, when $\lsucc>\lcol$, nodes in the $\AO$ as shown in Figure~\ref{fig:Ex_NE_10lc_eq_ls}, often access the medium aggressively, i.e., with $\tauD^{*} = 1$ during a stage. Such a behavior of nodes in the $\AO$ is due to the presence of the $\TO$. As the length of the collision slot decreases, the impact of collision on the age of the $\AO$ reduces. If nodes in the $\AO$ choose to refrain from transmission, the \SG{network age} of the $\AO$ will depend on the events -- successful transmission, collision or idle slot, happening in the $\TO$. Whereas if nodes in the $\AO$ choose to transmit aggressively with $\tauD^{*} = 1$, the \SG{network age} of the $\AO$ would only be impacted by the collision slot.  For instance, for a coexistence scenario with $\ND = \NW = 5$, $\lsucc = 1+\beta$, $\lcol = 0.1\lsucc$, $\lidle = \beta$, $\beta = 0.01$ and $\AvginitAoI = \lsucc$, if $\tauD=0$, the \SG{network age} in the stage computed using~(\ref{Eq:NetAoI}) is $1.4535$, whereas, if $\tauD = 1$, the \SG{network age} is $1.1110$. As a result, due to reduced impact of collision, nodes in the $\AO$ choose to contend with the $\TO$ aggressively for the medium and transmit with $\tauD^{*} = 1$ during a stage. 
\section{Cooperation between an $\AO$ and a $\TO$}
\label{sec:coop}
\begin{figure}[t]
\centering
\includegraphics[width=0.8\columnwidth]{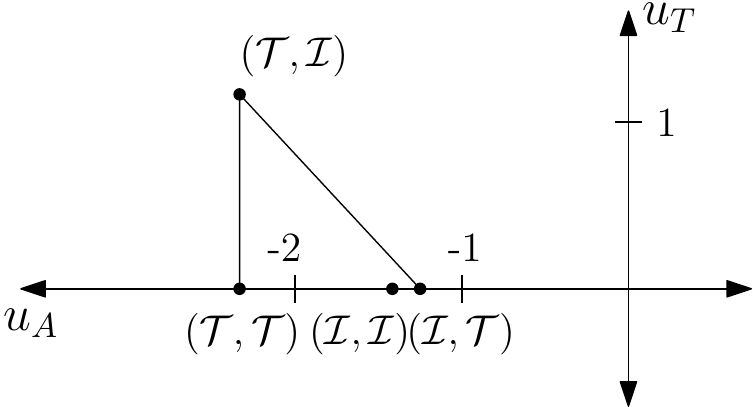}
\caption{\small The convex hull of payoffs for the 2-player one-shot game (see Figure~\ref{fig:payoff_1b}).}
\label{fig:convexhull}
\end{figure}
Consider the 2-player one-shot game shown in Figure~\ref{fig:payoff_1b}. Figure~\ref{fig:convexhull} shows the convex hull of payoffs corresponding to it. The game has three pure strategy Nash Equilibria, i.e., ($\T,\T$), ($\T,\I$) and ($\I,\T$), which have, respectively either both the networks transmit or one of them transmit and the other idle. The corresponding MSNE is given by $\phi_{A}^{*}=\{1,0\}$ and $\phi_{T}^{*} = \{1,0\}$. Both networks transmit with probability $1$. 

Now suppose that the players cooperate and comply with the recommendation of a coordination device, which probabilistically chooses exactly one player to transmit in a stage while the other idles. Say, with probability $0.5$, the device recommends that the $\AO$ transmit and the $\TO$ stays idle. The expected payoff of the $\AO$ is $(-1.01 \times \PR - 2.02 \times (1-\PR)) = -1.515$ and that of the $\TO$ is $(0 \times \PR + 1.01 \times (1-\PR)) = 0.505$, which is more than what the $\AO$ and the $\TO$ would get had they played the MSNE, i.e., payoffs of $-2.02$ and $0$, respectively.

As exemplified above, players may achieve higher expected one-shot payoffs in case they cooperate instead of playing the MSNE~(\ref{Eq:MSNE}). This motivates us to enable cooperation between an $\AO$ and a $\TO$ in the following manner. Consider a coordination device that picks one of the two networks to access ($\mathcal{A}$) the shared spectrum in a slot and the other to backoff ($\mathcal{B}$). To arrive at its recommendation, the device tosses a coin with the probability of obtaining heads ($\mathbb{H}$), P$[\mathbb{H}]$ = $\PR$, and that of obtaining tails ($\mathbb{T}$), P$[\mathbb{T}]$ = $(1-\PR)$. In case $\mathbb{H}$ (resp. $\mathbb{T}$) is observed on tossing the coin, the device picks the $\AO$ (resp. the $\TO$) to access the medium and the $\TO$ (resp. the $\AO$) to backoff. 

Note that the recommendation of the device allows interference free access to the spectrum and eliminates the impact of competition, leaving the networks to deal with self-contention alone. We assume that the probabilities and the recommendations are common knowledge to players.
\subsection{Stage game with cooperating networks}
\label{sec:one-shot-coop}
We begin by modifying the network model defined in Section~\ref{sec:model} to incorporate the recommendation of the coordination device $\PR$. The $\AO$ gets interference free access to the spectrum with probability $\PR$. Let $\tauDC$ denote the optimal probability with which nodes in the $\AO$ must attempt transmission, given that the $\AO$ has access to the spectrum. Let $\tauWC$ be the corresponding probability for the $\TO$.
\begin{proposition}
The optimal strategy of the one-shot game $G_{\mathbbm{C}}$ when networks cooperate is given by the probabilities $\tauDC$ and $\tauWC$. We have
\begin{subequations}
{\small
\begin{align}
\tauDC &= 
    \begin{cases}
     \hspace{-0.5em}
      \begin{aligned}
 	  \frac{\AvginitAoI-\ND(\lsucc-\lidle)}{\ND(\AvginitAoI+(\lidle-\lcol)-\ND(\lsucc-\lcol))}
 	  \end{aligned}
	  \vspace{1em}
      &\hspace{-0.75em}
      \begin{aligned}
      &\AvginitAoI> \AvginitAoITh{}{\text{th}},
      \end{aligned}\\
      1&\hspace{-8em}
      \begin{aligned}
      \AvginitAoI\leq \AvginitAoITh{}{\text{th}}\text{ \& }\AvginitAoITh{}{\text{th}} = \AvginitAoITh{}{\text{th},1},
      \end{aligned}\\
      0&\hspace{-8em}
      \begin{aligned}
      \AvginitAoI\leq \AvginitAoITh{}{\text{th}}\text{ \& }\AvginitAoITh{}{\text{th}} = \AvginitAoITh{}{\text{th},0}.
      \end{aligned}
    \end{cases}\label{Eq:DSRC_optimal}\\
\tauWC &= \frac{1}{\NW}\label{Eq:WiFi_optimal}.
\end{align}}
%\normalsize
\end{subequations}
\label{Eq:optimal}
where, $\AvginitAoITh{}{\text{th}} = \max\{\AvginitAoITh{}{\text{th},0},\AvginitAoITh{}{\text{th},1}\}$, $\AvginitAoITh{}{\text{th},0} = \ND(\lsucc - \lidle)$ and $\AvginitAoITh{}{\text{th},1} = \ND(\lsucc - \lcol)$.\\
\textbf{Proof:} The proof is given in Appendix~\ref{sec:appendix_2}. 
\label{prop:access_coop}
\end{proposition}
\SG{Similar to~(\ref{Eq:Age_MSNE_RTS_CTS}), the optimal strategy $\tauDC$~(\ref{Eq:DSRC_optimal}) of the $\AO$ in any slot is a function of $\AvginitAoI$.} However, in contrast to~(\ref{Eq:Age_MSNE_RTS_CTS}), in the cooperative game $G_{\mathbbm{C}}$, $\tauDC$ is a function of only the number of nodes in its own network, since the coordination device allows networks to access the medium one at a time. Similarly, the optimal strategy $\tauWC$ of the $\TO$ is a function of the number of nodes in its own network and is independent of the number of nodes in the $\AO$. The threshold value $\AvginitAoITh{}{\text{th}}$ can either take a value equal to $\AvginitAoITh{}{\text{th},0}$ or $\AvginitAoITh{}{\text{th},1}$. For instance, when $\ND = 1$, $\NW = 1$ and $\lsucc = \lcol$, $\AvginitAoITh{}{\text{th}}$ takes a value equal to $\AvginitAoITh{}{\text{th},0} = (\lsucc-\lidle)$, and the $\AO$ chooses $\tauDC = 1$. 

As defined in Section~\ref{sec:model}, for the cooperative game $G_{\mathbbm{C}}$, let $\pidleM{\mathbbm{C}}$ be the probability of an idle slot, $\psuccM{\mathbbm{C}}$ be the probability of a successful transmission in a slot, $\psucciM{\mathbbm{C}}{i}$ be the probability of a successful transmission by node $i$, $\pbusyM{\mathbbm{C}}{i}$ be the probability of a busy slot and $\pcolM{\mathbbm{C}}$ be the probability of collision. We have
\begin{subequations}
\begin{align}
\pidleM{\mathbbm{C}}& = \PR(1-\tauDC)^{\ND} + (1-\PR)(1-\tauWC)^{\NW},
\label{Eq:probidlecoop}\\
\psuccM{\mathbbm{C}}& = \PR\ND\tauDC(1-\tauDC)^{(\ND-1)}\nonumber\\
& \quad+ (1-\PR)\NW\tauWC(1-\tauWC)^{(\NW-1)},\label{Eq:probsucccoop}\\
\psucciM{\mathbbm{C}}{i}& = 
\begin{cases}
\PR\tauDC(1-\tauDC)^{(\ND-1)},\enspace \forall i\in \nD,\\
(1-\PR)\tauWC(1-\tauWC)^{(\NW-1)},\enspace \forall i\in \nW,\\
\end{cases}\label{Eq:probsuccicoop}\\
\pbusyM{\mathbbm{C}}{i}&= 
\begin{cases}
\begin{aligned}
&(1-\PR)\NW\tauWC(1-\tauWC)^{(\NW-1)}\\
&+\PR(\ND-1)\tauDC(1-\tauDC)^{(\ND-1)},\enspace\forall i\in \nD,
\end{aligned}\\
\begin{aligned}
&(1-\PR)(\NW-1)\tauWC(1-\tauWC)^{(\NW-1)}\\
&+\PR\ND\tauDC(1-\tauDC)^{(\ND-1)},\enspace\forall i \in \nW,
\end{aligned}
\end{cases}\label{Eq:probbusycoop}\\
\pcolM{\mathbbm{C}}& = 1-\psuccM{\mathbbm{C}}-\pidleM{\mathbbm{C}}.
\label{Eq:probcolcoop}
\end{align}
\end{subequations}
\SG{By substituting~(\ref{Eq:probidlecoop})-(\ref{Eq:probcolcoop}) in~(\ref{Eq:NetThr}) and~(\ref{Eq:NetAoI}), we can obtain the stage utility of the $\TO$ and the $\AO$, defined in~(\ref{Eq:thrpayoff}) and~(\ref{Eq:agepayoff}), respectively, when networks cooperate.} 
%\SG{The probabilities in~(\ref{Eq:probidlecoop})-(\ref{Eq:probcolcoop}) can be substituted to obtain the PMF of throughput $\Thr{i}$ of any $\TO$ node $i\in \nW$ in~(\ref{Eq:ThrPMF}) and conditional PMF of age $\AoIT{i}{}$ of any $\AO$ node $i \in \nD$ in~(\ref{Eq:AoIPMF}). Further, (\ref{Eq:ThrPMF}) and~(\ref{Eq:AoIPMF}) can be used to obtain the throughput $\AvgThr{i}$ of node $i$ in~(\ref{Eq:Thr}) and the conditional expected age $\AvgAoIT{i}{}$ of $\AO$ node $i$ in~(\ref{Eq:AoI}), which can later be used to compute the throughput $\AvgThr{}$ of the $\TO$ in a slot and age $\AvgAoIT{}{}$ of the $\AO$ at the end of the slot in~(\ref{Eq:NetThr}) and~(\ref{Eq:NetAoI}), respectively. Using~(\ref{Eq:NetThr}) and~(\ref{Eq:NetAoI}), we can obtain the stage utility of the $\TO$ and the $\AO$, defined in~(\ref{Eq:thrpayoff}) and~(\ref{Eq:agepayoff}), respectively, when networks cooperate.} The resulting stage utilities of the $\TO$ and the $\AO$ are
\begin{align}
\uW{\mathbbm{C}}(\tauDC,\tauWC) &= \AvgThr{}{}(\tauDC,\tauWC)\label{Eq:thrpayoff_coop},\\
\uD{\mathbbm{C}}(\tauDC,\tauWC) &= -\AvgAoIT{}{}(\tauDC,\tauWC)\label{Eq:agepayoff_coop}.
\end{align}
The networks would like to maximize their payoffs.

\begin{table*}[b]
\begin{align}
\frac{\widetilde{\Delta}^{-}\psucciM{\mathbbm{NC}}{i}-(\sigma_{I}-\sigma_{C})
(\pidleM{\mathbbm{NC}}-(1-\widehat{\tau}_{T})^{N_{T}})-(\sigma_{S}-\sigma_{C})
(\psuccM{\mathbbm{NC}}-N_{T}
\widehat{\tau}_{T}(1-\widehat{\tau}_{T})^{N_{T}-1})}{\splitfrac{\widetilde{\Delta}^{-}\widehat{\tau}_{A}(1-\widehat{\tau}_{A})^{N_{A}-1}
-(\sigma_{I}-\sigma_{C})((1-\widehat{\tau}_{A})^{N_{A}}-(1-\widehat{\tau}_{T})^{N_{T}})}{-(\sigma_{S}-\sigma_{C})(N_{A}\widehat{\tau}_{A}
(1-\widehat{\tau}_{A})^{N_{A}-1}-N_{T}\widehat{\tau}_{T}(1-\widehat{\tau}_{T})^{N_{T}-1})}}
\leq P_{R} \leq 1-(1-\tau_{A}^{*})^{N_{A}} 
\label{Eq:PR} 
\end{align}
\end{table*}
\subsection{Cooperating vs. competing in a stage}
We consider when both networks find cooperation to be beneficial over competition in a stage game. That is $\uW{\mathbbm{C}}(\tauDC,\tauWC)\geq\uW{\mathbbm{NC}}(\tauD^{*},\tauW^{*})$ and $\uD{\mathbbm{C}}(\tauDC,\tauWC)\geq\uD{\mathbbm{NC}}(\tauD^{*},\tauW^{*})$. Using these inequalities, we determine the range of $\PR$, given in~(\ref{Eq:PR}), over which networks prefer cooperation in the stage game.

Consider when $\ND = 1$ and $\NW = 1$. The range of $\PR$ in~(\ref{Eq:PR}) depends on the length $\lcol$ of the collision slot. As discussed earlier in Section~\ref{sec:game}, when networks compete and $\lsucc\geq\lcol$, $\tauD^{*} = \tauW^{*} = 1$, whereas, when $\lsucc<\lcol$, $\tauD^{*} = 0$ and $\tauW^{*} = 1$. In contrast, when $\ND = 1$ and $\NW = 1$, irrespective of the length of collision slot $\lcol$, when networks cooperate, $\tauDC = 1$ and $\tauWC = 1$. As a result, cooperation is beneficial for $\PR\in[0,1]$ when $\lsucc\geq\lcol$, and only beneficial at $\PR = 0$ when $\lsucc<\lcol$. This is because when $\lsucc\geq\lcol$, while networks see a collision when they compete, they see a successful transmission if they choose to cooperate. In contrast, when $\lsucc<\lcol$, since the $\AO$ chooses not to access the medium when networks compete, the $\TO$ gets a \SG{competition free} access to the medium and hence always sees a successful transmission. As a result, the $\TO$ suffers from cooperation unless the $\AO$ doesn't get a chance to access the medium, which is when $\PR = 0$. Note that while the analysis for $\ND=1$ and $\NW=1$ as discussed above, is simple,~(\ref{Eq:PR}) becomes intractable for $\ND>1$ and $\NW>1$. Hence, we resort to computational analysis and show that as the number of nodes increases, when $\lsucc\leq\lcol$, cooperation is beneficial only at $\PR = 0$, whereas, when $\lsucc>\lcol$, it is beneficial only for higher values of $\PR$, i.e., for $\PR$ close to $1$. We discuss this in detail in Section~\ref{sec:results}.

\subsection{Repeated game with cooperating networks}
\label{sec:repeated_game_coop}
\SG{We define an infinitely repeated game $G^{\infty}_{\mathbbm{C}}$ given the coordination device $\PR$. The course of action followed by the networks is: Players in the beginning of stage $n$ receive a recommendation $\text{R}_{n}\in\{\mathbb{H},\mathbb{T}\}$ from the coordination device $\PR$ and, following on the recommendation, the players either access ($\mathcal{A}$) the shared spectrum or backoff ($\mathcal{B}$).} We define the strategy profile of players in stage $n$ as
\begin{align}
a_{n} = 
\begin{cases}
\text{($\mathcal{A},\mathcal{B}$)}&\text{ if } \text{R}_{n} = \mathbb{H},\\
\text{($\mathcal{B},\mathcal{A}$)}& \text{ if } \text{R}_{n} = \mathbb{T}.
\end{cases}
\label{Eq:PathofPlay}
\end{align}

\SG{The evolution of ages, as the players go from playing one stage to another in the repeated game, is governed by the conditional PMF given in Equation~(\ref{Eq:AoIPMF}), with the probabilities of idle, successful transmission, busy, and collision in the equation, appropriately substituted by those corresponding to the cooperation stage game and given by~(\ref{Eq:probidlecoop})-(\ref{Eq:probcolcoop}).}

We have player $k$'s average discounted payoff for the game $G^{\infty}_{\mathbbm{C}}$, where $k \in \mathcal{N}$ is
\SG{\begin{align}
\mathrm{U}^{k}_{\mathbbm{C}}= E_{\phi}\left\{(1-\alpha)\sum\limits_{n=1}^{\infty}\alpha^{n-1}u^{k,n}_{\mathbbm{C}}(\phi)\right\},
\label{Eq:AvgDisPayoff_coop}
\end{align}}
where the expectation is taken with respect to the strategy profile $\phi$, \SG{$u^{k,n}_{\mathbbm{C}}(\phi)$} is player $k$'s payoff in stage $n$ and $0<\alpha<1$ is the discount factor. By substituting~(\ref{Eq:thrpayoff_coop}) and~(\ref{Eq:agepayoff_coop}) in~(\ref{Eq:AvgDisPayoff_coop}), we can obtain the average discounted payoffs $\UWC{\mathbbm{C}}$ and $\UDC{\mathbbm{C}}$, of the $\TO$ and the $\AO$, respectively.

\begin{figure*}[t]
\subfloat[]{\includegraphics[width = 0.32\textwidth]{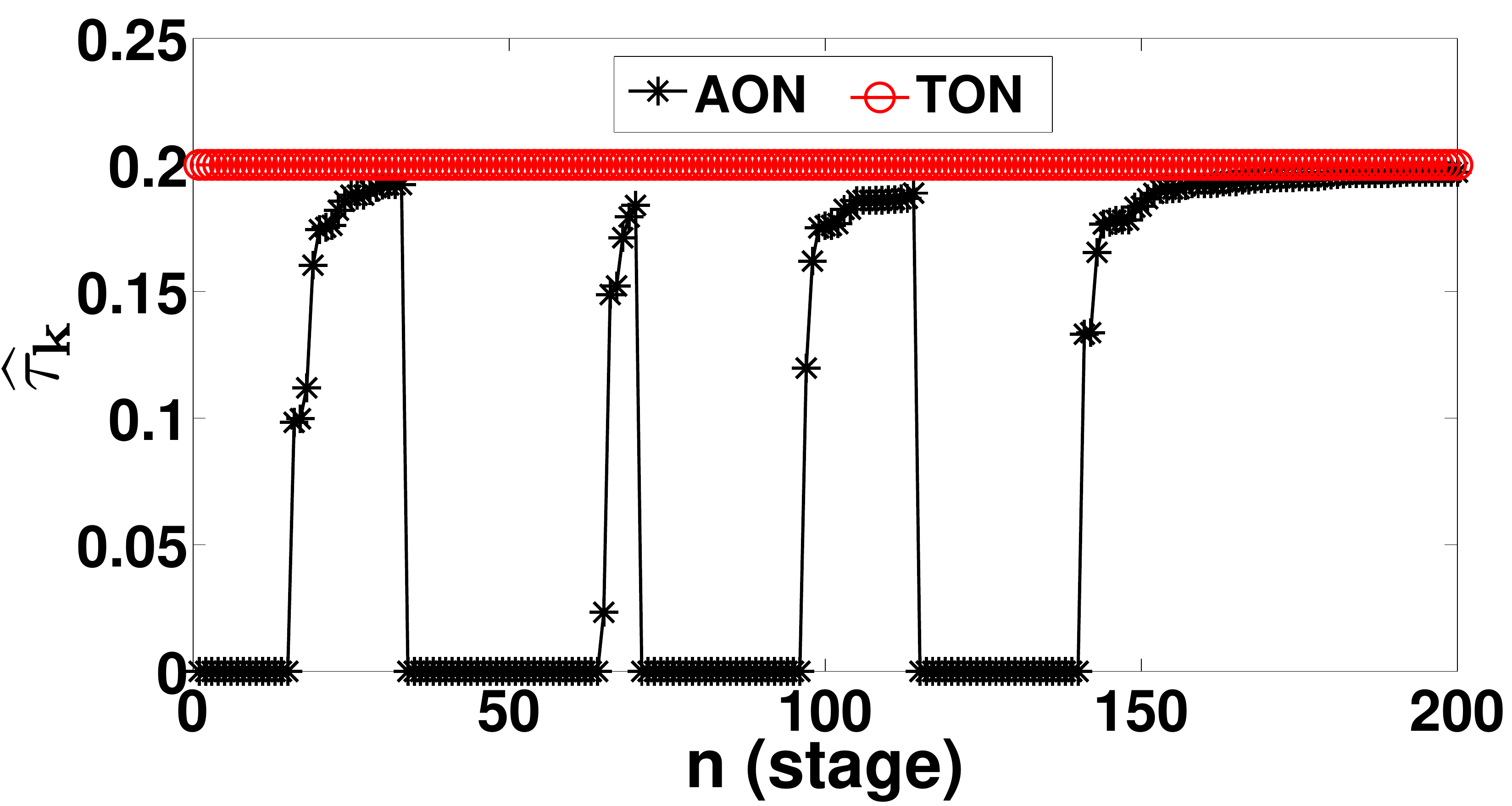}\label{fig:Opt_Ex_NE_10lc_eq_ls}}
\subfloat[]{\includegraphics[width = 0.32\textwidth]{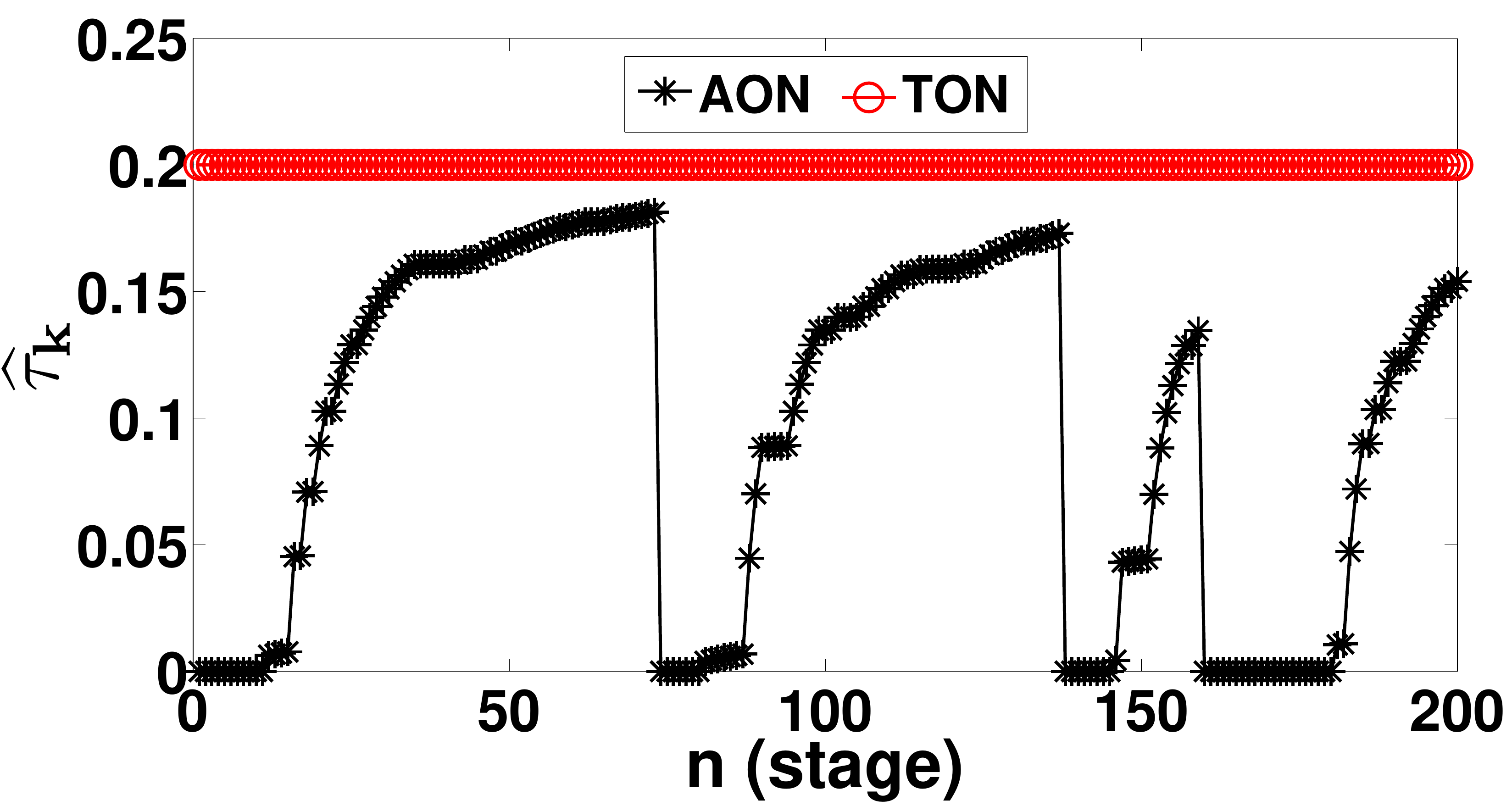}\label{fig:Opt_Ex_NE_lc_eq_ls}}
\subfloat[]{\includegraphics[width = 0.32\textwidth]{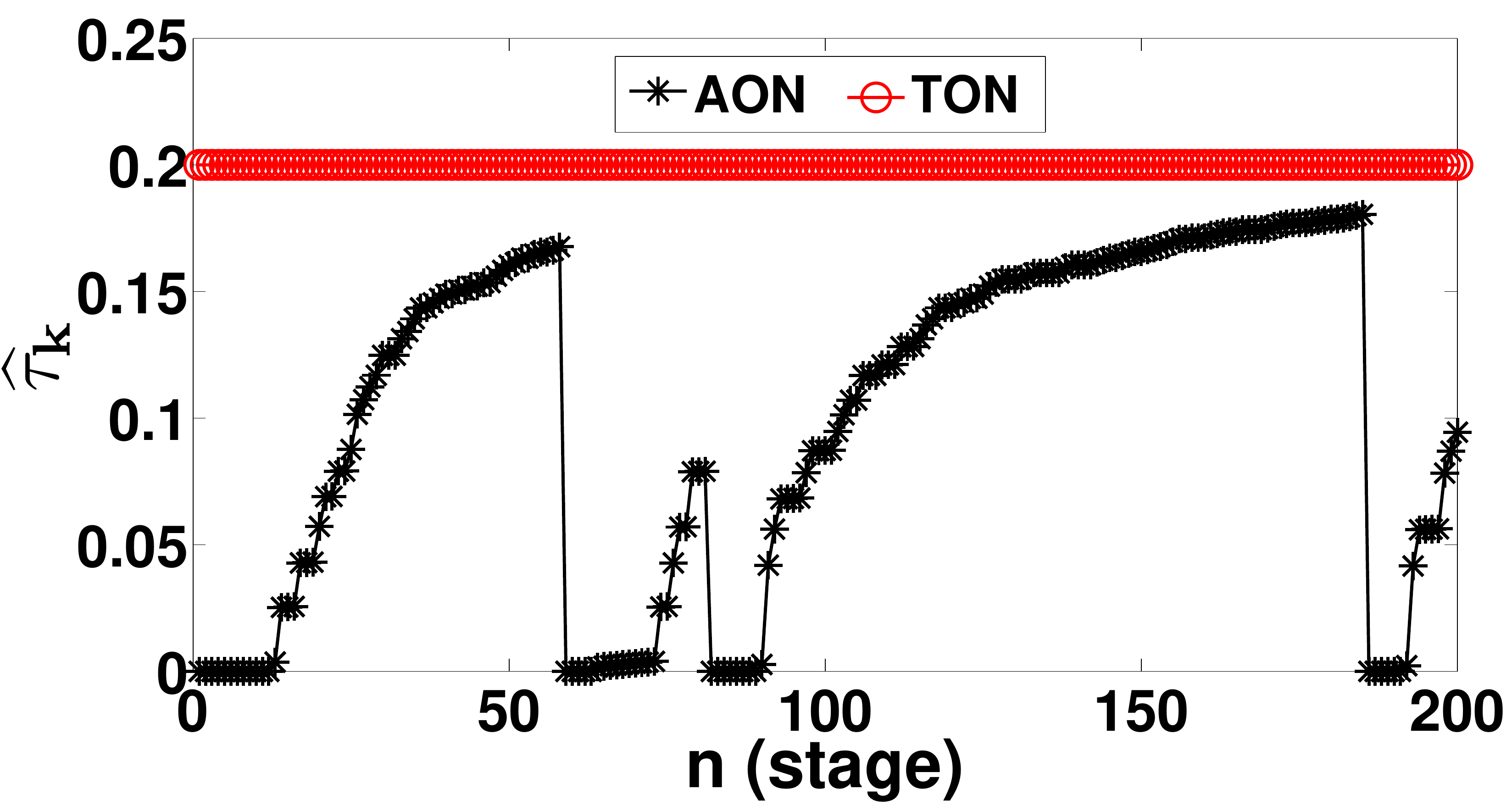}\label{fig:Opt_Ex_NE_lc_eq_2ls}}
\caption{Illustration of per stage access probability $\widehat{\tau}_k$ ($k\in\{\mathrm{A},\mathrm{T}\}$) of the $\AO$ and the $\TO$, as a function of the stage, obtained from an independent run when (a) $\lcol = 0.1\lsucc$, (b) $\lcol = \lsucc$, and (c) $\lcol = 2\lsucc$. The results correspond to $\ND = 5, \NW = 5, \lsucc = 1+\beta, \lidle = \beta, \beta = 0.01$ and $\PR = 0.5$.}
\label{fig:OptExRun}
\end{figure*}

Figure~\ref{fig:OptExRun} shows the access probabilities of $\TO$ and $\AO$ for the repeated game $G^{\infty}_{\mathbbm{C}}$ when (a) $\lsucc > \lcol$ (Figure~\ref{fig:Opt_Ex_NE_10lc_eq_ls}), and (b) when $\lsucc \leq \lcol$ (Figure~\ref{fig:Opt_Ex_NE_lc_eq_ls} corresponds to $\lsucc = \lcol$ and Figure~\ref{fig:Opt_Ex_NE_lc_eq_2ls} corresponds to $\lsucc < \lcol$) . The results correspond to $\AO$-$\TO$ coexistence with $\ND = \NW = 5$ and $\PR = 0.5$.

In contrast to the repeated game in Section~\ref{sec:repeated_game} where nodes in the $\AO$ choose to occasionally access the medium aggressively when $\lsucc>\lcol$, in the repeated game $G^{\infty}_{\mathbbm{C}}$, nodes in the $\AO$ as shown in Figure~\ref{fig:OptExRun}, irrespective of the length of collision slot, never access the medium aggressively, i.e., do not choose $\tauDC = 1$, instead they occasionally refrain from transmission and choose $\tauDC = 0$ during a stage. This is due to the absence of \SG{competition} from the $\TO$ when networks obey the recommendation of the coordination device. \SG{In the absence of competition from the $\TO$ when the coordination device chooses the $\AO$ to access and the $\TO$ to backoff, if the nodes in the $\AO$ refrain from transmission (that is access with $\tauDC = 0$), the age of the $\AO$ only increases by the length of an idle slot. Since the benefit of idling surpasses that of contending aggressively, nodes in the $\AO$ occasionally choose to refrain from transmission irrespective of the relative length of collision slot.}
\section{The Coexistence Etiquette}
\label{sec:coex_etiquette}
The networks are selfish players and may find it beneficial to disobey the recommendations of the device. We enforce a coexistence etiquette which ensures that in the long run the networks either cooperate or compete forever. We have the networks adopt the grim trigger strategy~\cite{bookrepeated} in case the other network doesn't follow the recommendation of the coordination device in a certain stage of the repeated game. Specifically, if in any stage, a network does not comply with the recommendation of the coordination device, the networks play their respective Nash equilibrium strategies~(\ref{Eq:MSNE}) in each stage that follows. 

The penalty of a network not following the coordination device in a stage is to have to compete in every stage thereafter. \SG{While this strategy is commonly understood to be a hardly plausible mode of cooperation and other strategies such as Tit-for-Tat strategy~\cite{bookrepeated} in which players keep switching between cooperative and competitive mode are preferred more, we choose this strategy because it explores the theoretical feasibility of cooperation with arbitrarily patient players. Also, if even the strongest possible threat of perpetual competition posed under the grim trigger strategy cannot induce cooperation, then it is unlikely that players would cooperate under less severe strategies such as Tit-for-Tat.}

To enable the etiquette, in addition to the recommendation of the device, we assume that the players at the beginning of any stage $n$ have information about the actions that the players chose in stage $(n-1)$. Since the players may disobey the device, the action profile $a_n$ is not restricted to that in~(\ref{Eq:PathofPlay}). Let $\psi_{n}\in\{0,1\}$ be an indicator variable such that $\psi_{n}=1$  if the networks obey the coordination device $\PR$ in stage $n$, and $\psi_{n}=0$ corresponds to them deviating. We set $\psi_{n}=1$ when $\text{R}_{n} = \mathbb{H}$ and action profile $a_{n} = (\mathcal{A},\mathcal{B})$ or when $\text{R}_{n} = \mathbb{T}$ and action profile $a_{n} = (\mathcal{B},\mathcal{A})$. Else, $\psi_{n}=0$.

If $\psi_{n-1}=0$, networks play their respective Nash strategies $(\phi_{A}^{*},\phi_{T}^{*})$ in stage $n$ and all stages that follow.

\subsection{Is cooperation self-enforceable?}
\label{sec:coopfeasibility}
\sg{We check if the cooperation strategy profile defined in~(\ref{Eq:PathofPlay}) is self-enforceable when using grim trigger, that is, if the networks always comply with the recommendations of the coordination device and do not have any incentive to deviate. Nash Equilibrium~\cite{nash} is often referred to as self-enforcing in any non-cooperative strategic game because once players expectations are coordinated on such behavior, players left to act on their own accord find that there is no incentive for them to deviate. In repeated games, such self-enforcing behavior after any history is true of a subgame-perfect equilibrium. Therefore, for the repeated game under study, we check whether the cooperation strategy profile is a subgame-perfect equilibrium (SPE)~\cite{bookrepeated}. That is, whether either player would benefit from deviating unilaterally from the recommendation of the randomization device at any stage of the game.
	
For the cooperation strategy profile, when using grim trigger, to be a subgame perfect equilibrium it has to remain a Nash Equilibrium in the repeated game that follows every history of play. While the repeated game under study has some initial age $\AvginitAoI$ associated with it; it is otherwise the same as the repeated game starting at any point. Therefore, without loss of generality, we consider stage $1$ and check whether networks always comply with the recommendations of the coordination device or if they have any incentive to deviate.

At the beginning of the stage the coordination device observes either $\mathbf{R_{1}} = \mathbb{H}$ or $\mathbf{R_{1}} = \mathbb{T}$. For both, we must consider the two deviations: (a) the $\AO$ adheres to the recommendation but the $\TO$ deviates and (b) the $\TO$ adheres but the $\AO$ deviates. We consider the resulting four possibilities in turn.}
\subsubsection{\textbf{$\mathbf{R_{1}}$ = $\mathbb{H}$}}
\SG{Suppose the networks follow the recommended action profile ($\mathcal{A},\mathcal{B}$) in the stage $1$}. The resulting payoffs, respectively, of the $\AO$ and the $\TO$, conditioned on $\mathbf{R_{1}}$ = $\mathbb{H}$ and the action profile in stage $1$, are given by
\begin{subequations}
{\small
\begin{align}
    &\UW{\mathbbm{C}}|_{\{\mathbb{H},\text{($\mathcal{A},\mathcal{B}$)}\}} = (1-\alpha)E\left[\sum_{n=2}^{\infty}\alpha^{n-1}\uW{\mathbbm{C}}\right],\label{Eq:H_AB_TON}\\
    &\UD{\mathbbm{C}}|_{\{\mathbb{H},\text{($\mathcal{A},\mathcal{B}$)}\}} = -(1-\alpha)\left(E[\AvginitAoIT{2}{}]+E\left[\sum_{n=2}^{\infty}\alpha^{n-1}\uD{\mathbbm{C}}\right]\right),\label{Eq:H_AB_AON}\\
    &\text{where }E[\AvginitAoIT{2}{}] = \AvginitAoIT{1}{}(1-\tauDC(1-\tauDC)^{(\ND-1)}) + \lcol +\nonumber\\ &(1-\tauDC)^{\ND}(\lidle-\lcol) + \ND\tauDC(1-\tauDC)^{(\ND-1)}(\lsucc-\lcol).\nonumber
\end{align}}
\end{subequations}
Since the $\TO$ backs-off its stage $1$ throughput is $0$.

In case the $\AO$ unilaterally deviates, that is it backs-off, the age increases by the idle slot length. The action profile is ($\mathcal{B},\mathcal{B}$). Given the grim trigger etiquette, stage $2$ onward both networks play the MSNE. The discounted payoff obtained by the $\AO$, denoted by $\UD{\mathbbm{NC}}|_{\{\mathbb{H},\text{($\mathbfcal{B},\mathcal{B}$)}\}}$, where the bold $\mathbfcal{B}$ emphasizes the deviation, is given by~(\ref{Eq:H_BB_AON}). On the other hand, if the $\TO$ unilaterally deviates, the resulting action profile is ($\mathcal{A},\mathcal{A}$), and the $\TO$ gets an \SG{network throughput} larger than $0$ in stage $1$. Given the grim trigger etiquette, its resulting discounted payoff is given by~(\ref{Eq:H_AA_TON}). 
\begin{subequations}
{\small
\begin{align}
    \UD{\mathbbm{NC}}|_{\{\mathbb{H},\text{($\mathbfcal{B},\mathcal{B}$)}\}} &= -(1-\alpha)\left((\AvginitAoIT{1}{}+\lidle)+\right.\nonumber\\
    &\left.\qquad\qquad E\left[\sum_{n=2}^{\infty}\alpha^{n-1}\uD{\mathbbm{NC}}\right]\right)\label{Eq:H_BB_AON},\\
    \UW{\mathbbm{NC}}|_{\{\mathbb{H},\text{($\mathcal{A},\mathbfcal{A}$)}\}} &= (1-\alpha)\left( \tauWC(1-\tauWC)^{(\NW-1)}(1-\tauDC)^{\ND} \lsucc r+\right.\nonumber\\
    &\left.E\left[\sum_{n=2}^{\infty}\alpha^{n-1}\uW{\mathbbm{NC}}\right]\right).\label{Eq:H_AA_TON}
\end{align}
}
\end{subequations}

The $\AO$ and the $\TO$ would want to deviate only if their resulting payoffs while competing were larger than when obeying the device. The equations~(\ref{Eq:Inq_1})-(\ref{Eq:Inq_2}) next capture the conditions under which both networks will always obey the coordination device. 
\begin{subequations}
\begin{align}
\UD{\mathbbm{C}}|_{\{\mathbb{H},\text{($\mathcal{A},\mathcal{B}$)}\}}\geq\UD{\mathbbm{NC}}|_{\{\mathbb{H},\text{($\mathbfcal{B},\mathcal{B}$)}\}},\label{Eq:Inq_1}\\
\UW{\mathbbm{C}}|_{\{\mathbb{H},\text{($\mathcal{A},\mathcal{B}$)}\}}\geq\UW{\mathbbm{NC}}|_{\{\mathbb{H},\text{($\mathcal{A},\mathbfcal{A}$)}\}}.\label{Eq:Inq_2}
\end{align}
\end{subequations}

\subsubsection{\textbf{$\mathbf{R_{1}}$ = $\mathbb{T}$}} The coordination device recommends the networks to play ($\mathcal{B},\mathcal{A}$). The resulting payoffs, respectively, of the $\AO$ and the $\TO$, conditioned on the action profile in stage $1$, are given by
%\begin{subequations}
\begin{subequations}
\small
\begin{align}
	&\UW{\mathbbm{C}}|_{\{\mathbb{T},\text{($\mathcal{B},\mathcal{A}$)}\}} = (1-\alpha)\left( \tauWC(1-\tauWC)^{(\NW-1)}\lsucc r+\right.\nonumber\\
	&\qquad\qquad\qquad\qquad\qquad\left.E\left[\sum_{n=2}^{\infty}\alpha^{n-1}\uW{\mathbbm{C}}\right]\right),\\
	&\UD{\mathbbm{C}}|_{\{\mathbb{T},\text{($\mathcal{B},\mathcal{A}$)}\}} = -(1-\alpha)\left(E[\AvginitAoIT{2}{}] + E\left[\sum_{n=2}^{\infty}\alpha^{n-1}\uD{\mathbbm{C}}\right]\right),\\
	&\text{where } E[\AvginitAoIT{2}{}] = \AvginitAoIT{1}{} + \lcol + (1-\tauDC)^{\ND}(\lidle-\lcol)\nonumber\\ &\qquad\qquad\qquad+\NW\tauWC(1-\tauWC)^{(\NW-1)}(\lsucc-\lcol).\nonumber
\end{align}
\normalsize
\end{subequations}

Similarly to the earlier case when $\mathbf{R_{1}}$ = $\mathbb{H}$, we can calculate the payoffs obtained by the $\AO$ and the $\TO$, respectively, when they unilaterally deviate as
\begin{subequations}
\small
\begin{align}
    &\UW{\mathbbm{NC}}|_{\{\mathbb{T},\text{($\mathcal{B},\mathbfcal{B}$)}\}} = (1-\alpha)E\left[\sum_{n=2}^{\infty}\alpha^{n-1}\uW{\mathbbm{NC}}\right],\\
    &\UD{\mathbbm{NC}}|_{\{\mathbb{T},\text{($\mathbfcal{A},\mathcal{A}$)}\}} = -(1-\alpha)\left(E[\AvginitAoIT{2}{}] + E\left[\sum_{n=2}^{\infty}\alpha^{n-1}\uD{\mathbbm{NC}}\right]\right),\\
    &\text{where }E[\AvginitAoIT{2}{}] = \AvginitAoIT{1}{}(1-\tauDC(1-\tauDC)^{(\ND-1)}(1-\tauWC)^{\NW})\nonumber\\
    &\qquad\qquad+ \lcol +(1-\tauWC)^{\NW}(1-\tauDC)^{\ND}(\lidle-\lcol)\nonumber\\
    &\qquad\qquad+ \ND\tauDC(1-\tauDC)^{\ND-1}(1-\tauWC)^{\NW}\nonumber\\
    &\qquad\qquad+\NW\tauWC(1-\tauWC)^{(\NW-1)}(1-\tauDC)^{\ND}(\lsucc-\lcol).\nonumber
\end{align}
\normalsize
\end{subequations}

The equations~(\ref{Eq:Inq_3})-(\ref{Eq:Inq_4}) capture the conditions under which both networks will always obey the coordination device.
\begin{subequations}
\begin{align}
\UD{\mathbbm{C}}|_{\{\mathbb{T},\text{($\mathcal{B},\mathcal{A}$)}\}}\geq\UD{\mathbbm{NC}}|_{\{\mathbb{T},\text{($\mathbfcal{A},\mathcal{A}$)}\}},\label{Eq:Inq_3}\\
\UW{\mathbbm{C}}|_{\{\mathbb{T},\text{($\mathcal{B},\mathcal{A}$)}\}}\geq\UW{\mathbbm{NC}}|_{\{\mathbb{T},\text{($\mathcal{B},\mathbfcal{B}$)}\}}.\label{Eq:Inq_4}
\end{align}
\end{subequations}

We state the requirement for cooperation to be self-enforceable.
\SG{\begin{statement}
Cooperation is self-enforceable in the repeated game via grim trigger strategies if there exists $\overline{\alpha}\in(0,1)$, such that $\forall \alpha>\overline{\alpha}$, $\exists \PR\in(0,1)$, such that the grim trigger strategy profile in~(\ref{Eq:PathofPlay}) is a subgame perfect equilibrium (SPE).
\label{stmt:coop}
\end{statement}}
\SG{The set of $(\overline{\alpha}, \PR)$ for which the Statement~\ref{stmt:coop} is true can be obtained using the equilibrium incentive constraints specified by~(\ref{Eq:Inq_1})-(\ref{Eq:Inq_2}) and~(\ref{Eq:Inq_3})-(\ref{Eq:Inq_4}). We resort to computational analysis. In Section~\ref{sec:results} we show that the existence of a non-empty set of $(\overline{\alpha}, \PR)$ is dependent on the size of the $\AO$ and the $\TO$.}
\SG{\begin{observation}
Cooperation is self-enforceable (Statement \ref{stmt:coop}) for smaller networks. However, as the networks grow in size, competition becomes more favorable than cooperation, the SPE ceases to exist, and cooperation is not self-enforceable.
\label{prop:coop}
\end{observation}}
\section{Evaluation Methodology and Results}
\label{sec:results}
\sg{We study two scenarios (a) when $\lsucc > \lcol$ and (b) when $\lsucc \leq \lcol$. In practice, the idle slot is much smaller than a collision or a successful transmission slot. We set $\lidle = \beta << 1$. For the shown results, when $\lsucc > \lcol$, we set $\lsucc = (1+\beta)$ and $\lcol = 0.1(1+\beta)$. When evaluating $\lsucc < \lcol$, we set $\lsucc = (1+\beta)$ and $\lcol = 2(1+\beta)$. Lastly, we set $\lsucc = \lcol = (1+\beta)$ when $\lsucc = \lcol$. The results presented later use $\beta = 0.01$.\footnote{\sg{The selection of slot lengths is such that the ratio $(\lsucc-\lidle)/\lsucc$ for the simulation setup is approximately the same as that for $802.11$ac~\cite{80211ac} based WiFi devices and $802.11$p~\cite{80211p} based vehicular network.}}} To illustrate the impact of self-contention and competition, we simulated $\ND \in \{1, 2, 5, 10, 50\}$ and $\NW \in \{1, 2, 5, 10, 50\}$. To show when the networks cooperate, we simulated the discount factor $\alpha \in [0.01,0.99]$ and the coordination device $\PR \in [0.01,0.99]$. We used Monte Carlo simulations to compute the average discounted payoff of the $\AO$ and the $\TO$. Averages were calculated over $100,000$ independent runs each comprising of $1000$ stages. We set the rate of transmission $r=1$ bit/sec for each node in the WiFi network.

We begin by studying the impact of the length of collision slot $\lcol$ on the average discounted payoff when (a) networks play the MSNE in each stage and compete for the medium (payoffs $\mathrm{U}_{\mathrm{T},\mathbbm{NC}}$ and $\mathrm{U}_{\mathrm{A},\mathbbm{NC}}$), and (b) networks obey the recommendation of the coordination device $\PR$ in each stage and hence cooperate, (payoffs $\mathrm{U}_{\mathrm{T},\mathbbm{C}}$ and $\mathrm{U}_{\mathrm{A},\mathbbm{C}}$). We show that when networks compete, while nodes in the $\AO$ occasionally choose to refrain from transmitting during a stage when $\lsucc\leq\lcol$, they choose to access the medium aggressively when $\lsucc>\lcol$. Note that the nodes in the $\TO$, however, access the shared spectrum independently of the ordering of the $\lsucc$ and $\lcol$ (see~(\ref{Eq:Thr_MSNE_RTS_CTS}) and~(\ref{Eq:Thr_MSNE_BA})). Such behavior when competing impacts the desirability of cooperation over competition.

We show the region of cooperation, i.e., the range of $\alpha$ and $\PR$ for which the inequalities~(\ref{Eq:Inq_1})-(\ref{Eq:Inq_2}) and (\ref{Eq:Inq_3})-(\ref{Eq:Inq_4}) are satisfied and the repeated game has a SPE supported with the coordination device $\PR$. We discuss why cooperation isn't enforceable and the SPE ceases to exist in the repeated game, as the number of nodes in the networks increases.

\begin{figure}[t]
  \centering
  \subfloat[Empirical frequency of $\tauD^{*} = 1$, $\mathbf{f_{\tauD^{*} = 1}}$ ($\lsucc>\lcol$)]{\includegraphics[width=0.49\columnwidth]{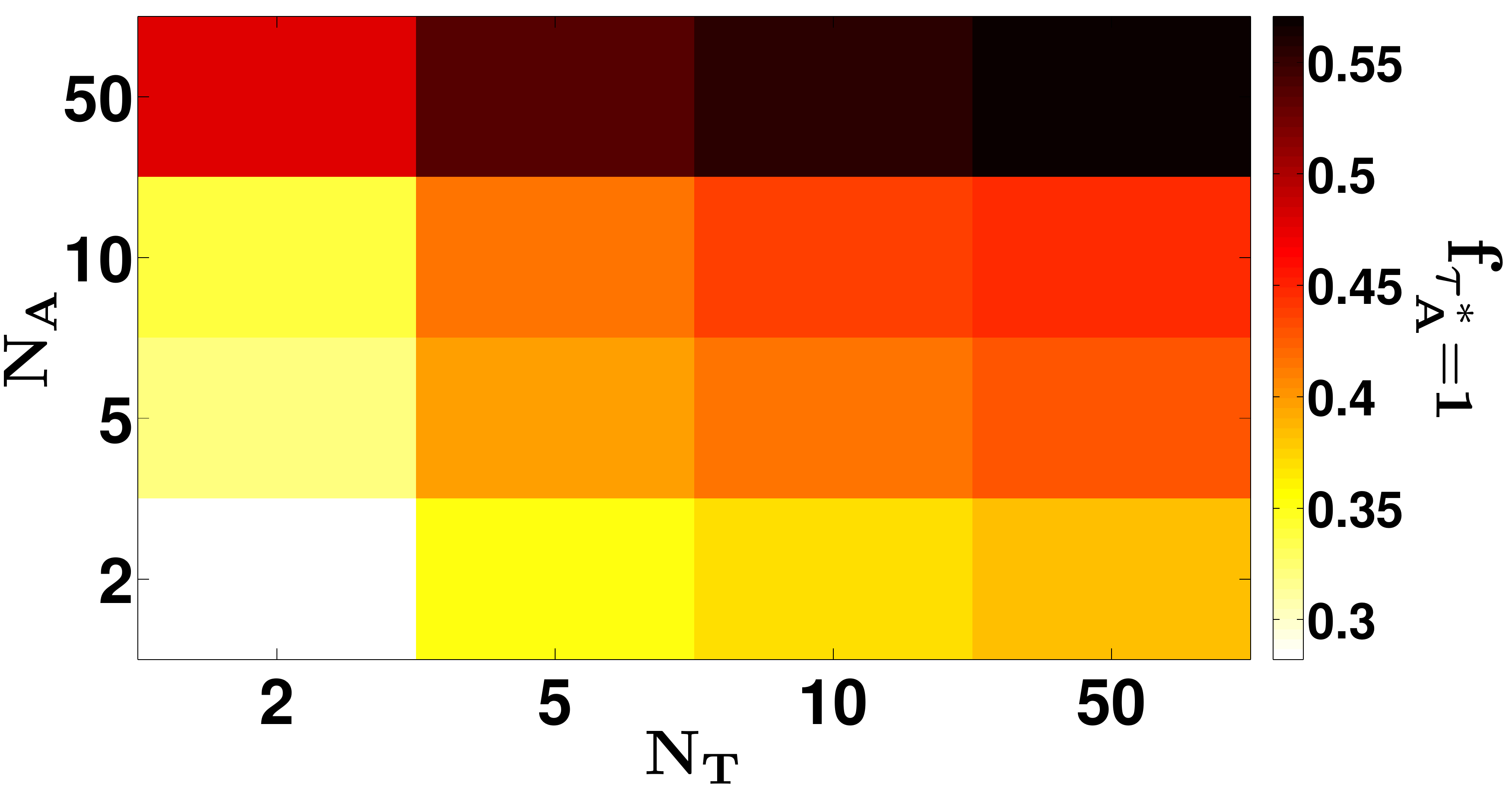}\label{fig:freqTd1}}
  \subfloat[\sg{Empirical frequency of $\tauD^{*} = 0$, $\mathbf{f_{\tauD^{*} = 0}}$ ($\lsucc=\lcol$)}]{\includegraphics[width=0.48\columnwidth]{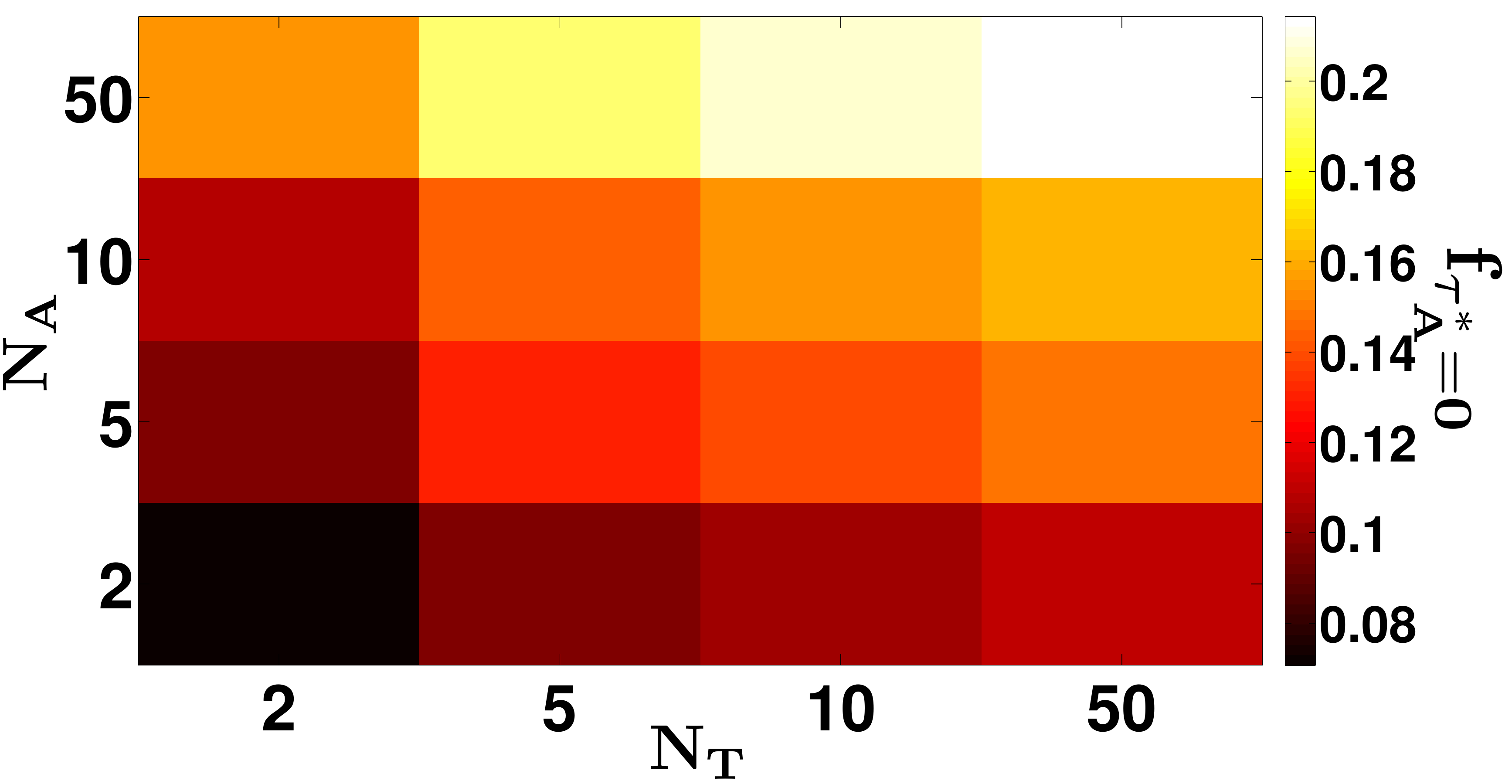}\label{fig:freqTd0}}
\caption{\small \sg{Average empirical frequency of occurrence of $\tauD^{*}=1$ ($\mathbf{f_{\tauD^{*}=1}}$) and $\tauD^{*}= 0$ ($\mathbf{f_{\tauD^{*}=0}}$) for different scenarios computed over $100,000$ independent runs with $1000$ stages each when networks choose to play the MSNE in each stage. Figure~\ref{fig:freqTd1} and Figure~\ref{fig:freqTd0} correspond to when $\lsucc>\lcol$ and $\lsucc=\lcol$, respectively. The results correspond to $\lsucc = 1+\beta$, $\lidle = \beta$, $\lcol = \{0.1\lsucc,\lsucc\}$ and $\beta = 0.01$.}}
\label{fig:FreqOfTauD}
\end{figure}

\textbf{\textit{Impact of $\lcol$ on network payoffs in the repeated game with competition:}} 
Let $\mathbf{f_{\tauD^{*}=1}}$ and $\mathbf{f_{\tauD^{*}=0}}$ denote the \SG{average empirical frequency} of occurrence of $\tauD^{*}=1$, $\tauD^{*}= 0$, respectively. We computed these over the independent runs of the repeated game. Figure~\ref{fig:FreqOfTauD} shows these frequencies for different sizes of the $\AO$ and the $\TO$ when networks choose to play the MSNE in each stage, for the cases $\lsucc > \lcol$ and $\lsucc=\lcol$. We skip $\lsucc<\lcol$ as the observations are similar to $\lsucc = \lcol$.

Figure~\ref{fig:freqTd1} shows how $\mathbf{f_{\tauD^{*} = 1}}$ varies as a function of the number of nodes in the $\AO$ and the $\TO$ for when $\lsucc>\lcol$. Observe the increase in $\mathbf{f_{\tauD^{*} = 1}}$ as $\ND$ increases. This is explained by the resulting increase in the threshold age $\ND(\lsucc - \lcol)$ (see~(\ref{Eq:Age_MSNE_RTS_CTS})). On the other hand, when $\lsucc = \lcol$, the $\AO$ refrains from transmission more often as the number of nodes $\ND$ in it increases. See Figure~\ref{fig:freqTd0} that shows the increase in $\mathbf{f_{\tauD^{*} = 0}}$. 

\begin{figure}[t]
  \centering
  \subfloat[$\TO$ average discounted payoff]{\includegraphics[width=0.49\columnwidth]{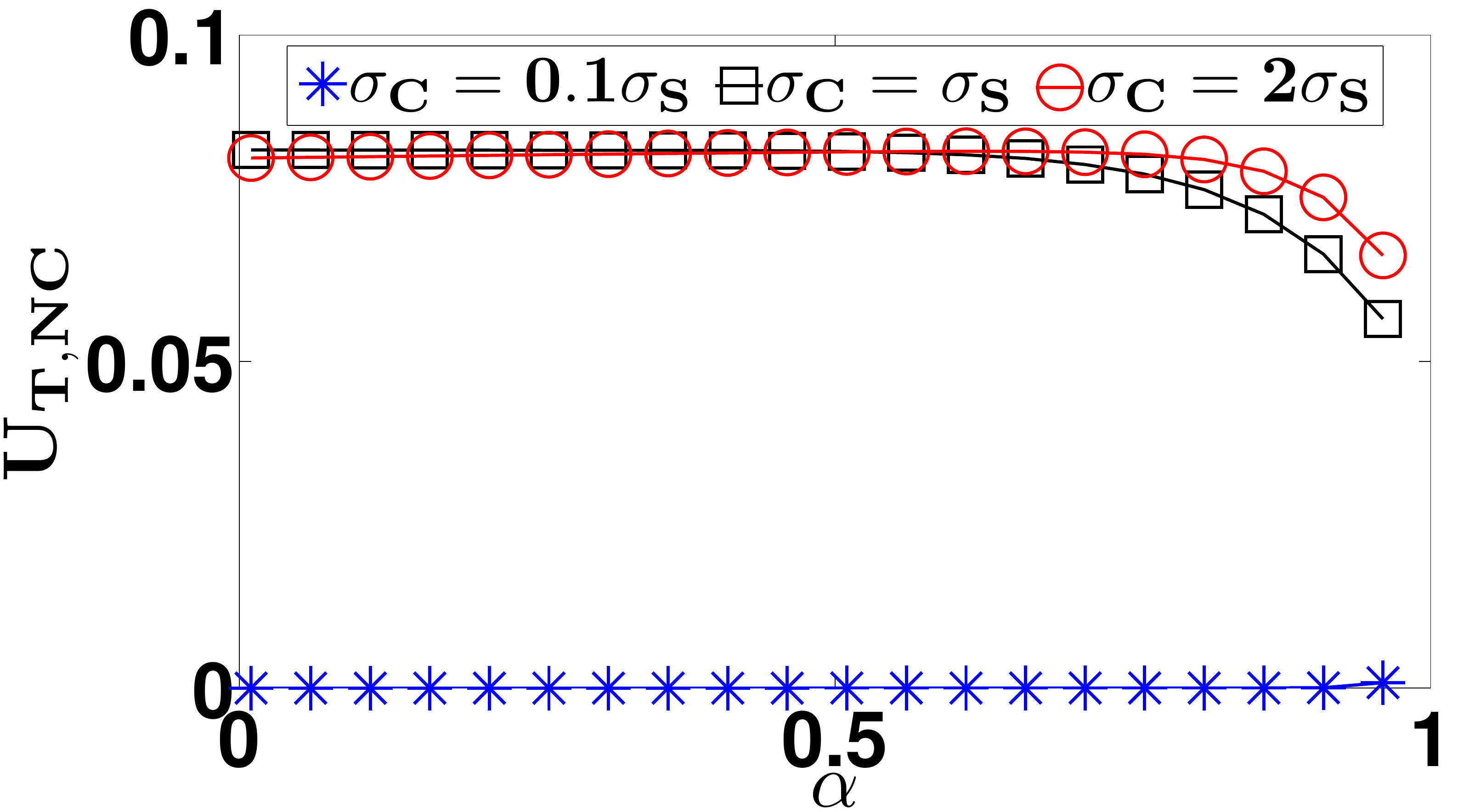}\label{fig:uw_avg_payoff}}
  \enspace
  \subfloat[$\AO$ average discounted payoff]{\includegraphics[width=0.48\columnwidth]{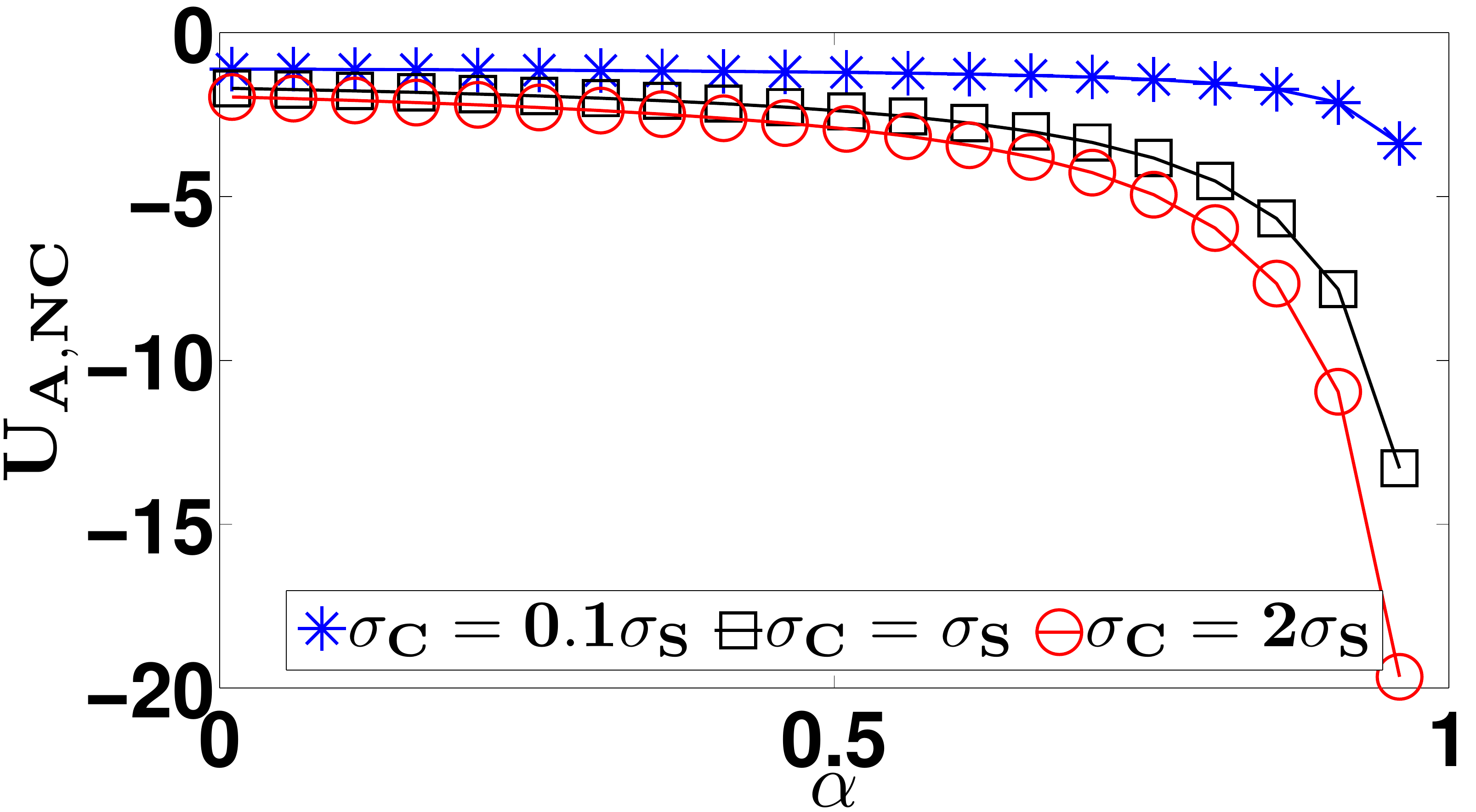}\label{fig:ua_avg_payoff}}
\caption{\small Average discounted payoff of the $\TO$ and the $\AO$ for $\NW = \ND = 5$ when networks choose to play MSNE in each stage. We set $\lsucc = 1+\beta$, $\lidle = \beta$, $\lcol = \{0.1\lsucc,\lsucc,2\lsucc\}$ and $\beta = 0.01$.}
\label{fig:AvgDisPayoffvsalpha}
\end{figure}

The increase in $\mathbf{f_{\tauD^{*} = 1}}$ with $\ND$, when $\lsucc>\lcol$, increases the fraction of slots occupied by the $\AO$. The resulting increased competition from the $\AO$ for the shared access adversely impacts the $\TO$. In contrast, the increase in $\mathbf{f_{\tauD^{*} = 0}}$ with $\ND$, when $\lsucc\le \lcol$, results in larger fraction of contention free slots for the $\TO$ and works in its favour. The impact of slot sizes on the average discounted payoff of the $\TO$ is summarized in Figure~\ref{fig:uw_avg_payoff}, which shows this payoff for different selections of $\lcol$. In accordance with the above observations, the payoff increases with the length of the collision slot.

Further note that an increase in $\mathbf{f_{\tauD^{*} = 1}}$ with $\ND$ should result in the $\AO$ seeing collision slots more often. However, as shown in Figure~\ref{fig:ua_avg_payoff}, despite this fact the average discounted payoff of the $\AO$ is larger when collision slots are smaller than the successful transmission slots. This is because when $\lsucc > \lcol$ and the $\AO$ chooses to transmit aggressively leading to collision, the increase in age due to a collision slot is smaller than when the $\AO$ chooses not to transmit. The latter choice has the AON see a slot that is either successful ($\TO$ transmits successfully), a collision (more than one node in the $\TO$ transmits), or an idle slot, and for a longer $\lsucc$, can be on an average longer than a collision slot.

\begin{figure}[t]
  \centering
  \subfloat[]{\includegraphics[width=0.49\columnwidth]{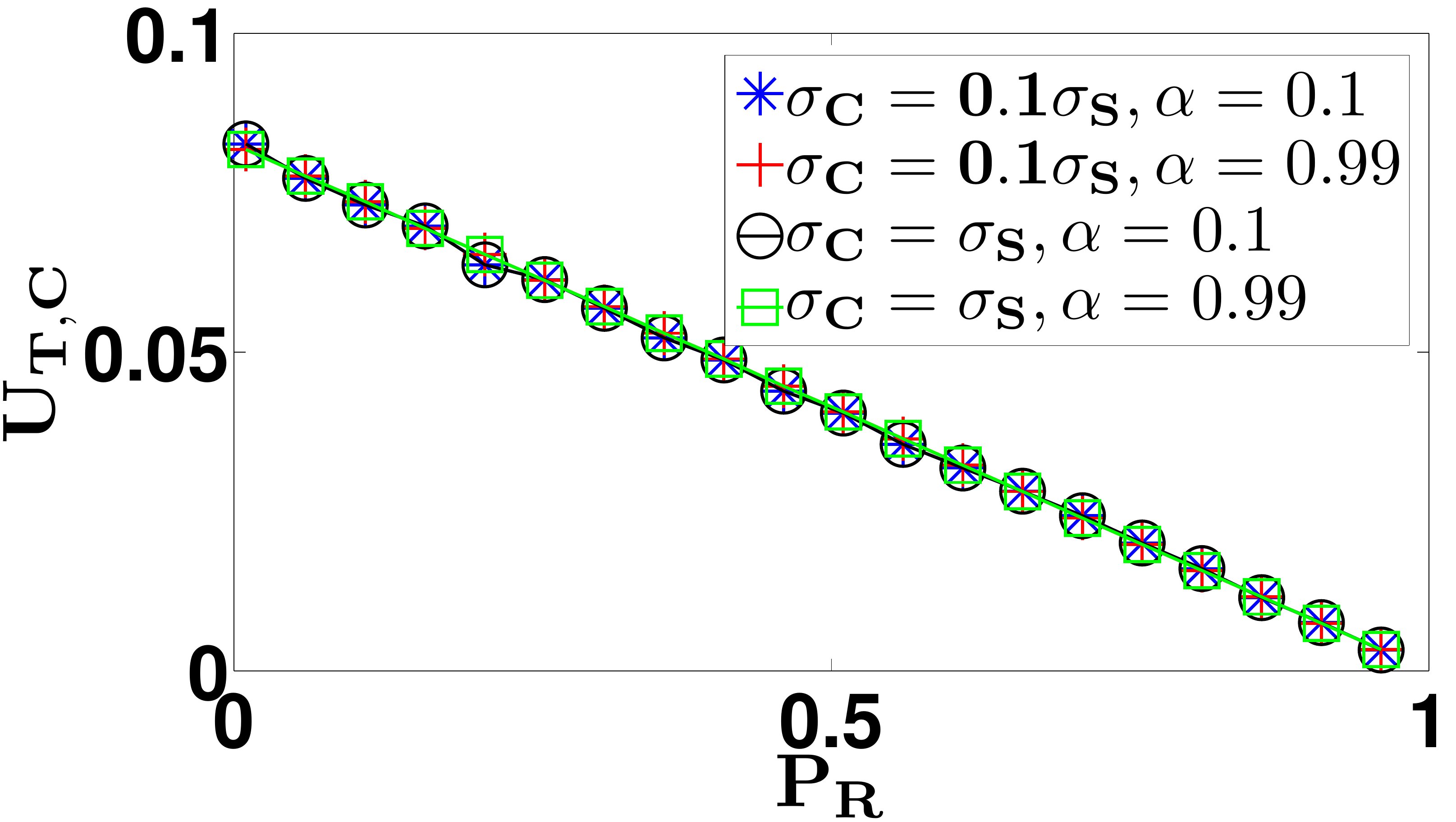}\label{fig:UWvaryalpha}}
  \enspace
  \subfloat[]{\includegraphics[width=0.49\columnwidth]{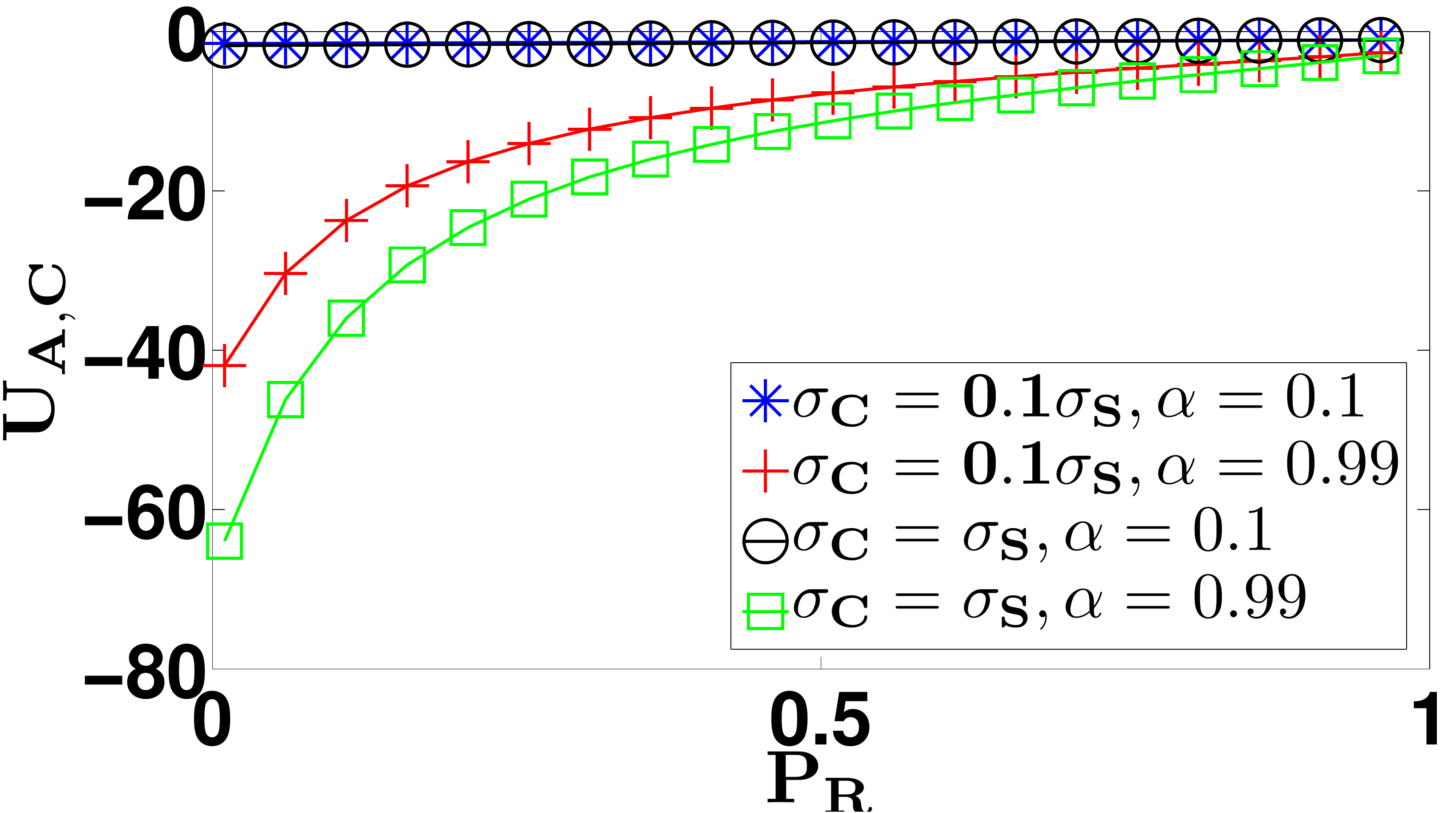}\label{fig:UDvaryalpha}}
\caption{\small Average discounted payoff of the $\TO$ and the $\AO$ for $\NW = \ND = 5$ when networks cooperate and follow the recommendation of the coordination device $\PR$ in each stage. Shown for $\alpha=0.1$, $\alpha=0.99$, $\lsucc = 1+\beta$, $\lidle = \beta$ and $\beta = 0.01$.}
\label{fig:AvgDisPayoffvsalphaPD}
\end{figure}

\textbf{\textit{Impact of $\lcol$ on network payoffs in the repeated game with cooperation:}} Figure~\ref{fig:AvgDisPayoffvsalphaPD} shows the average discounted payoff of the $\TO$ and the $\AO$ when networks cooperate. As shown in Figure~\ref{fig:UWvaryalpha}, the payoff of the $\TO$ when networks obey the recommendation of the coordination device $\PR$, is the same, irrespective of the choice of length of collision slot $\lcol$. This is because the optimal strategy of the $\TO$ (see~(\ref{Eq:WiFi_optimal})) is independent of $\lcol$. 

Figure~\ref{fig:UDvaryalpha} shows the payoff of the $\AO$ as a function of $\PR$. The payoff increases with $\PR$. This is expected as a larger $\PR$ implies that the $\AO$ gets to access the medium in a larger fraction of slots. Also seen in the figure is that a small collision slot (compare payoffs for $\lcol = 0.1\lsucc$ and $\lcol = \lsucc$) results in larger payoffs, especially at smaller values of $\PR \le 0.5$. At any given value of $\PR$, an increase in $\lcol$ for a given $\lsucc$, increases the average length of slots occupied by the $\TO$ and thus the \SG{network age}. At smaller $\PR$, a larger fraction of slots have the $\TO$ access, which makes the increase in age more significant.

\begin{figure}[t]
  \centering
  \subfloat[]{\includegraphics[width=0.48\columnwidth]{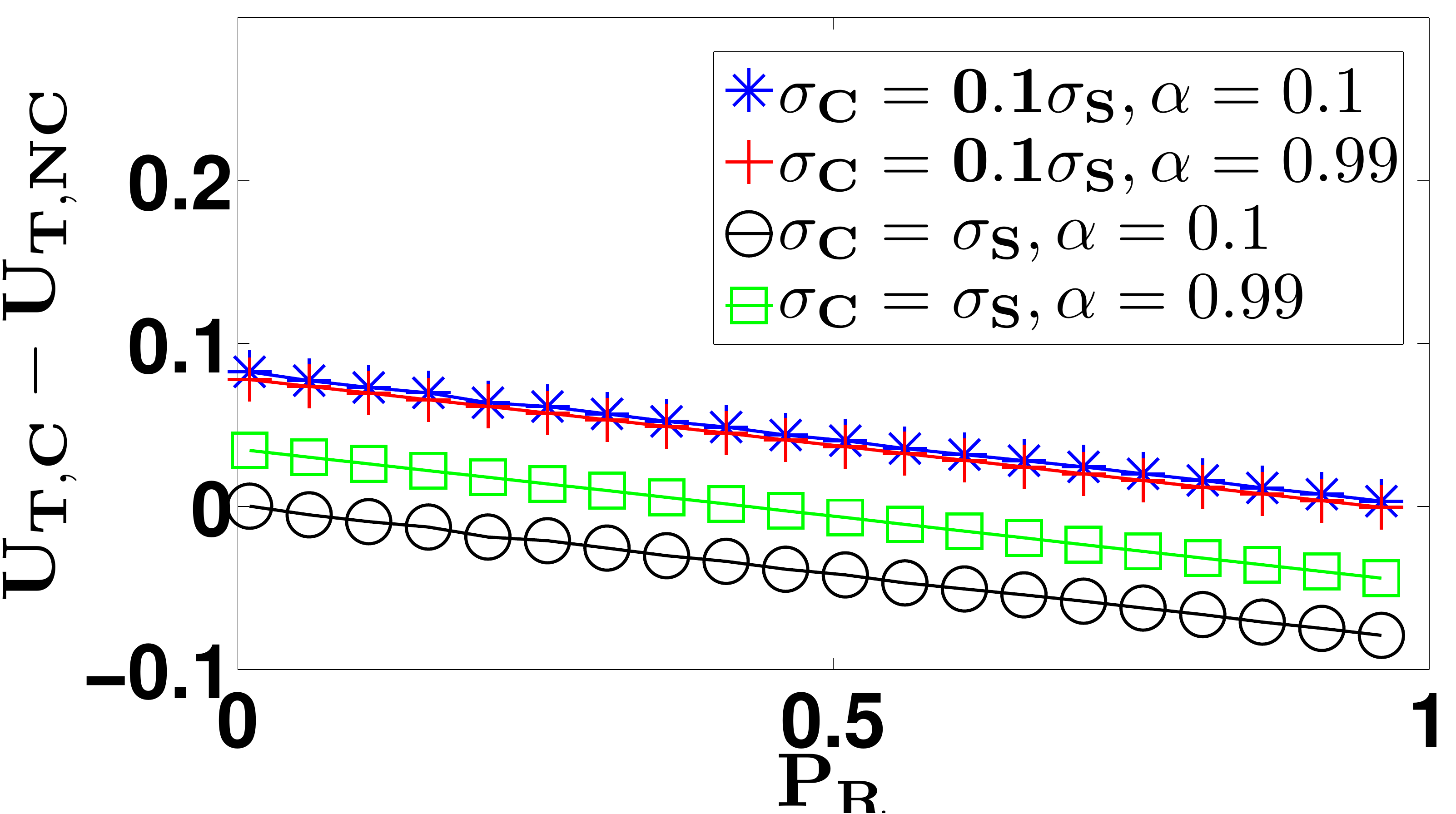}\label{fig:gainUW}}
  \enspace
  \subfloat[]{\includegraphics[width=0.49\columnwidth]{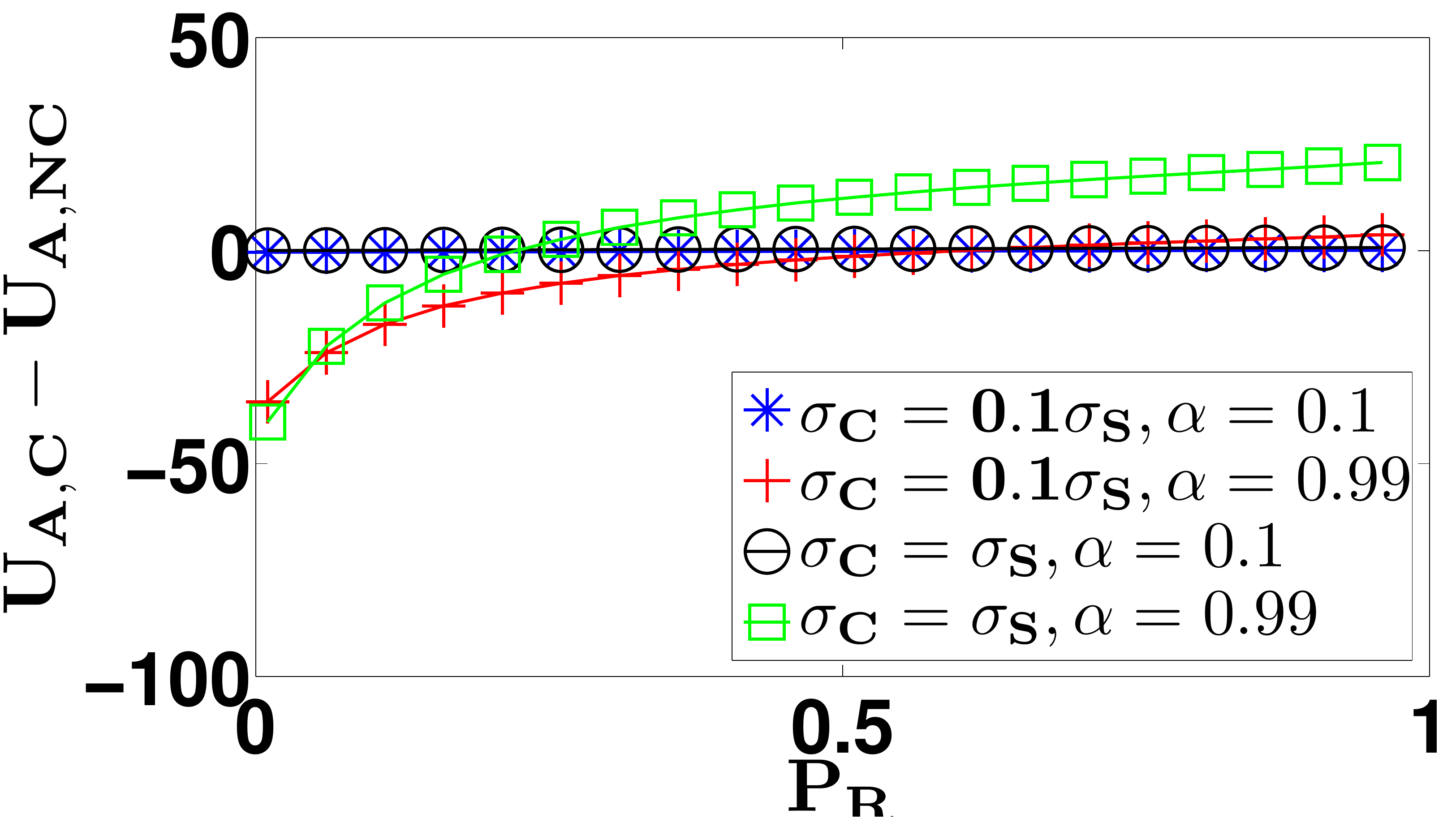}\label{fig:gainUD}}\\ 
\caption{\small Gain of cooperation over competition for the $\TO$ and the $\AO$ for $\NW = \ND = 5$. The results correspond to $\alpha=0.1$, $\alpha=0.99$, $\lsucc = 1+\beta$, $\lidle = \beta$, $\lcol=\{0.1\lsucc,\lsucc\}$ and $\beta = 0.01$.}
\label{fig:Gain}
\end{figure}

Figure~\ref{fig:Gain} shows the gains in payoff on choosing cooperation over competition for the $\AO$ and $\TO$. While the $\TO$ prefers cooperation to competition for smaller collision slots, the $\AO$ prefers cooperation for larger collision slots. As seen in Figure~\ref{fig:gainUW}, when $\lsucc>\lcol$, for all values of $\alpha$ and $\PR$, the payoff of the $\TO$ is higher when networks cooperate than when they compete. This is because, when $\lsucc>\lcol$, nodes in the $\AO$ transmit aggressively (see Figure~\ref{fig:FreqOfTauD}) when competing, making it less favorable for the $\TO$. 
On the other hand, for larger collision slots, as seen in Figure~\ref{fig:FreqOfTauD} for $\lcol\ge \lsucc$, the $\AO$ often refrains from transmission when competing. The resulting increase in slots free of contention from the $\AO$ makes competing favorable for the $\TO$. Finally, observe in Figure~\ref{fig:gainUW} that the gains from cooperation reduce as $\PR$ increases. As the fraction of slots available via the recommendation device decreases, the $\TO$ increasingly prefers competing over all slots.

Unlike the $\TO$, as shown in Figure~\ref{fig:gainUD}, as $\lcol$ increases, $\AO$ prefers cooperation. Also, the desirability of cooperation increases with $\PR$. As explained earlier, for $\lcol\geq\lsucc$, when competing the $\AO$ refrains from transmitting in a stage in case the age at the beginning is small enough. Such a slot has the length of one of successful, collision or idle slots, and is determined by the $\TO$. When cooperating such slots are always of length $\lidle$ of an idle slot.

Lastly, as shown in Figure~\ref{fig:Gain}, the gains from cooperation for both the networks are larger for higher value of $\alpha$ indicating that cooperation is more beneficial when the player is farsighted, i.e., it cares about long run payoff. For instance, as shown in Figure~\ref{fig:gainUW}, when $\lcol\ge \lsucc$, cooperation is more beneficial for the $\TO$ when $\alpha=0.99$ as compared to when $\alpha=0.01$. Similarly, as shown in Figure~\ref{fig:gainUD}, the benefits of cooperation for the $\AO$ increases with increase in $\alpha$.

\begin{figure}[t]
\centering
\subfloat[$N_{D}=2,N_{W}=2$, Region $\TO$ prefers cooperation.]{\includegraphics[width = 0.32\columnwidth]{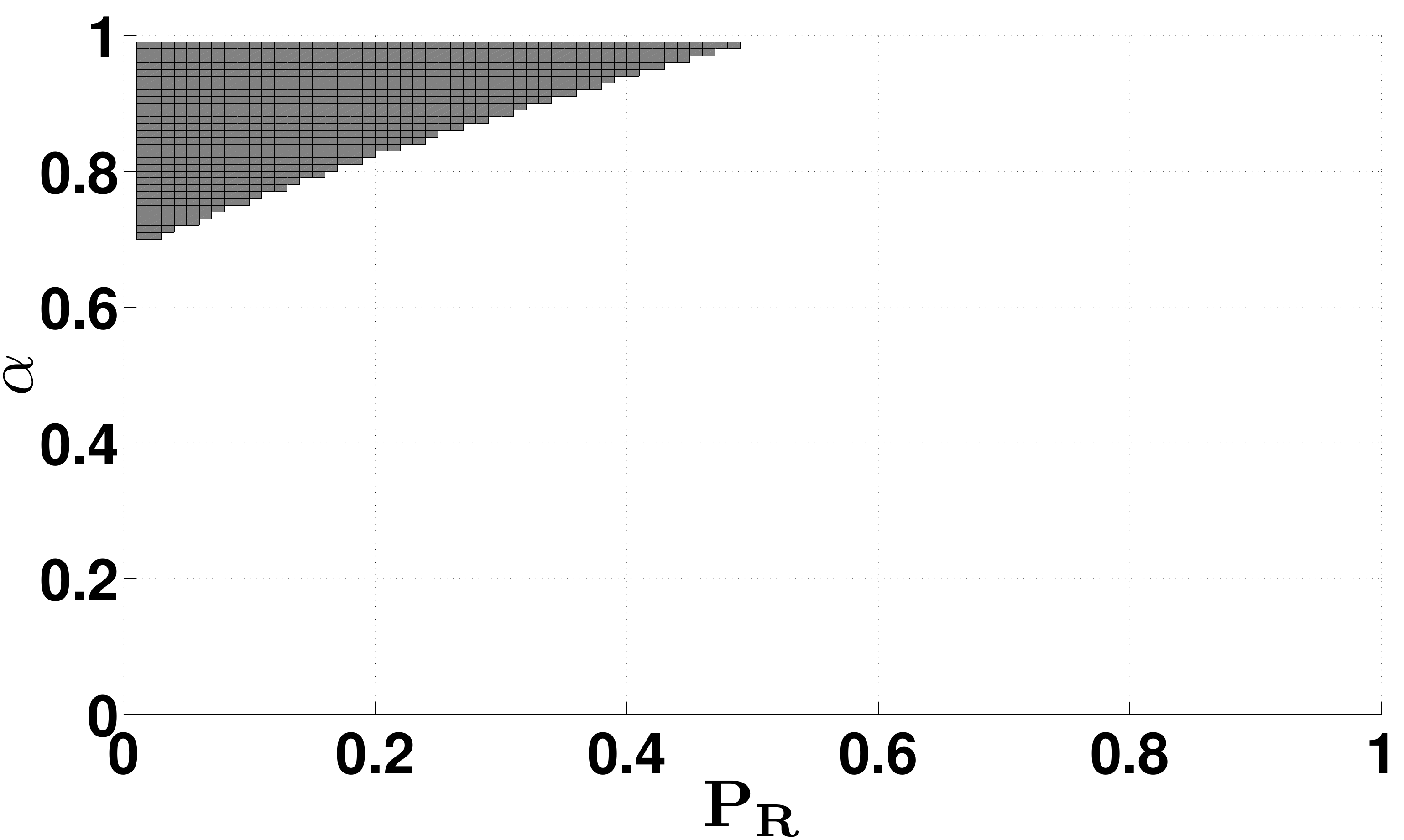}\label{fig:baTON2}}
\enspace
\subfloat[$N_{D}=2,N_{W}=2$, Region $\AO$ prefers cooperation.]{\includegraphics[width = 0.32\columnwidth]{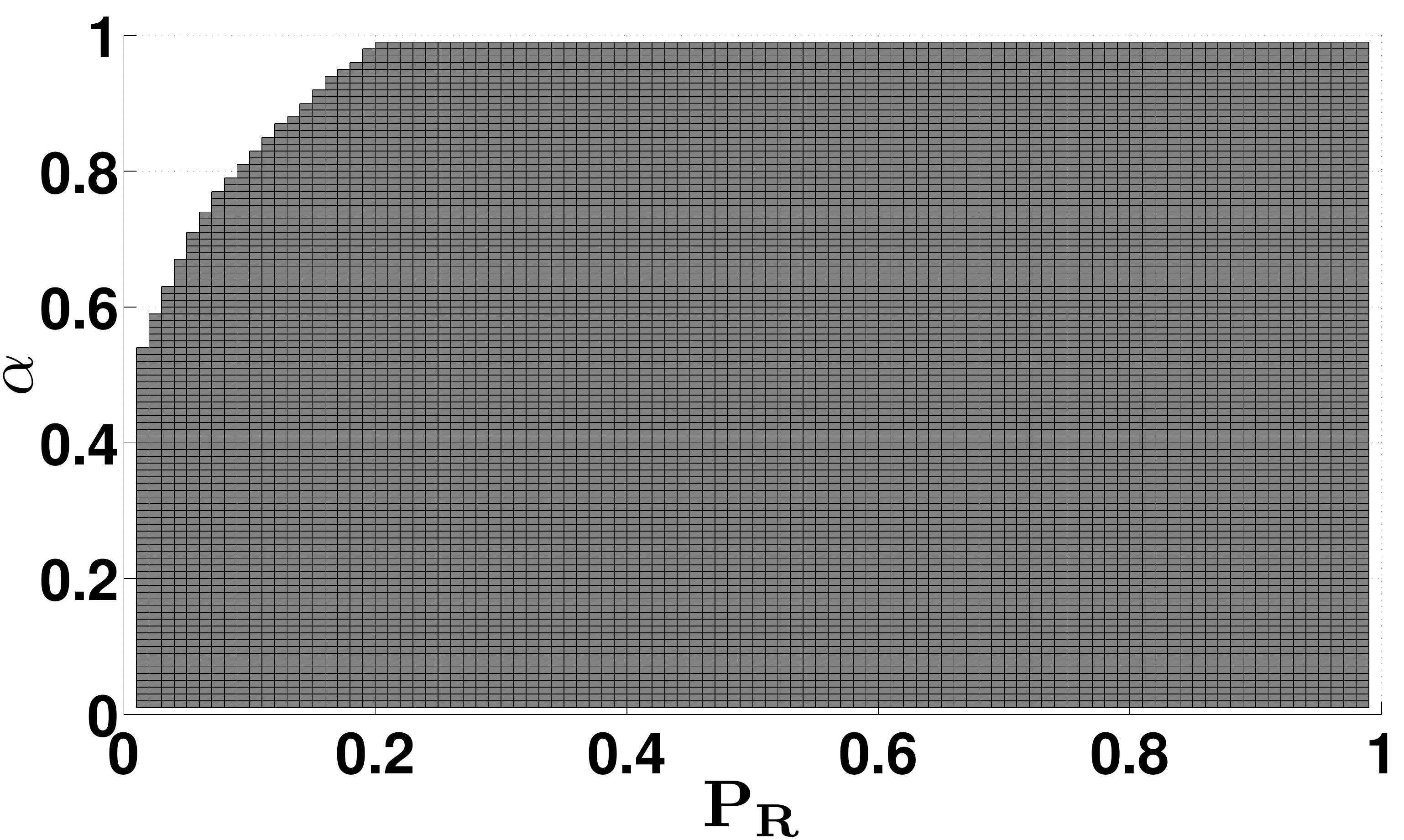}\label{fig:baAON2}}
\enspace
\subfloat[$N_{D}=2,N_{W}=2$, Region cooperation is self-enforceable.]{\includegraphics[width = 0.32\columnwidth]{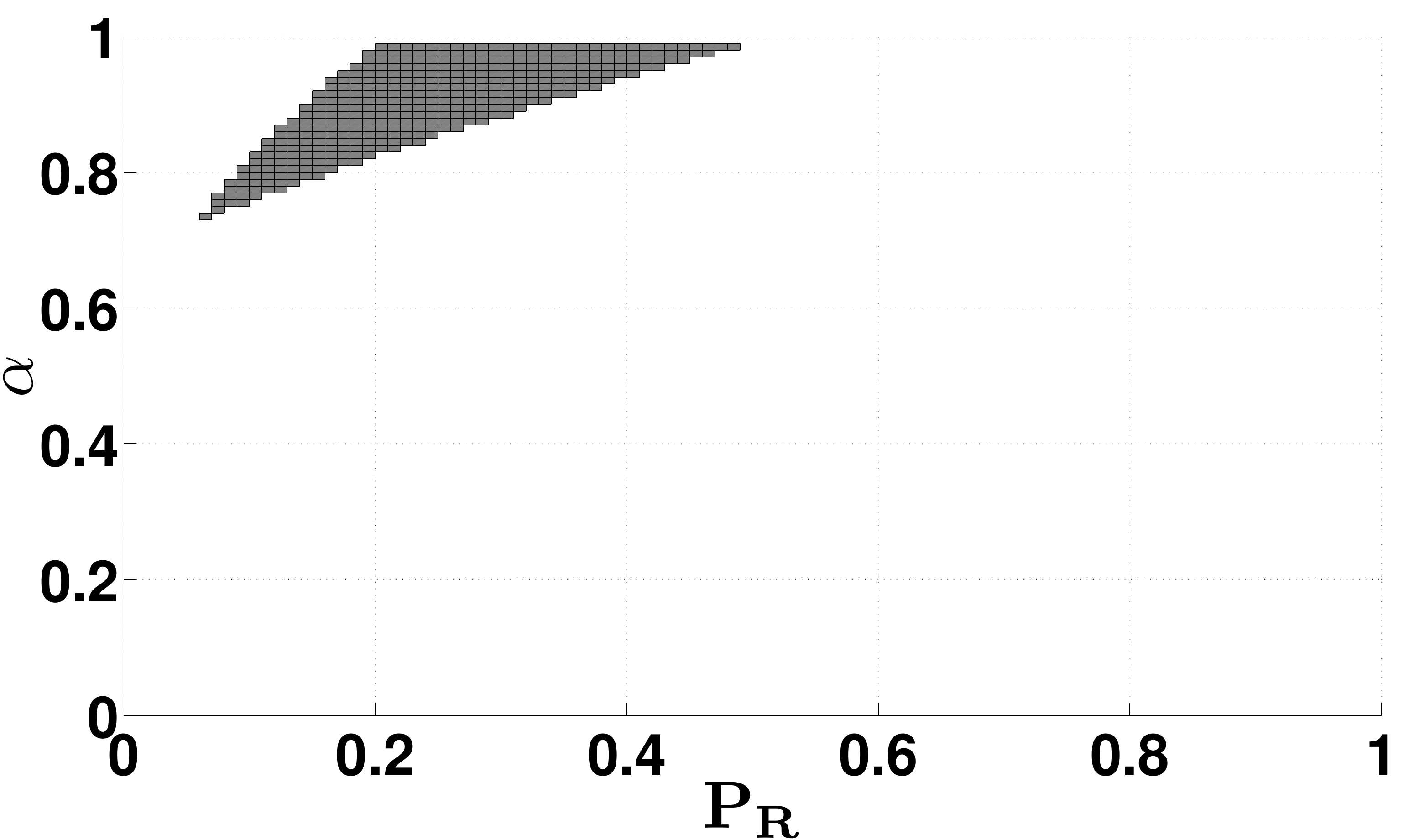}\label{fig:ba2}}\\
\subfloat[$N_{D}=5,N_{W}=5$, Region $\TO$ prefers cooperation.]{\includegraphics[width = 0.32\columnwidth]{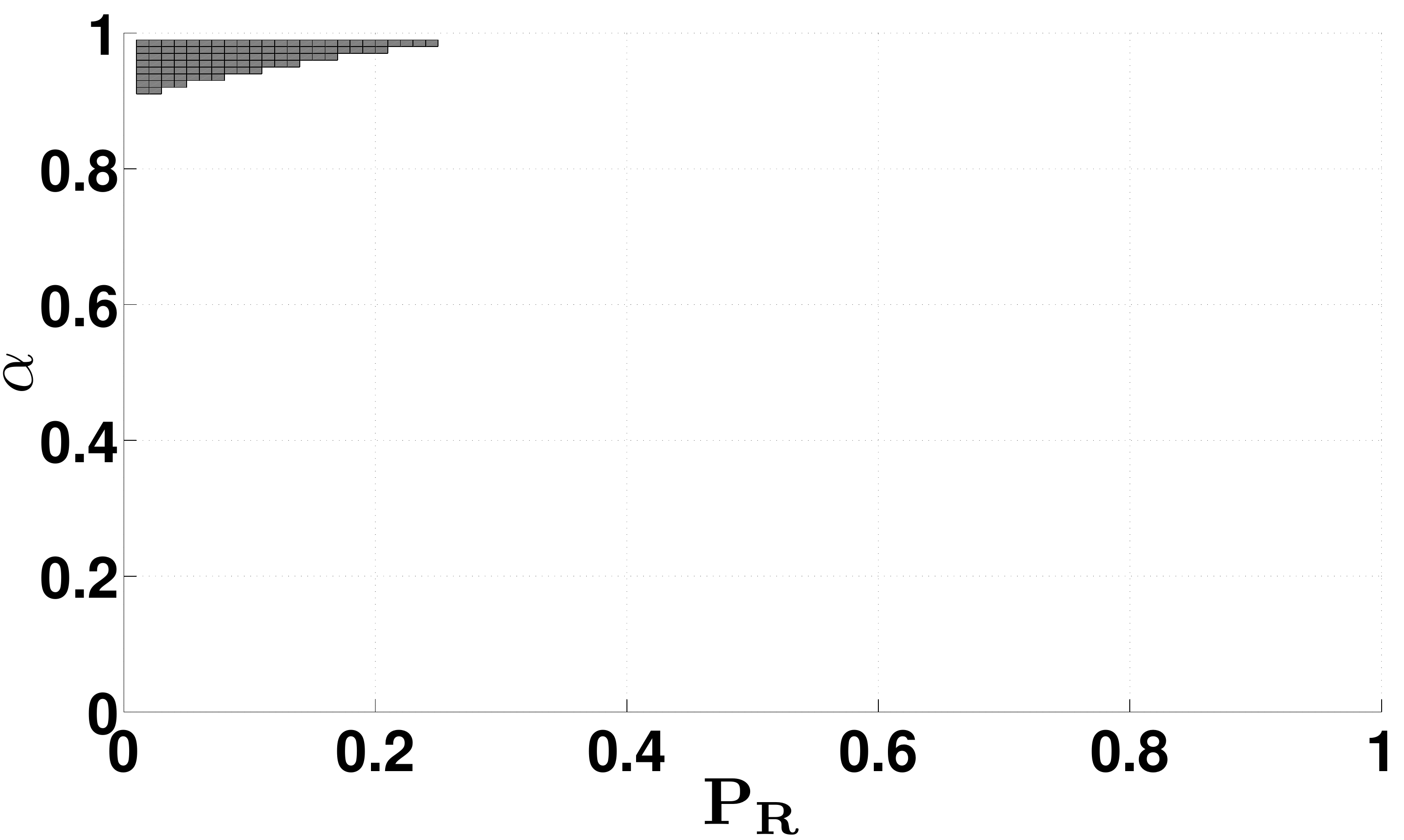}\label{fig:baTON5}}
\enspace
\subfloat[$N_{D}=5,N_{W}=5$, Region $\AO$ prefers cooperation.]{\includegraphics[width = 0.32\columnwidth]{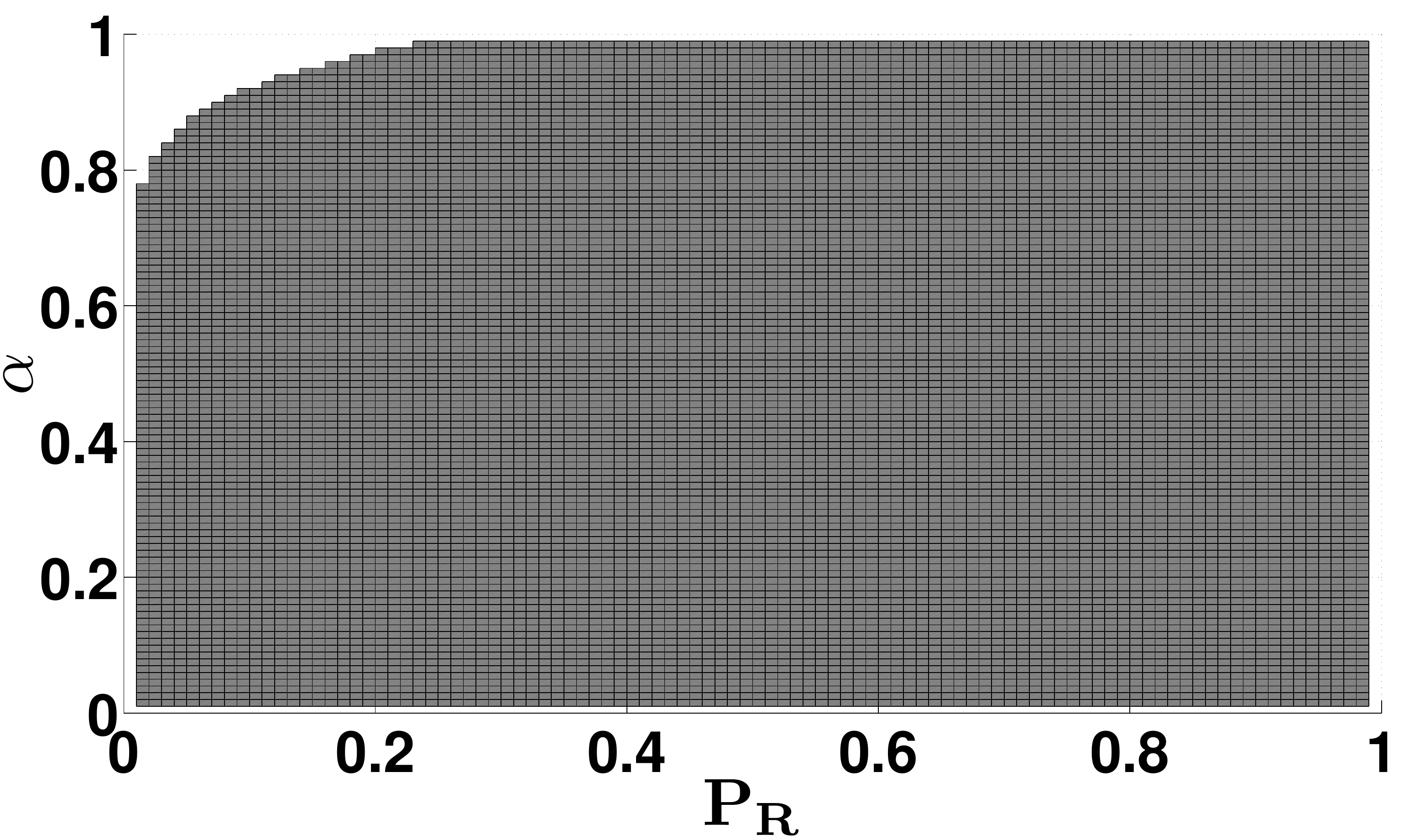}\label{fig:baAON5}}
\enspace
\subfloat[$N_{D}=5,N_{W}=5$, Region cooperation is self-enforceable.]{\includegraphics[width = 0.32\columnwidth]{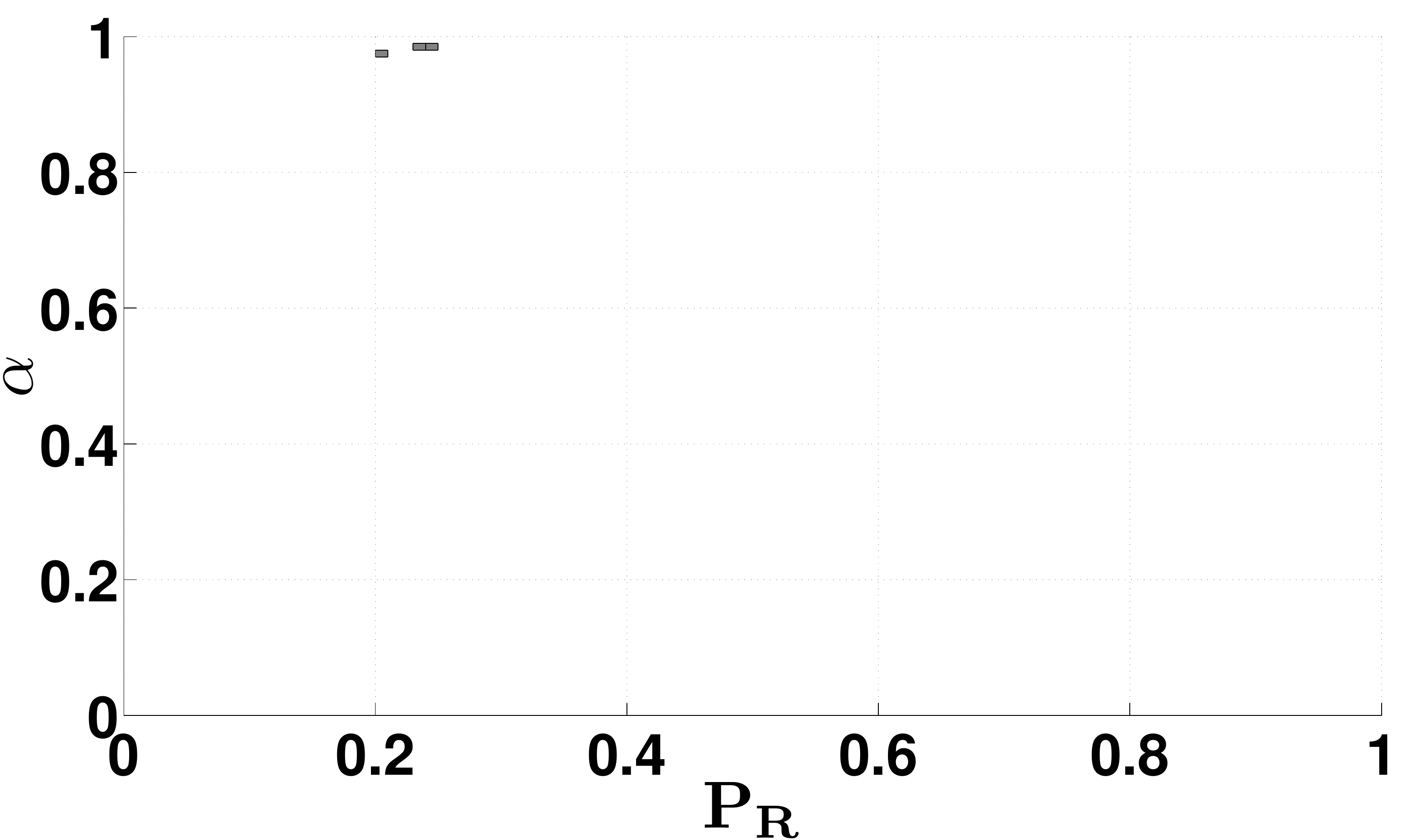}\label{fig:ba5}}\\
\subfloat[$N_{D}=10,N_{W}=10$, Region $\TO$ prefers cooperation.]{\includegraphics[width = 0.32\columnwidth]{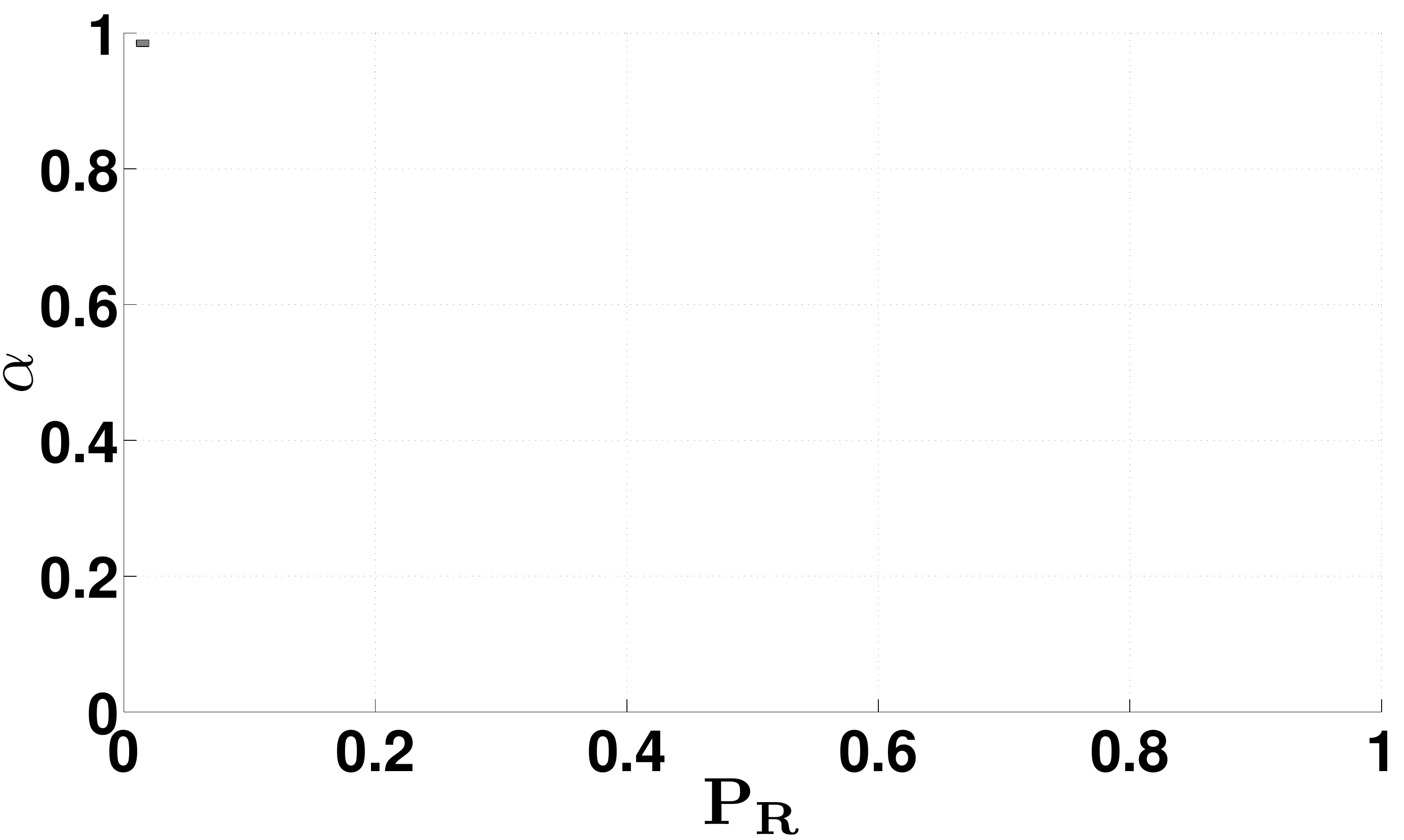}\label{fig:baTON10}}
\enspace
\subfloat[$N_{D}=10,N_{W}=10$, Region $\AO$ prefers cooperation.]{\includegraphics[width = 0.32\columnwidth]{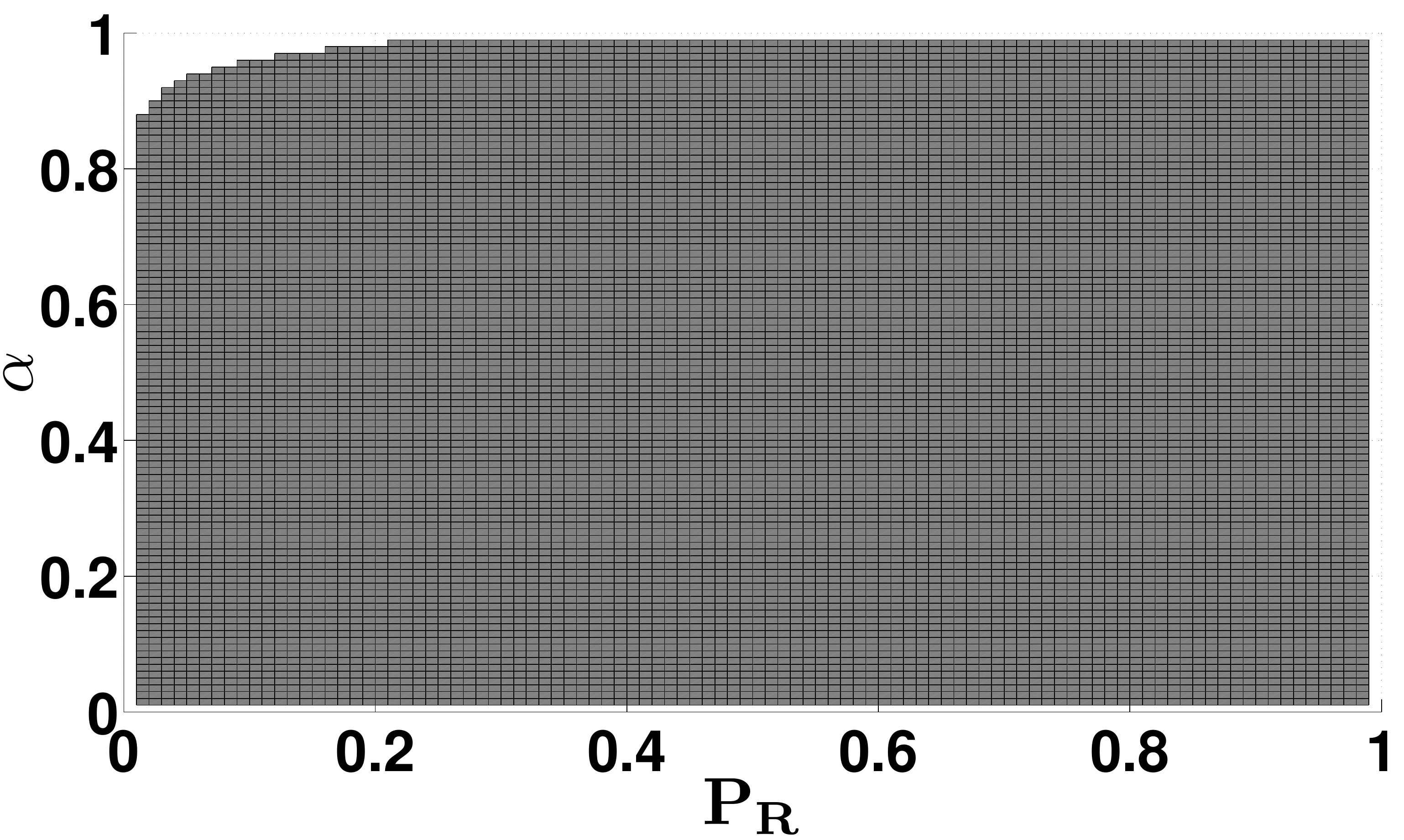}\label{fig:baAON10}}
\enspace
\subfloat[$N_{D}=10,N_{W}=10$, Region cooperation is self-enforceable.]{\includegraphics[width = 0.32\columnwidth]{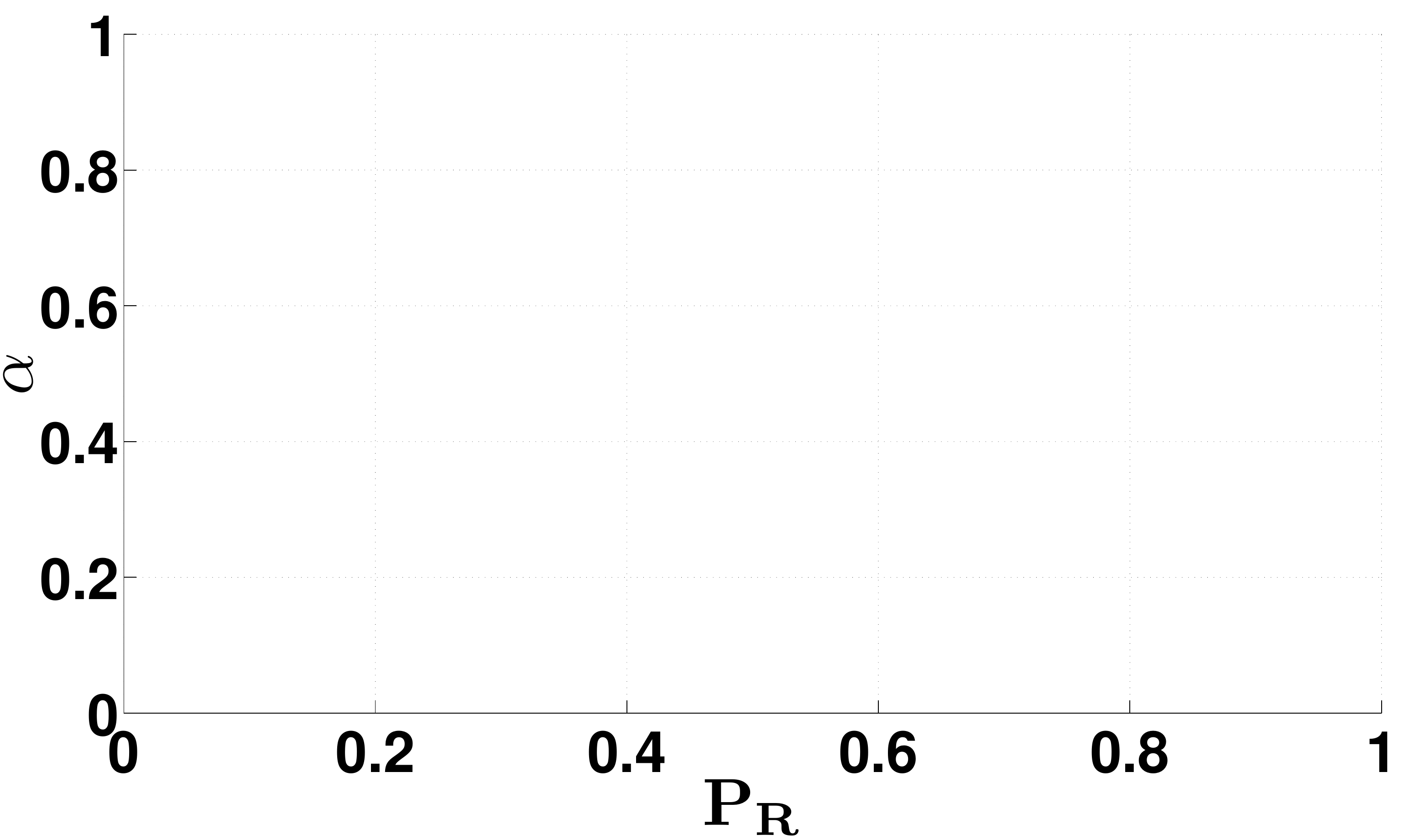}\label{fig:ba10}}\\
\caption{\small Range of $\alpha$ and $\PR$ for different selections of $\ND$ and $\NW$ when $\lcol = \lsucc$. The ranges are qualitatively similar for $\lcol > \lsucc$.}
\label{fig:PDandAlpha_ls_eq_lc}
\end{figure}

\begin{figure}[t]
\subfloat[$N_{D}=2,N_{W}=2$, Region $\TO$ prefers cooperation.]{\includegraphics[width = 0.32\columnwidth]{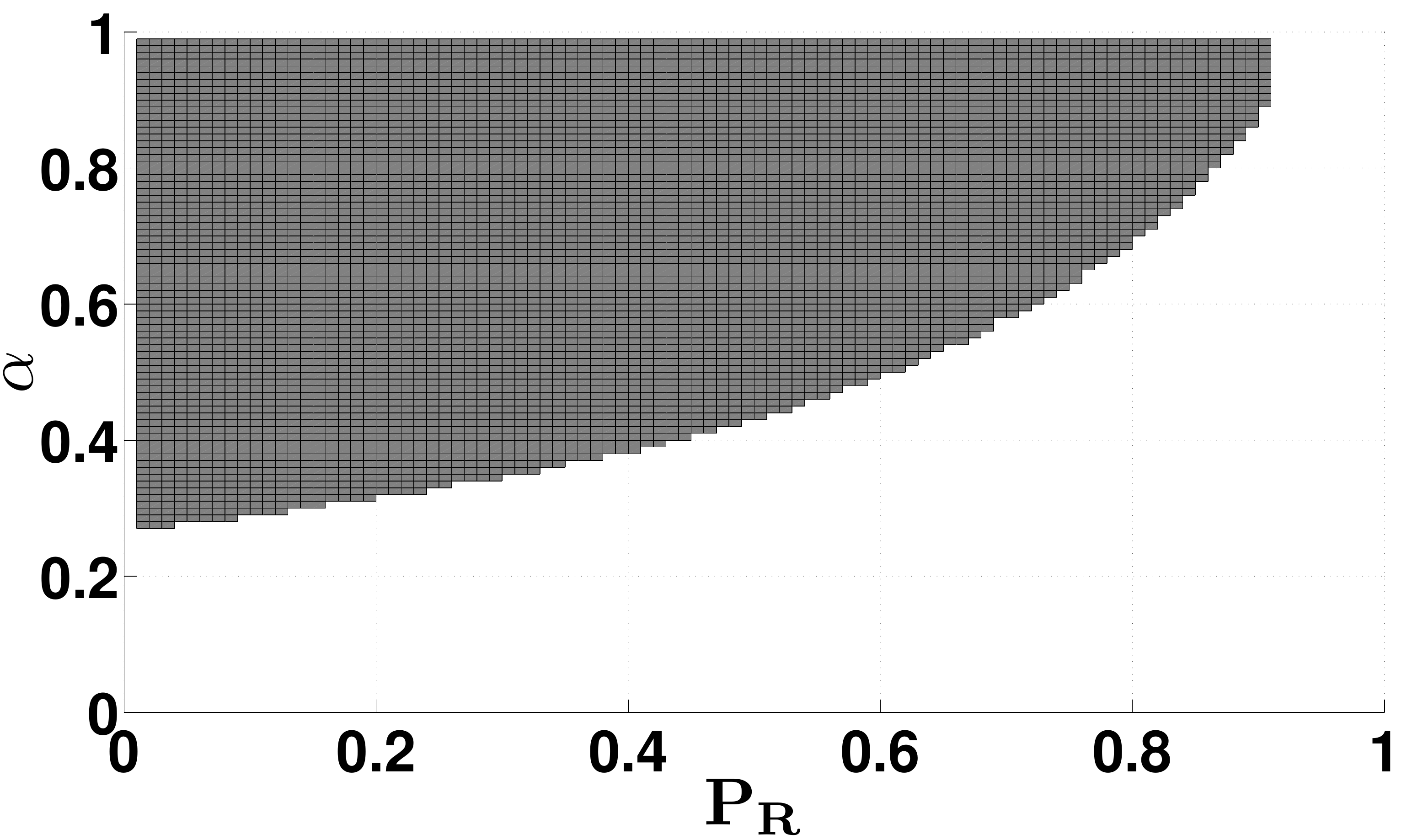}\label{fig:rtsTON2}}
\enspace
\subfloat[$N_{D}=2,N_{W}=2$, Region $\AO$ prefers cooperation.]{\includegraphics[width = 0.32\columnwidth]{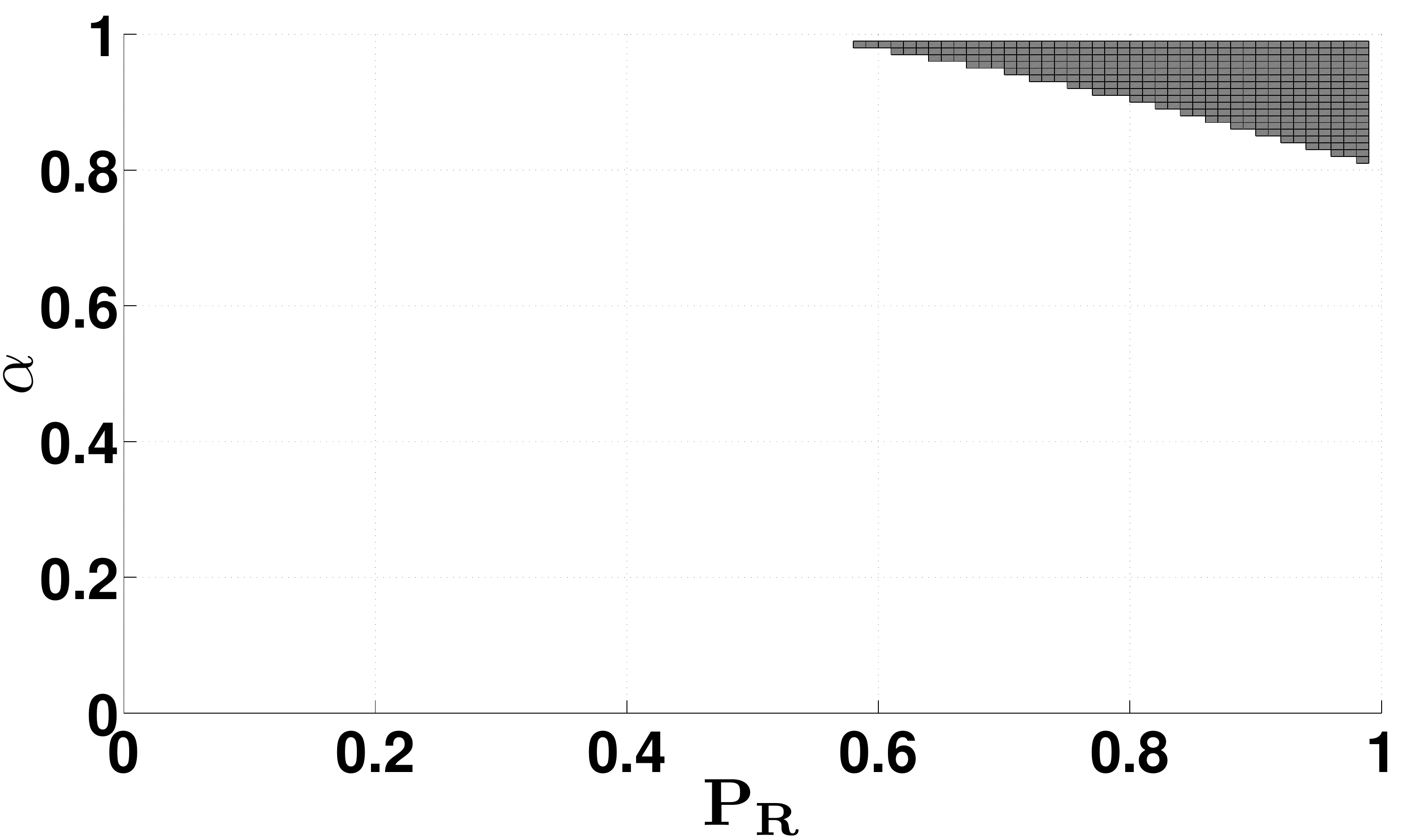}\label{fig:rtsAON2}}
\enspace
\subfloat[$N_{D}=2,N_{W}=2$, Region cooperation is self-enforceable.]{\includegraphics[width = 0.32\columnwidth]{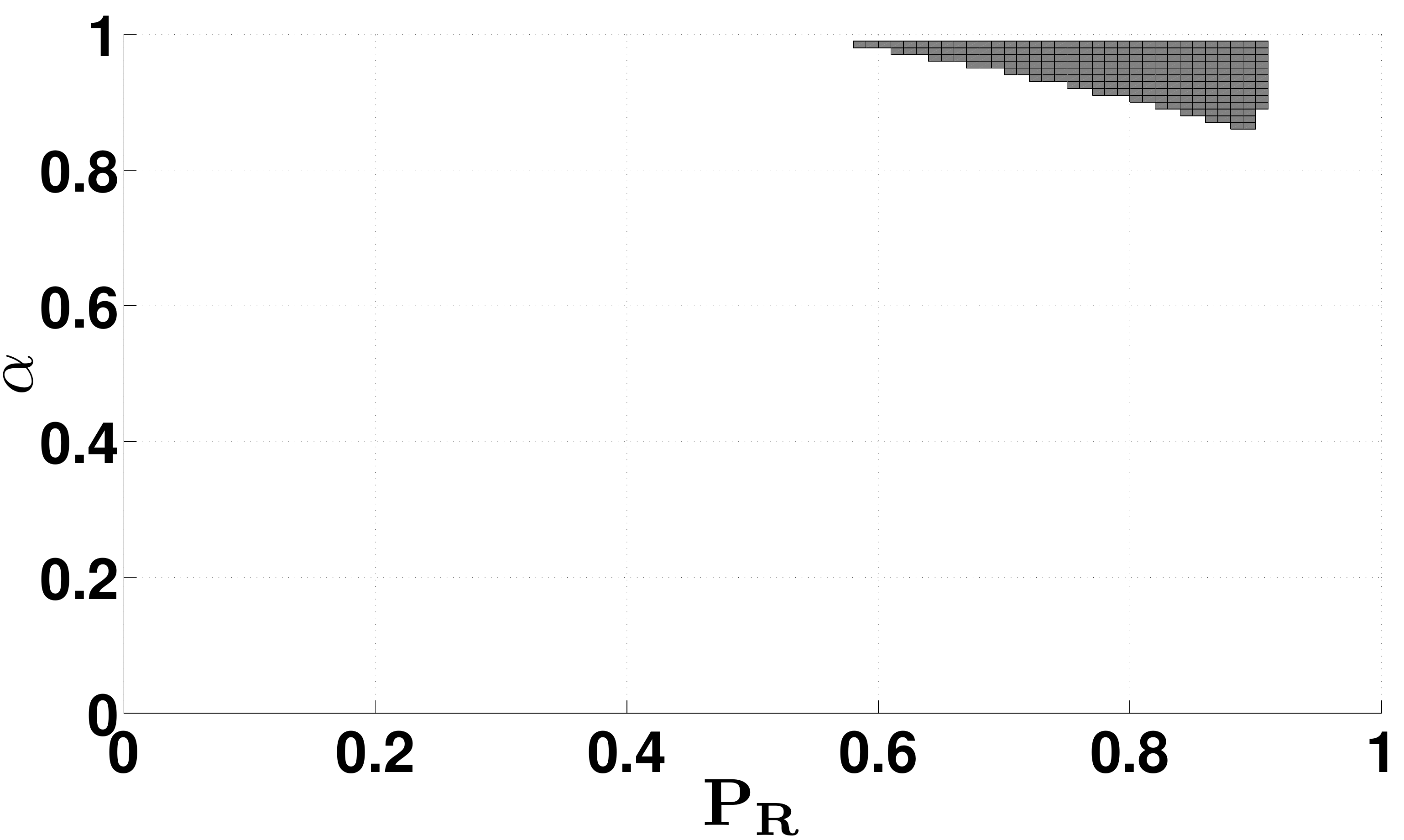}}\\
\subfloat[$N_{D}=10,N_{W}=10$, Region $\TO$ prefers cooperation.]{\includegraphics[width = 0.32\columnwidth]{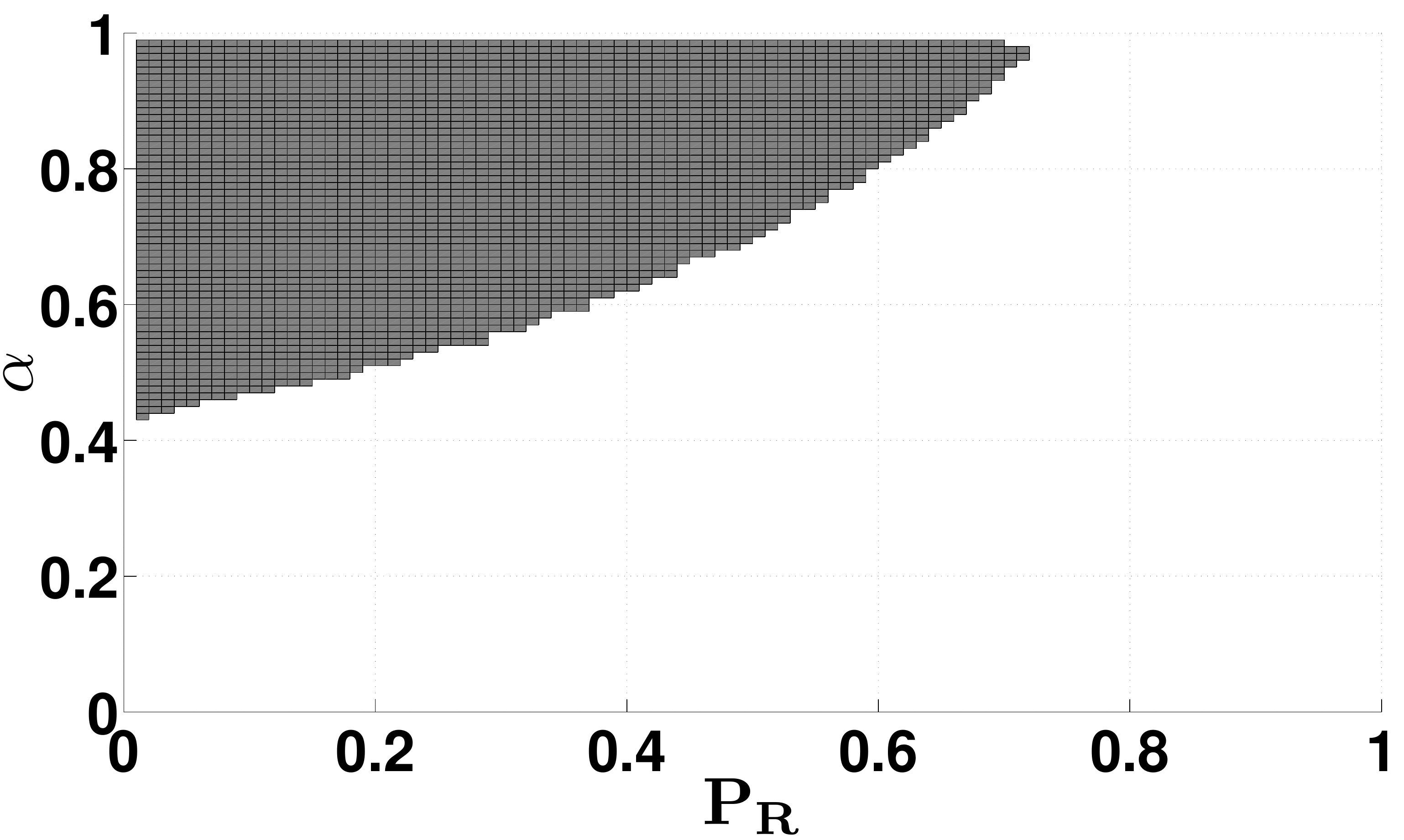}\label{fig:rtsTON10}}
\enspace
\subfloat[$N_{D}=10,N_{W}=10$, Region $\AO$ prefers cooperation.]{\includegraphics[width = 0.32\columnwidth]{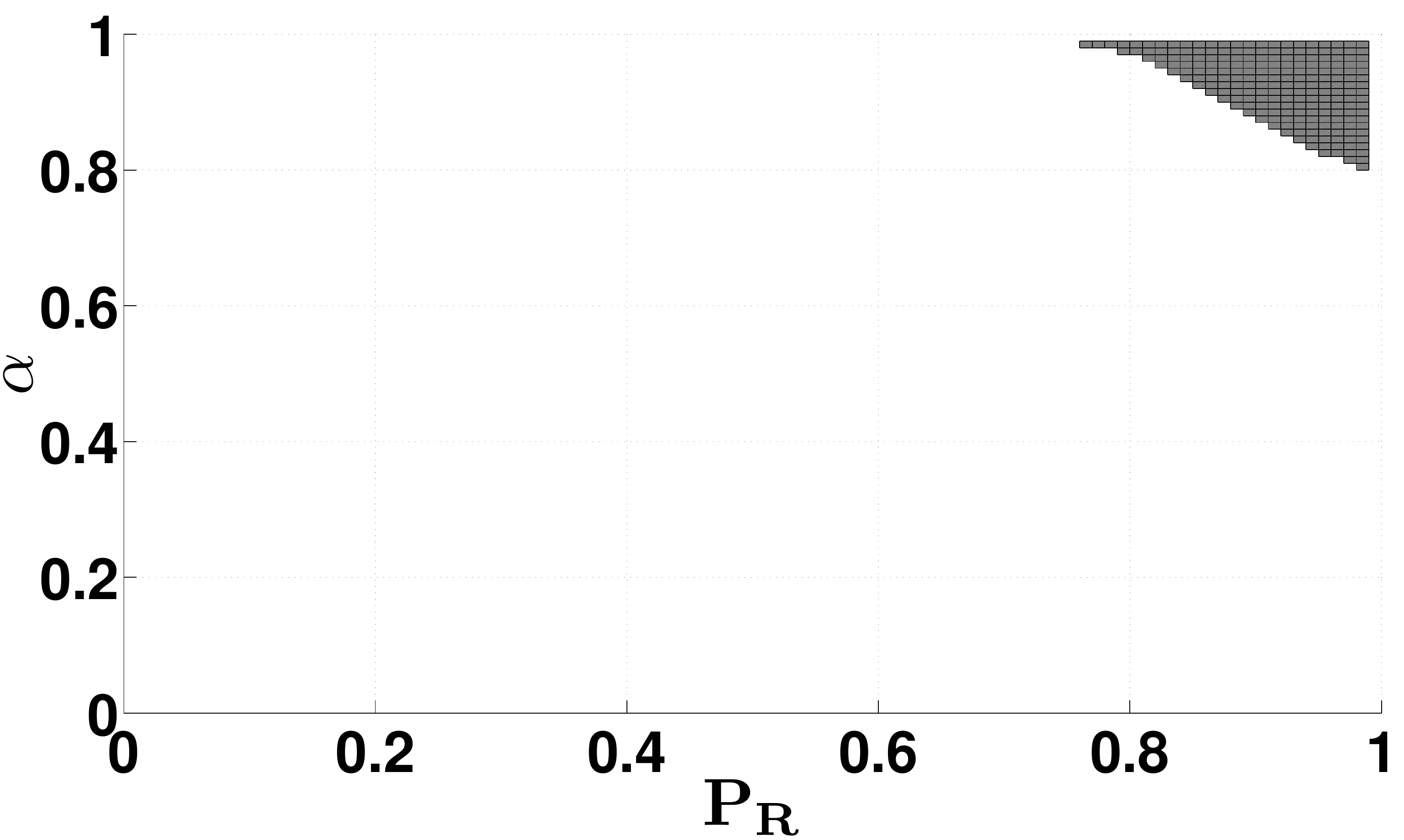}\label{fig:rtsAON10}}
\enspace
\subfloat[$N_{D}=10,N_{W}=10$, Region cooperation is self-enforceable.]{\includegraphics[width = 0.32\columnwidth]{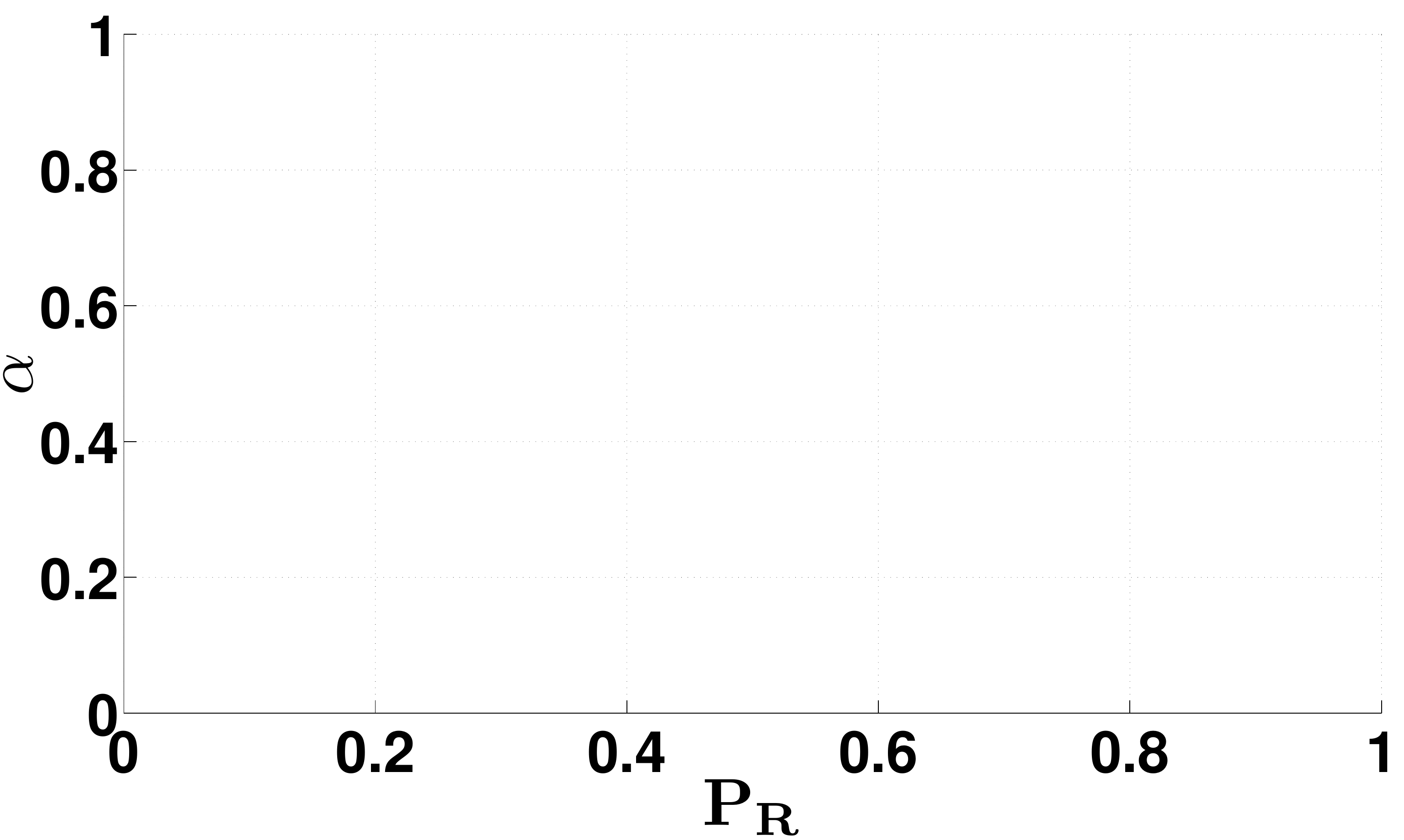}}\\
\caption{\small Range of $\alpha$ and $\PR$ for different selections of $\ND$ and $\NW$ when $\lsucc > \lcol$ ($\lcol = 0.1\lsucc$).}
\label{fig:PDandAlpha_ls_gr_lc}
\end{figure}
\textbf{\textit{When is cooperation self-enforceable?}} Figures~\ref{fig:PDandAlpha_ls_eq_lc} and~\ref{fig:PDandAlpha_ls_gr_lc} show the values of $\alpha$ and $\PR$ for which cooperation is self-enforceable, for when $\lsucc=\lcol$ and $\lsucc>\lcol$, respectively. We consider different selections of $\ND$ and $\NW$. We are interested in the values of $\alpha$ and $\PR$ that satisfy the inequalities~(\ref{Eq:Inq_1})-(\ref{Eq:Inq_2}) and~(\ref{Eq:Inq_3})-(\ref{Eq:Inq_4}). We observe that the range of $\alpha$ and $\PR$ over which cooperation is self-enforceable reduces as the numbers of nodes in the networks increase. Next we discuss the cases $\lsucc\leq\lcol$ and $\lsucc > \lcol$ in detail.

\textit{Case I: When $\lsucc\leq\lcol$:} Figures~\ref{fig:baTON2} and~\ref{fig:baAON2} show the values of $\alpha$ and $\PR$ for which the $\TO$ and the $\AO$, respectively, prefer cooperation to competition. Both networks have two nodes each. The values in Figure~\ref{fig:baTON2} are the set of $(\alpha,\PR)$ that satisfy~(\ref{Eq:Inq_2}) and (\ref{Eq:Inq_4}) and those in Figure~\ref{fig:baAON2} satisfy~(\ref{Eq:Inq_1}) and~(\ref{Eq:Inq_3}). As discussed earlier in the context of Figure~\ref{fig:Gain}, the $\AO$ prefers cooperation when $\lsucc\leq\lcol$ while the $\TO$ prefers competition. This explains the larger region of $(\alpha,\PR)$ in Figure~\ref{fig:baAON2} when compared to Figure~\ref{fig:baTON2}. Figure~\ref{fig:ba2} shows the values $(\alpha,\PR)$ for which both the networks prefer cooperation. The resulting region is an intersection of the regions in Figures~\ref{fig:baTON2} and~\ref{fig:baAON2}. For the values in Figure~\ref{fig:ba2}, all the Equations~(\ref{Eq:Inq_1}),~(\ref{Eq:Inq_2}),~(\ref{Eq:Inq_3}), and~(\ref{Eq:Inq_4}) are satisfied.

Similar to the figures described above, Figures~\ref{fig:baTON5},~\ref{fig:baAON5}, and~\ref{fig:ba5} show the regions of values, respectively, for which the $\TO$ prefers cooperation, the $\AO$ prefers cooperation, and both networks prefer cooperation. Each network now has five instead of two nodes. The larger number of nodes makes cooperation attractive for the $\AO$ over a larger range of $\alpha$ and $\PR$ (compare Figures~\ref{fig:baAON2} and~\ref{fig:baAON5}). The range of values, however, shrinks for the $\TO$. This is explained by the fact that as the number of $\AO$ nodes increases, as shown in Figure~\ref{fig:freqTd0}, the frequency of $\tauD^{*}=0$ increases, giving the $\TO$ greater contention free access when competing and making cooperation less favourable. The result is a smaller region of values, shown in Figure~\ref{fig:ba5}, over which cooperation is self-enforceable.

Figures~\ref{fig:baTON10},~\ref{fig:baAON10}, and~\ref{fig:ba10} show the regions for when the networks have ten nodes each. As is clear, the region corresponding to $\AO$ further increases, while that corresponding to the $\TO$ almost disappears, and so does the region over which cooperation is self-enforceable.

\textit{Case II: When $\lsucc>\lcol$:} Figure~\ref{fig:PDandAlpha_ls_gr_lc} shows the regions over which the two networks prefer cooperation and the resulting region of values $(\alpha,\PR)$ for which cooperation is self-enforceable. We show the regions for when both the networks have two and ten nodes each. In contrast to when $\lsucc\le \lcol$, we see that the region over which the $\AO$ prefers cooperation shrinks. Also the $\TO$ prefers cooperation over a range of values, which decreases as the number of nodes increases. This is explained by the fact that the $\AO$, as shown in Figure~\ref{fig:freqTd1}, attempts access with probability $1$ with higher frequency as networks grow in size, making competition better for the $\AO$.
\section{Conclusion}
\label{sec:conclusion}
We formulated a repeated game to model coexistence between an $\AO$ and a $\TO$. The $\AO$ desires a small age of updates while the $\TO$ desires a large throughput. The networks could either compete, that is play the mixed strategy Nash equilibrium in every stage of the repeated game, or cooperate by following recommendations in every stage from a randomized signalling device to access the spectrum in a non-interfering manner. The networks when cooperating employed the grim trigger strategy, which had both the networks play the MSNE in all stages following a stage in which a network disobeyed the device. This ensured that the networks would disobey the device only if they found competing to be more beneficial than cooperating in the long run.

Having modeled competition and cooperation, together with the grim trigger strategy, we investigated if cooperation between the networks was self-enforceable. For this we checked if and when the cooperation strategy profile was a subgame perfect equilibrium. We considered two cases of practical interest (a) when $\lsucc \leq \lcol$ and (b) when $\lsucc > \lcol$. We showed that while cooperation is self-enforceable when networks have a small number of nodes, networks prefer competing when they grow in size.
\begin{spacing}{}
\bibliographystyle{IEEEtran}
\bibliography{references}
\end{spacing}
\appendix
\subsection{Mixed Strategy Nash Equilibrium (MSNE)}
\label{sec:appendix_1}
We define $\vect{\tau^{*}} = [\tauD^{*},\tauW^{*}]$ as the parameter required to compute the mixed strategy Nash equilibrium of the one-shot game. We begin by finding the $\tauD^{*}$ of the $\AO$ by solving the optimization problem 
\begin{equation}
\begin{aligned}
\textbf{OPT I:}\quad& \underset{\tauD}{\text{minimize}}
& & \uD{\mathbbm{NC}} \\
& \text{subject to}
& & 0 \leq \tauD \leq 1.
\end{aligned}
\label{Eq:opt1}
\end{equation}
where, $\uD{\mathbbm{NC}}$ is the payoff of the $\AO$ defined as
\begin{align}
\uD{\mathbbm{NC}} & = (1-\tauD(1-\tauD)^{(\ND-1)}(1-\tauW)^{\NW})\AvginitAoI\nonumber\\
&+ (1-\tauD)^{\ND}(1-\tauW)^{\NW}(\lidle-\lcol) + \lcol\nonumber\\
&+ (\ND\tauD(1-\tauD)^{(\ND-1)}(1-\tauW)^{\NW}\nonumber\\
&+ \NW\tauW(1-\tauW)^{(\NW-1)}(1-\tauD)^{\ND})(\lsucc-\lcol).
\label{Eq:stage_payoff_aon}
\end{align}
The Lagrangian of the optimization problem~(\ref{Eq:opt1}) is
\begin{align*}
\mathcal{L}(\tauD,\mu) = & \uD{\mathbbm{NC}} -\mu_{1}\tauD + \mu_{2}(\tauD-1).
\end{align*}
where $\vect{\mu} = [\mu_{1},\mu_{2}]^{T}$ is the Karush-Kuhn-Tucker (KKT) multiplier vector. The first derivative of the objective function in~(\ref{Eq:opt1}) is
\footnotesize
{\begin{align*}
\uD{\mathbbm{NC}}{}' & = -\AvginitAoI(1-\tauW)^{\NW}[(1-\tauD)^{(\ND-1)} - (\ND-1)\nonumber\\
&\hspace{-1.5em}\tauD(1-\tauD)^{(\ND-2)}] + (\lsucc-\lcol)[(1-\tauW)^{\NW}(\ND(1-\tauD)^{(\ND-1)}\nonumber\\
&\hspace{-1.5em}-\ND(\ND-1)\tauD(1-\tauD)^{(\ND-2)})-\NW\tauW(1-\tauW)^{(\NW-1)}\nonumber\\
&\hspace{-1.5em}(1-\tauD)^{\ND-1}]-(\lidle-\lcol)\ND(1-\tauW)^{\NW}(1-\tauD)^{(\ND-1)}.
\end{align*}}
\normalsize
The KKT conditions can be written as
\begin{subequations}
\begin{align}
\uD{\mathbbm{NC}}{}'-\mu_{1} + \mu_{2} &= 0,\label{Eq:stationarity}\\
-\mu_{1}\tauD &= 0,\label{Eq:complementarySlackness_1}\\
\mu_{2}(\tauD-1) &= 0,\label{Eq:complementarySlackness_2}\\
-\tauD &\leq 0,\\
\tauD-1 &\leq 0,\\
\vect{\mu} = [\mu_{1},\mu_{2}]^{T} &\geq 0.\label{Eq:complementarySlackness_3}
\end{align}
\end{subequations}
We consider three cases. In case (i), we consider $\mu_{1} = \mu_{2} = 0$. From the stationarity condition~(\ref{Eq:stationarity}), we get
\begin{align}
\tauD = \frac{(1-\tauW)(\AvginitAoI-\ND(\lsucc-\lidle))+\ND\NW\tauW(\lsucc-\lcol)}{\left(\splitfrac{(1-\tauW)\ND(\AvginitAoI+(\lidle-\lcol)-\ND(\lsucc-\lcol))}{+\ND\NW\tauW(\lsucc-\lcol)}\right)}.
\label{Eq:tauD}
\end{align}
In case (ii) we consider $\mu_{1} \geq 0, \mu_{2} = 0$. Again, using~(\ref{Eq:stationarity}), we get $\mu_{1} = \uD{\mathbbm{NC}}{}'$. From~(\ref{Eq:complementarySlackness_3}), we have $\mu_{1}\geq 0$, therefore, $\uD{\mathbbm{NC}}{}' \geq 0$. On solving this inequality on $\uD{\mathbbm{NC}}{}'$ we get, $\AvginitAoI \leq \AvginitAoITh{}{\text{th},0}$, where $\AvginitAoITh{}{\text{th},0} = \ND(\lsucc - \lidle) - \frac{\ND\NW\tauW(\lsucc-\lcol)}{(1-\tauW)}$.

Finally, in case (iii) we consider $\mu_{1} = 0, \mu_{2} \geq 0$. On solving~(\ref{Eq:stationarity}), we get $\AvginitAoI \leq \AvginitAoITh{}{\text{th},1}$, where $\AvginitAoITh{}{\text{th},1} = \ND(\lsucc - \lcol)$.

Therefore, the solution from the KKT condition is %~(\ref{Eq:Age_MSNE_RTS_CTS}).
\scriptsize{
\begin{align}
&\tauD^{*} = 
    \begin{cases}
     \hspace{-0.5em}
      \begin{aligned}
 	  \frac{(1-\tauW^{*})(\AvginitAoI-\ND(\lsucc-\lidle))+\ND\NW\tauW^{*}(\lsucc-\lcol)}{\left(\splitfrac{(1-\tauW^{*})\ND(\AvginitAoI+(\lidle-\lcol)-\ND(\lsucc-\lcol))}{+\ND\NW\tauW^{*}(\lsucc-\lcol)}\right)}
 	  \end{aligned}
	  \vspace{1em}
      &\hspace{-0.75em}
      \begin{aligned}
      &\AvginitAoI> \AvginitAoITh{}{\text{th}},
      \end{aligned}\\
      1&\hspace{-8em}
      \begin{aligned}
      \AvginitAoI\leq \AvginitAoITh{}{\text{th}}\text{ \& }\AvginitAoITh{}{\text{th}} = \AvginitAoITh{}{\text{th},1},
      \end{aligned}\\
      0&\hspace{-8em}
      \begin{aligned}
      \AvginitAoI\leq \AvginitAoITh{}{\text{th}}\text{ \& }\AvginitAoITh{}{\text{th}} = \AvginitAoITh{}{\text{th},0}.
      \end{aligned}
    \end{cases}
\label{Eq:KKT_OPT1_temp}
\end{align}}
\normalsize
where, $\AvginitAoITh{}{\text{th}} = \max\{\AvginitAoITh{}{\text{th},0},\AvginitAoITh{}{\text{th},1}\}$. 
Under the assumption that length of successful transmission is equal to the length of collision, i.e., $\lsucc = \lcol$,~(\ref{Eq:KKT_OPT1_temp}) reduces to %~(\ref{Eq:Age_MSNE_BA}).
\begin{align}
\tauD^{*} = 
\begin{cases}
  \begin{aligned}
  \frac{\ND(\lidle-\lsucc)+\AvginitAoI}{\ND(\lidle-\lcol+\AvginitAoI)}
  \end{aligned}
  &
  \begin{aligned}
  &\AvginitAoI>\ND(\lsucc-\lidle),
  \end{aligned}\\
  0 &\text{ otherwise }.
\end{cases}
\label{Eq:KKT_OPT1}
\end{align}

Similarly, we find $\tauW^{*}$ for the $\TO$ by solving the optimization problem
\begin{equation}
\begin{aligned}\textbf{OPT II:}\quad& \underset{\tauW}{\text{minimize}}
& & -\uW{\mathbbm{NC}} \\
& \text{subject to}
& & 0 \leq \tauW \leq 1.
\end{aligned}
\label{Eq:opt2}
\end{equation}
where, $\uW{\mathbbm{NC}}$ is the payoff of the $\TO$ defined as
\begin{align}
\uW{\mathbbm{NC}} & = \tauW(1-\tauW)^{(\NW-1)}(1-\tauD)^{\ND}\lsucc.\label{Eq:stage_payoff_ton}
\end{align}
The Lagrangian of the optimization problem~(\ref{Eq:opt2}) is
\begin{align*}
\mathcal{L}(\tauW,\mu) = & -\uW{\mathbbm{NC}} -\mu_{1}\tauW + \mu_{2}(\tauW-1).
\end{align*}
where $\vect{\mu} = [\mu_{1},\mu_{2}]^{T}$ is the KKT multiplier vector. The first derivative of $\uW{\mathbbm{NC}}$ is
\begin{align*}
\uW{\mathbbm{NC}}{}' &= (1-\tauD)^{\ND}(1-\tauW)^{(\NW-1)}\lsucc\\
& - (\NW-1)\tauW(1-\tauW)^{(\NW-2)}(1-\tauD)^{\ND}\lsucc.
\end{align*}
The KKT conditions can be written as
\begin{subequations}
\begin{align}
-\uW{\mathbbm{NC}}{}' -\mu_{1} + \mu_{2} &= 0,\label{Eq:stationarityWiFi}\\
-\mu_{1}\tauW &= 0,\label{Eq:complementarySlacknessWiFi_1}\\
\mu_{2}(\tauW-1) &= 0,\label{Eq:complementarySlacknessWiFi_2}\\
-\tauW &\leq 0,\\
\tauW-1 &\leq 0,\\
\vect{\mu} = [\mu_{1},\mu_{2}]^{T} &\geq 0.\label{Eq:complementarySlacknessWiFi_3}
\end{align}
\end{subequations}
%We consider the case when $\mu_{1} = \mu_{2} = 0$. From the~(\ref{Eq:stationarityWiFi}), we get $\uW{\mathbbm{NC}}{}' = 0$. On solving the stationarity condition, we get $\tauW^{*} = 1/\NW$, which is also the solution of the KKT conditions.

We consider three cases. In case (i), we consider $\mu_{1} = \mu_{2} = 0$. From the stationarity condition in~(\ref{Eq:stationarityWiFi}), we get $\uW{\mathbbm{NC}}{}' = 0$. On solving~(\ref{Eq:stationarityWiFi}), we get $\tauW^{*} = 1/\NW$. In case (ii) we consider $\mu_{1}\geq 0, \mu_{2} = 0$. Again, using~(\ref{Eq:stationarityWiFi}), we get $\mu_{1} = -\uW{\mathbbm{NC}}{}'$. From~(\ref{Eq:complementarySlacknessWiFi_3}), we have $\mu_{1}\geq 0$, therefore, $\uW{\mathbbm{NC}}{}'\leq 0$. On solving this inequality on $\uW{\mathbbm{NC}}{}'$, we get $\tauW^{*} \geq 1/\NW$. Finally, in case (iii) we consider $\mu_{1} = 0, \mu_{2} \geq 0$ and on solving~(\ref{Eq:stationarityWiFi}) we get $\mu_{2} = \uW{\mathbbm{NC}}{}'$. Since $\mu_{2}\geq 0$ from~(\ref{Eq:complementarySlacknessWiFi_3}), we have $\uW{\mathbbm{NC}}{}'\geq 0$. On solving this inequality, we get $\tauW^{*} \leq 1/\NW$. 

Therefore, the solution from the KKT conditions is $\tauW^{*} = 1/\NW$.

Note that any Mixed Strategy Nash Equilibrium (MSNE) $(\tauD^{*},\tauW^{*})$ is a solution of the optimization problems OPT I and OPT II. All solutions to the OPT I and OPT II must satisfy the necessary KKT conditions. Since these conditions yield a unique solution $(\tauD^{*},\tauW^{*})$, this is the only MSNE.

\subsection{Optimal Strategy under Cooperation}
\label{sec:appendix_2}
We define $\vect{\widehat{\tau}} = [\tauDC,\tauWC]$ as the optimal strategy of the one-shot game when networks cooperate. We begin by finding the $\tauDC$ of the $\AO$ by solving the optimization problem 
\begin{equation}
\begin{aligned}
\textbf{OPT I:}\quad& \underset{\tauDC}{\text{minimize}}
& & \uD{\mathbbm{C}} \\
& \text{subject to}
& & 0 \leq \tauDC \leq 1.
\end{aligned}
\label{Eq:opt1_coop}
\end{equation}
where, $\uD{\mathbbm{C}}$ is the payoff of $\AO$ defined as
\begin{align*}
\uD{\mathbbm{C}}& = (1-\PR\tauD(1-\tauD)^{(\ND-1)})\AvginitAoI+\lcol\\
&+ (\PR(1-\tauD)^{\ND}+(1-\PR)(1-\tauW)^{\NW})(\lidle-\lcol)\\
&+ ((1-\PR)\NW\tauW(1-\tauW)^{(\NW-1)}\\
&+ \PR\ND\tauD(1-\tauD)^{(\ND-1)})(\lsucc-\lcol).
\end{align*}
The Lagrangian of the optimization problem~(\ref{Eq:opt1_coop}) is
\begin{align*}
\mathcal{L}(\tauD,\mu) = & \uD{\mathbbm{C}} -\mu_{1}\tauD + \mu_{2}(\tauD-1).
\end{align*}
where $\vect{\mu} = [\mu_{1},\mu_{2}]^{T}$ is the Karush-Kuhn-Tucker (KKT) multiplier vector. The first derivative of the objective function in~(\ref{Eq:opt1_coop}) is
\small{\begin{align*}
\uD{\mathbbm{C}}{}' & = -\PR\AvginitAoI[(1-\tauD)^{(\ND-1)} - (\ND-1)\tauD(1-\tauD)^{(\ND-2)}]\\
&\hspace{-2em}+(\lsucc-\lcol)\PR\ND[(1-\tauD)^{\ND-1}-(\ND-1)\tauD(1-\tauD)^{(\ND-2)}]\\
&\hspace{-2em}-(\lidle-\lcol)\PR\ND(1-\tauD)^{(\ND-1)}.
\end{align*}}
\normalsize
The KKT conditions can be written as
\begin{subequations}
\begin{align}
\uD{\mathbbm{C}}{}'-\mu_{1} + \mu_{2} &= 0,\label{Eq:stationarity_coop}\\
-\mu_{1}\tauD &= 0,\label{Eq:complementarySlackness_1_coop}\\
\mu_{2}(\tauD-1) &= 0,\label{Eq:complementarySlackness_2_coop}\\
-\tauD &\leq 0,\\
\tauD-1 &\leq 0,\\
\vect{\mu} = [\mu_{1},\mu_{2}]^{T} &\geq 0.\label{Eq:complementarySlackness_3_coop}
\end{align}
\end{subequations}
We consider three cases. In case (i), we consider $\mu_{1} = \mu_{2} = 0$. From the stationarity condition~(\ref{Eq:stationarity_coop}), we get
\begin{align}
\tauD = \frac{\AvginitAoI-\ND(\lsucc-\lidle)}{\ND(\AvginitAoI+(\lidle-\lcol)-\ND(\lsucc-\lcol))}.
\label{Eq:tauDC}
\end{align}
In case (ii) we consider $\mu_{1} \geq 0, \mu_{2} = 0$. Again, using~(\ref{Eq:stationarity_coop}), we get $\mu_{1} = \uD{\mathbbm{C}}{}'$. From~(\ref{Eq:complementarySlackness_3_coop}), we have $\mu_{1}\geq 0$, therefore, $\uD{\mathbbm{C}}{}' \geq 0$. On solving this inequality on $\uD{\mathbbm{C}}{}'$ we get, $\AvginitAoI \leq \AvginitAoITh{}{\text{th},0}$, where $\AvginitAoITh{}{\text{th},0} = \ND(\lsucc - \lidle)$.

Finally, in case (iii) we consider $\mu_{1} = 0, \mu_{2} \geq 0$. On solving~(\ref{Eq:stationarity_coop}), we get $\AvginitAoI \leq \AvginitAoITh{}{\text{th},1}$, where $\AvginitAoITh{}{\text{th},1} = \ND(\lsucc - \lcol)$.

Therefore, the solution from the KKT condition is %~(\ref{Eq:DSRC_optimal}).
\begin{align}
&\tauDC = 
    \begin{cases}
     \hspace{-0.5em}
      \begin{aligned}
 	  \frac{\AvginitAoI-\ND(\lsucc-\lidle)}{\ND(\AvginitAoI+(\lidle-\lcol)-\ND(\lsucc-\lcol))}
 	  \end{aligned}
	  \vspace{1em}
      &\hspace{-0.5em}
      \begin{aligned}
      &\AvginitAoI> \AvginitAoITh{}{\text{th}},
      \end{aligned}\\
      1&\hspace{-7em}
      \begin{aligned}
      \AvginitAoI\leq \AvginitAoITh{}{\text{th}}\text{ \& }\AvginitAoITh{}{\text{th}} = \AvginitAoITh{}{\text{th},1},
      \end{aligned}\\
      0&\hspace{-7em}
      \begin{aligned}
      \AvginitAoI\leq \AvginitAoITh{}{\text{th}}\text{ \& }\AvginitAoITh{}{\text{th}} = \AvginitAoITh{}{\text{th},0}.
      \end{aligned}
    \end{cases}
\label{Eq:KKT_OPT1_temp_coop}
\end{align}

where, $\AvginitAoITh{}{\text{th}} = \max\{\AvginitAoITh{}{\text{th},0},\AvginitAoITh{}{\text{th},1}\}$. 
Under the assumption that length of successful transmission is equal to the length of collision i.e. $\lsucc = \lcol$,~(\ref{Eq:KKT_OPT1_temp_coop}) reduces to
\begin{align}
\tauDC = 
\begin{cases}
  \begin{aligned}
  \frac{\ND(\lidle-\lsucc)+\AvginitAoI}{\ND(\lidle-\lcol+\AvginitAoI)}
  \end{aligned}
  &
  \begin{aligned}
  &\AvginitAoI>\ND(\lsucc-\lidle),
  \end{aligned}\\
  0 &\text{ otherwise }.
\end{cases}
\label{Eq:KKT_OPT1_coop}
\end{align}

Similarly, we find $\tauWC$ for the $\TO$ by solving the optimization problem
\begin{equation}
\begin{aligned}\textbf{OPT II:}\quad& \underset{\tauW}{\text{minimize}}
& & -\uW{\mathbbm{C}} \\
& \text{subject to}
& & 0 \leq \tauW \leq 1.
\end{aligned}
\label{Eq:opt2_coop}
\end{equation}
where, $\uW{\mathbbm{C}}$ is the payoff of $\TO$ defined as
\begin{align*}
\uW{\mathbbm{C}} & = (1-\PR)\tauW(1-\tauW)^{(\NW-1)}\lsucc.
\end{align*}
The Lagrangian of the optimization problem~(\ref{Eq:opt2_coop}) is
\begin{align*}
\mathcal{L}(\tauW,\mu) = & -\uW{\mathbbm{C}} -\mu_{1}\tauW + \mu_{2}(\tauW-1).
\end{align*}
where $\vect{\mu} = [\mu_{1},\mu_{2}]^{T}$ is the KKT multiplier vector. The first derivative of $\uW{\mathbbm{C}}$ is
\begin{align*}
\uW{\mathbbm{C}}{}' &= (1-\PR)\lsucc[(1-\tauW)^{(\NW-1)}\\
&\qquad\qquad- (\NW-1)\tauW(1-\tauW)^{(\NW-2)}].
\end{align*}
The KKT conditions can be written as
\begin{subequations}
\begin{align}
-\uW{\mathbbm{C}}{}' -\mu_{1} + \mu_{2} &= 0,\label{Eq:stationarityWiFi_coop}\\
-\mu_{1}\tauW &= 0,\label{Eq:complementarySlacknessWiFi_1_coop}\\
\mu_{2}(\tauW-1) &= 0,\label{Eq:complementarySlacknessWiFi_2_coop}\\
-\tauW &\leq 0,\\
\tauW-1 &\leq 0,\\
\vect{\mu} = [\mu_{1},\mu_{2}]^{T} &\geq 0.\label{Eq:complementarySlacknessWiFi_3_coop}
\end{align}
\end{subequations}
We consider three cases. In case (i), we consider $\mu_{1} = \mu_{2} = 0$. From the stationarity condition in~(\ref{Eq:stationarityWiFi_coop}), we get $\uW{\mathbbm{C}}{}' = 0$. On solving~(\ref{Eq:stationarityWiFi_coop}), we get $\tauWC = 1/\NW$. In case (ii) we consider $\mu_{1}\geq 0, \mu_{2} = 0$. Again, using~(\ref{Eq:stationarityWiFi_coop}), we get $\mu_{1} = -\uW{\mathbbm{C}}{}'$. From~(\ref{Eq:complementarySlacknessWiFi_3_coop}), we have $\mu_{1}\geq 0$, therefore, $\uW{\mathbbm{C}}{}'\leq 0$. On solving this inequality on $\uW{\mathbbm{C}}{}'$, we get $\tauWC \geq 1/\NW$. Finally, in case (iii) we consider $\mu_{1} = 0, \mu_{2} \geq 0$ and on solving~(\ref{Eq:stationarityWiFi_coop}) we get $\mu_{2} = \uW{\mathbbm{C}}{}'$. Since $\mu_{2}\geq 0$~(\ref{Eq:complementarySlacknessWiFi_3_coop}), we have $\uW{\mathbbm{C}}{}'\geq 0$. On solving this inequality, we get $\tauWC \leq 1/\NW$. 

Therefore, the solution from the KKT conditions is $\tauWC = 1/\NW$.

%We consider the case when $\mu_{1} = \mu_{2} = 0$. From the~(\ref{Eq:stationarityWiFi_coop}), we get $\uW{\mathbbm{C}}{}' = 0$. On solving the stationarity condition, we get $\tauWC = 1/\NW$, which is also the solution of the KKT conditions.

%\SG{Note that any Mixed Strategy Nash Equilibrium (MSNE) $(\tauDC,\tauWC)$ is a solution of the optimization problems OPT I and OPT II. All solutions to the OPT I and OPT II must satisfy the necessary KKT conditions. Since these conditions yield a unique solution $(\tauDC,\tauWC)$, this is the only MSNE.}
Note that all solutions to the optimization problems, OPT I and OPT II, must satisfy the necessary KKT conditions and since these conditions yield a unique solution $(\tauDC,\tauWC)$, this is the unique global solution.
\end{document}